\documentclass[12pt,aps,nofootinbib]{revtex4}

\usepackage{amsmath}
\usepackage{amssymb}
\usepackage{mathtools}
\usepackage{bbold}
\usepackage{verbatim}
\usepackage[titletoc,toc,page]{appendix}

\usepackage[utf8]{inputenc}
\usepackage{lmodern}
\usepackage{graphicx}

\usepackage{cancel}

\usepackage{mathrsfs}

\usepackage[hidelinks]{hyperref}

\usepackage{relsize,exscale}
\usepackage{tensor}
\usepackage[normalem]{ulem}

\usepackage[low-sup]{subdepth}
\usepackage[]{subdepth}

\usepackage[colorinlistoftodos]{todonotes}

\newcommand{\hypgeo}[2]{%
  \operatorname{%
    {\vphantom{\mathnormal{F}}}_{#1}%
    \kern-\scriptspace
    \mathnormal{F}_{#2}%
  }%
}

\usepackage{xcolor}

\usepackage{array}

\usepackage{pgfplots}
\usetikzlibrary{decorations.pathmorphing}
\usetikzlibrary{arrows}
\usetikzlibrary{patterns}
\usetikzlibrary{snakes}

\begin{document}

\title{Gradient corrections to the quantum effective action}

\author{Sofia Canevarolo
}
\email[]{s.canevarolo@uu.nl}

\author{Tomislav Prokopec}
\email[]{t.prokopec@uu.nl}
\affiliation{Institute for Theoretical Physics, Spinoza Institute
\& $\rm EMME\Phi$,
Faculty of Science, Utrecht University, Postbus 80.195, 3508 TD Utrecht,
The Netherlands \looseness=-1}


\maketitle
\date{\today}




\centerline{\textbf{Abstract}}

We derive the quantum effective action up to second order in gradients and up to two-loop order for an interacting scalar field theory. This expansion of the effective action is useful to study problems in cosmological settings where spatial or time gradients are important, such as bubble nucleation in first-order phase transitions. Assuming spacetime dependent background fields, we work in Wigner space and perform a midpoint gradient expansion, which is consistent with the equations of motion satisfied by the propagator. In particular, we consider the fact that the propagator is non-trivially constrained by an additional equation of motion, obtained from symmetry requirements. At one-loop order, we show the calculations for the case of a single scalar field and then generalise the result to the multi-field case. While we find a vanishing result in the single field case, the one-loop second-order gradient corrections can be significant when considering multiple fields. As an example, we apply our result to a simple toy model of two scalar fields with canonical kinetic terms and mass mixing at tree-level.  Finally, we calculate the two-loop one-particle irreducible (1PI) effective action in the single scalar field case, and obtain a nonrenormalisable result. The theory is rendered renormalisable by adding two-particle irreducible (2PI) counterterms, making the 2PI formalism the right framework for renormalization when resummed 1PI two-point functions are used in perturbation theory.


\tableofcontents
\section{Introduction}
\label{Intro}

In order to address unresolved questions regarding the Early Universe, it is often crucial to go beyond the classical description of the phenomena and account for their quantum physical nature. To do so, the quantum effective action is an important tool to consider. Indeed, it is often used to study phenomena,  such as spontaneous symmetry breaking mechanisms, as it allows to account for quantum corrections to the classical action while respecting all the symmetries of the theory. The effective action is the generating functional of the 1PI correlation functions, and thus it encodes the full quantum physics of a theory. It is a functional of the background fields, and its minima correspond to the vacuum expectation value of the quantum fields. It is therefore regarded as a powerful tool to estimate the quantum corrected stationary points of the theory. On the other hand, since the quantum effective action cannot be calculated in an exact way, we have to resort to perturbation theory where it can be evaluated by functional methods in a loop expansion~\cite{Coleman-Weinberg,Jackiw,Buchbinder:1992rb}. 

When the background fields are assumed to be spacetime independent, the effective action reduces to the effective potential, which at tree-level is simply given by all the non-derivative terms in the Lagrangian. The effective potential is widely used to study spontaneous symmetry breaking in the semi-classical approximation. Indeed, it allows to probe the structure of the minima of a theory, accounting for the radiative corrections up to desired order in loop expansion. This is of particular interest for theories such as the scale-free (conformal) extensions of the standard model (SM) \cite{Chataignier:2018aud,Chataignier:2018kay,Rezacek, Carone, Englert,Mohamadnejad}. These models are well-motivated candidates in which a classical scaling symmetry prohibits any mass-dimensionful term in the tree-level Lagrangian. Therefore, the negative mass term present in the scalar sector of the SM is absent in the Lagrangian of the conformal models. The breaking of the conformal symmetry is then realized by radiative corrections {\it via} the Coleman-Weinberg mechanism \cite{Coleman-Weinberg}. In these models, the effective potential is essential to find the non-vanishing expectation value of the scalar fields produced by the quantum corrections and, in turn, the mass terms of the other particles.

It is therefore clear that the effective potential is a powerful tool which, nonetheless, is not enough to describe all the dynamical phenomena in cosmology~\cite{Janssen, Koksma}. Indeed, in an evolving universe the assumption on constant background fields often shows to be too restrictive, as including spatial and/or time gradients is crucial for achieving a complete description of the dynamics. It is therefore of outstanding importance to compute quantum corrections to the kinetic terms and understand in which cases these corrections are quantitatively significant. 

An important example of application is provided by the Electroweak Phase Transition where the effective action is used to calculate the tunneling rate from the false to the true vacuum state of the theory. In this scenario, quantum corrections to the effective potential are necessary to correctly estimate the minima of the potential. However, this alone is not sufficient to compute the tunneling path of the fields. Indeed, the full effective action is needed to obtain the correct field configuration, known as the bounce, that describes the tunneling dynamics. Similarly, we expect a wide range of instances within the history of the Universe when the gradient terms of the scalar fields cannot be neglected: sphaleron and instantons
calculations~\cite{Klinkhamer,Klinkhamer 2}, baryogenesis~\cite{Sakharov}, 
 $\alpha-$attractor inflationary models~\cite{Kallosh-Linde-Roest,Galante-Linde-Roest,Carrasco-Linde-Roest}, 
 and multifield inflationary models~\cite{Dvali,Bernardeau:2002jy,Rigopoulos:2005us,Senatore:2010wk,Kaiser:2012ak,Liu-Prokopec,Barnaveli} are some other important examples of applications.
 
To properly account for the quantum corrections, a derivative expansion of the effective action can be performed. This is valid when the background fields are weakly dependent on the spacetime or, equivalently, for soft momenta flowing in the vertices. In the literature, attempts in this direction are already present~\cite{Fraser,Aitchison-Fraser,Aitchison-Fraser2,Chan,Iliopoulos-Itzykson-Martin}. Using a different technique, we perform a gradient expansion of the effective action at one- and two-loop orders. We expand the scalar propagator in Wigner space using the midpoint and relative coordinates and check that both the symmetric propagator equations of motion are satisfied. This allows us to obtain second-order gradient corrections for the scalar kinetic terms and check if it can compete with the quantum corrections of the effective potential of the same loop order. 

In Section~\ref{Sec.II}, we start with the simple case of a single real scalar field. In Section~\ref{Sec.III}, we then generalise the one-loop calculations to a theory with a generic number of real scalar fields with canonical kinetic terms, but with mass-mixing interactions in the potential. The one-loop corrections to the effective potential are divergent and ought to be renormalised, breaking scaling symmetry and introducing an unphysical scale dependence in the effective potential. It is therefore interesting
to understand if a similar behaviour is also found when considering the kinetic corrections. We apply the one-loop results to a simple toy model composed by two real scalar fields. Finally, in Section~\ref{Sec.IV} we use the gradient expanded propagator to obtain the two-loop effective action in the single scalar field case. We renormalise the theory considering both the 1PI and the two-particle irredicile (2PI) formalism. The theory is nonrenormalisable 
in the 1PI formalism, implying that the 2PI is the correct framework to use in our context. Our discussion and conclusions are in Section~\ref{Sec.V}.  A detailed comparison with the literature and some technical aspects 
of the calculations are presented in four extensive appendices.

\subsection*{Quantum effective action}

We consider the tree-level action to be a functional of $n$ real scalar fields $\Phi_a(x)\;(a=1,2,\cdots n)$.
The background field method posits that the effective action $\Gamma[\phi_a]$ is obtained by
inserting into the classical action the field written as $\Phi_a(x)=\phi_a(x) +\varphi_a(x)$, with $\phi_a(x)$ being the background fields, and integrating over 
the fluctuating fields $\varphi_a(x)$ according to
$\exp(i\Gamma[\phi_a]) = \int {\cal D}\varphi_a {\exp}(iS[\phi_a+\varphi_a])$. 
The thus-obtained effective action is a nonlocal functional of $\phi_a(x)$, and it is not known how to obtain it for general $\phi_a(x)$. In this work, we construct a framework which can be used to build an approximate quantum effective action by utilising standard perturbative loop expansion
and a gradient expansion, which is valid for slowly varying background fields. 

To compute the one-loop contribution to the effective action $\Gamma^{(1)}[\phi_{a}]$, we start from the following expression,
\begin{eqnarray}
    &\Gamma[\phi_{a}]\!=\! S[\phi_{a}]+\Gamma^{(1)}[\phi_{a}]+\Gamma^{(2)}[\phi_{a}]+ {\cal O}(\hbar ^3),
\label{effective action}\\
 &\Gamma^{(1)}[\phi_{a}] \!=\! \hbar\frac{i}{2} \text{Tr} 
             \log\big[ \mathcal{D}_{ab}[\phi_c](x;x')\big] 
\,,
\label{1 loop effective action}
\end{eqnarray}
with $S[\phi_{a}]$ the tree-level (classical) action, expressed in terms of the 
background fields $\phi_{a}$, and $\Gamma^{(2)}[\phi_{a}]$ the two-loop contribution which is explicitly given in Section~\ref{Sec.IV}. The tree-level two-point vertex function $\mathcal{D}_{ab}$
in equation~(\ref{1 loop effective action}) is obtained by varying the tree-level action,
\begin{equation}
    \mathcal{D}_{ab}[\phi_c](x;x')
    =\frac{\delta^2 S[\Phi_c]}{\delta \Phi_a(x) \delta \Phi_b (x')} \bigg|_{\Phi_{c}=\phi_{c}},
\label{tree level operator: def}
\end{equation}
where we stress that the tree-level two-point vertex function 
is evaluated on the background fields. The tree-level propagator $i\Delta_{cb}$ is defined as the 
operator inverse of the tree-level two-point vertex function,
\begin{equation}
    \int {\rm d}^Dy\; \mathcal{D}_{ac}[\phi_e](x;y) i\Delta_{cb}[\phi_e](y;x')
         =\delta_{ab} i \delta^{D} (x\!-\!x')
\,,
\label{prop eq of motion}
\end{equation}
where $D$ is the number of dimensions, summation over repeated field indices is assumed,
and we set $\hbar=1$.
Using the previous definition, equation~(\ref{1 loop effective action}) can be recast as, 
\begin{equation}
 \Gamma^{(1)}[\phi_{a}] = - \frac{i}{2} \text{Tr} 
             \log\big[ \Delta_{ab}[\phi_c](x;x')\big] 
\,.
\label{1 loop effective action 2}
\end{equation}
The Feynman propagator is a two-point function with time-ordered product of fields, 
and therefore it satisfies the following symmetry requirement:
\begin{equation}
    i\Delta_{ac}[\phi_e](x;x')=i\Delta_{ca}[\phi_e](x';x)
\,.
\label{propagator: symmetry requirement}
\end{equation}
In the multifield case, the propagator is invariant under the simultaneous exchange of points 
$x \leftrightarrow x'$ and transposition in flavour $a \leftrightarrow c$~\footnote{ 
Needless to say, in the single field case, the propagator is invariant under the exchange of the points.}.
This transposition property follows immediately from the definition of the propagator, 
\begin{equation}
i\Delta_{ab}[\phi_c](x;x') = \theta(x^0\!-\!{x'}^0) i\Delta_{ab}^{(+)}[\phi_c](x;x')
      +\theta({x'}^0\!-\!x^0) i\Delta_{ab}^{(-)}[\phi_c](x;x')
\,,
\label{Feynman propagator: definition}
\end{equation}
where $ i\Delta^{(\pm)}[\phi_c](x;x')$ are the positive and negative frequency Wightman functions,
\begin{eqnarray}
i\Delta^{(+)}_{ab}[\phi_c](x;x') 
        &\!=\!&\langle\Omega|\hat\phi_a(x)\hat \phi_b(x')|\Omega\rangle
\,,\quad 
\nonumber\\
i\Delta^{(-)}_{ab}[\phi_c](x;x') 
 &\!=\!&\langle\Omega|\hat\phi_b(x')\hat \phi_a(x)|\Omega\rangle
 = 
 i\Delta^{(+)}_{ba}[\phi_c](x';x)
\,,
\label{Wightman functions: definition}
\end{eqnarray}
and $|\Omega\rangle$ denotes the state of the system~\footnote{The definitions~(\ref{Wightman functions: definition}) are valid for pure states. For more general mixed
states, one replaces the expectation value in~(\ref{Wightman functions: definition}) with a trace weighted by the density operator.
}.
The symmetry property~(\ref{propagator: symmetry requirement}) differs from 
hermiticity, under which the Feynman propagator transforms into the Dyson propagator.

This symmetry property directly translates into another equation of motion,
\begin{equation}
    \int {\rm d}^Dy \; i\Delta_{bc}[\phi_e](x';y)\overleftarrow{\mathcal{D}}_{ca}[\phi_e](y;x) 
     =\delta_{ab} i\delta^{D}(x\!-\!x')
\,,
\label{symmetric prop eq of motion}
\end{equation}
which is the second equation which ought to be satisfied by the propagator.
To summarize, there are two equivalent ways of specifying the propagator:
(A) by Eqs.~(\ref{prop eq of motion}) and~(\ref{symmetric prop eq of motion}), or (B) by 
Eq.~(\ref{prop eq of motion}) and the symmetry 
requirement~(\ref{propagator: symmetry requirement}). 
In this paper we make ample use of this property of the propagator.

Taking the Lagrangian density to be a local function of the fields and its derivatives, then 
the vertex function is diagonal in position space, 
$ \mathcal{D}_{ac}[\phi_e](x;y)\rightarrow  \mathcal{D}_{ac}[\phi_e](x)\delta^D(x\!-\!y)$,
and the equations of motion~(\ref{prop eq of motion}) and~(\ref{symmetric prop eq of motion}) 
simplify to,
\begin{eqnarray}
      \mathcal{D}_{ac}[\phi_e](x)i\Delta_{cb}[\phi_e](x;x') &=& \delta_{ab} i \delta^{D} (x\!-\!x'),
\label{integrated prop eq of motion: 1}      \\
     i\Delta_{bc}[\phi_e](x;x')\overleftarrow{\mathcal{D}}_{ca}[\phi_e](x')  &=& \delta_{ab} i \delta^{D} (x\!-\!x')
     \,,
\label{integrated prop eq of motion}
\end{eqnarray}
where $\mathcal{D}_{ac}[\phi_e](x)=\mathcal{D}_{ca}[\phi_e](x)$ is a 
symmetric operator (invariant under transposition), which follows 
from its definition~(\ref{tree level operator: def}).
Clearly, in this simpler case, 
Eqs.~(\ref{integrated prop eq of motion: 1}--\ref{integrated prop eq of motion}) 
simultaneously hold true for the propagator, 
and are equivalent to Eqs.~(\ref{integrated prop eq of motion: 1})
and~(\ref{propagator: symmetry requirement}).


As we have already mentioned, the effective action is in general an unknown functional of the 
fields, and its one-loop correction
can be calculated exactly only in very special cases.
When the background fields vary slowly in space and time however, an expansion in 
derivatives of the fields can represent a useful technique for obtaining an accurate approximation
of the effective action. Considering for simplicity the case of a single scalar field $\Phi$, for which the canonically normalized bare action reads,
\begin{equation}
S[\Phi] = \int {\rm d}^D x \left(-\frac12 (\partial_\mu\Phi) (\partial_\nu\Phi) \eta^{\mu\nu}
           -V^{(0)}(\Phi)\right)
\,,
\label{single scalar field action}
\end{equation}
where $V^{(0)}(\Phi)$ denotes a local tree-level potential, the derivative expansion of the effective action in real space yields,
\begin{equation}
    \Gamma[\phi]=\int{\rm d}^4x\Big[-V_{\rm eff}(\phi)-\frac{1}{2}Z(\phi)(\partial_{\mu} \phi) (\partial_{\nu} \phi)\eta^{\mu \nu} +{\cal O} (\partial^4)\Big] ,
\end{equation}
where ${\cal O} (\partial^4)$ stands for the terms with four or more  derivatives, $V_{\rm eff}(\phi)$ the effective potential, and $Z(\phi)$ a field-dependent function that contains the quantum corrections to the kinetic term. The higher derivatives can be included, however their contribution must be treated perturbatively, for stability of the theory to be maintained. In this work we will only investigate second-order corrections, where stability issues do no arise.

The zeroth order term of this expansion is then given by the effective potential of the theory, whose analytic expression can be found by assuming to have spacetime independent background fields. Indeed, with such background fields, it is possible to solve for the functional inverse of $\mathcal{D}_{ab}$ in equation \eqref{prop eq of motion} and obtain the effective potential to a fixed loop order. At one-loop order, the effective potential is divergent and it ought to be renormalised, for example employing dimensional regularization. In the MS scheme, the well-known result for a single scalar 
field~(\ref{single scalar field action}) is~\cite{Coleman-Weinberg}:
\begin{equation}
    \Gamma^{(1)}_{\rm ren}[\phi,\mu]= - \int {\rm d}^4x V_{\rm eff}^{(1)}(\phi,\mu) 
\,,\qquad
V_{\rm eff}^{(1)}(\phi,\mu) = \frac{m^4(\phi)}{64 \pi^2} \bigg[ \log \bigg(\frac{m^2(\phi)}{4 \pi \mu^2} \bigg) +\gamma_E -\frac{3}{2} \bigg]
\,,\qquad
\label{effective action: single field: CW}
\end{equation}
with $m^2(\phi)={\rm d}^2V^{(0)}/{\rm d}\phi^2$
the tree-level field-dependent mass of the field, $\mu$ the renormalisation scale and 
$\gamma_E=-\psi(1)\simeq 0.57\dots$ the Euler-Mascheroni constant. 

When truncated at a fixed loop order, perturbative calculations produce 
results which depend on the scale $\mu$. This breaking of scaling symmetry is unphysical,
and can be traced back to local counterterms needed to renormalise the primitive results
of perturbative calculations,
which diverge in $D=4$. This scale dependence can be removed from the effective 
action~(\ref{effective action: single field: CW}) by constructing the renormalisation group (RG) improved
effective action $\Gamma^{(1)}_{\rm RG}[\phi,\mu]$, 
which resums the leading logarithms from all loops. One can obtain
the RG-improved effective action {\it e.g.} by imposing the Callan-Symanzik 
equation on the perturbative action. The resulting RG-improved effective action,
 $\Gamma^{(1)}_{\rm RG}[\phi,\mu]$, superficially contains dependence on $\mu$, but
this dependence just labels an equivalence class, {\it i.e.} a one-parameter family of
physically equivalent effective actions.


\section{One-loop effective action - single field case}
\label{Sec.II}

We start from the single field action \eqref{single scalar field action} and we perform an expansion in number of derivatives at one-loop order. Our procedure is based on a midpoint expansion of the propagator that is presented in the following section.

\subsection{Midpoint expansion of the propagator}
\label{subsec:Midpoint expansion of the propagator}
The approach that we employ consists in writing the propagator in Wigner space and then employing 
the derivative expansion. In translationally invariant systems (such as Minkowski vacuum or thermal
states), the propagator $i\Delta(x;y)$ only depends on the relative coordinate $r\equiv x\!-\!y$. 
Performing a Fourier transform with respect to $r$ -- also known 
as a Wigner transform -- yields a momentum space propagator, which can be used for 
an easy evaluation of the one-loop effective action in equation~(\ref{1 loop effective action 2}),
and the result can be expressed in terms of the effective potential, 
see equation~(\ref{effective action: single field: CW}). On the other hand, when the dependence of the propagator on spacetime coordinates is weak, it is still useful to perform the Wigner transform and use it as the starting point for a derivative expansion.

We start considering the set of linearly independent coordinates composed of the relative coordinate $r$ and the midpoint coordinate $X=(x+y)/2$. To transform the equation of motion of the propagator in Wigner space, we use the following general relation \cite{Prokopec-Diamond,Prokopec:2004ic}:
\begin{equation}
    \int {\rm d}^D(x-y) e^{ip\cdot(x-y)} \int {\rm d}^Dz A(x;z)B(z;y)=e^{-i \diamond}\{A(X;p),B(X;p)\},
    \label{wigner transform with diamond}
\end{equation}
where the diamond operator acts as:
\begin{equation}
    \diamond \{A(X;p),B(X;p)\}=\frac{1}{2}(\partial_X^A \cdot \partial_p^B-\partial_X^B \cdot \partial_p^A)A(X;p)B(X;p).
\end{equation}
This relation is useful since it gives us a starting point for the midpoint expansion of the propagator. Indeed, we first transform the propagator equation of motion to the linearly independent coordinates $(X;r)$ and then to the other set of linearly independent coordinates composed by the midpoint and Wigner momentum $(X;p)$.

We are considering the case of a single scalar field, thus we have:
\begin{equation}
    \begin{split}
        A(x;z)&=\big(\partial_x^2-m^2(x)\big)\delta^D(x\!-\!z),\\
        B(z;y)&=i\Delta(z;y),
    \end{split}
\end{equation}
where $m^2(x)$ is the tree-level mass which depends on the background value of the field $\phi(x)$. In case of a translationally invariant background field, we could easily use the equation of motion to find the functional inverse of the inverse propagator and, in turn, the one-loop effective potential. Considering instead a spacetime dependent background field, we can formally write the equation of motion of the propagator in Wigner space as,
\begin{equation}
    \big(\!-p^2-m^2(X)\big)e^{-\frac{i}{2} (\overleftarrow{\partial_X} \cdot \overrightarrow{\partial_p}-\overleftarrow{\partial_p} \cdot \overrightarrow{\partial_X})}i\Delta(X;p)=i.
\label{Derivative expansion: propagator}
\end{equation}
This representation of the propagator is useful when the propagator depends weakly on
the midpoint coordinate $X$. To make that more precise, let us introduce scales $\Delta X_0$ and $\Delta p_0$, 
which characterize the length scale over which the propagator changes significantly, and the typical 
momentum of these excitations~\footnote{Rigorously speaking, the propagator depends on all
momenta, with no upper bound. However, the propagator should be understood as a distribution,
meaning that its action on test functions is defined by integrating over the momenta, and in such integrals 
typically very high momenta do not significantly contribute, as their effects are generally canceled by 
rapid oscillations. When these oscillations are absent, the effects of high momentum modes 
are canceled by the counterterms needed to renormalise the object of interest.}. 
The gradient expansion in 
Eq.~(\ref{Derivative expansion: propagator}) is then well defined if 
$\Delta p_0 \Delta X_0 \gg \hbar$, and can be viewed as a generalization of the WKB expansion 
in quantum mechanics to quantum field theory. The gradient expansion fails to capture 
the effects that occur 
when $\Delta p_0 \Delta X_0 \sim \hbar$, the most prominent being the generation of particles 
by rapidly varying background fields.

Taylor-expanding the exponential in Eq.~(\ref{Derivative expansion: propagator}) 
and applying the derivatives on the left-hand side, the previous equation becomes,
\begin{equation}
    \begin{split}
        \Big[-&p^2\!-\!m^2(X)\!-\!i p\! \cdot\! \partial_X 
        \!+\!\frac{i}{2} (\partial_X m^2(X))\!\cdot\!\partial_p  \\
        &\hskip 3cm
      \!+\!\frac{1}{4} \partial^2_X  \!+\!\frac{1}{8} (\partial^X_\mu\partial^X_\nu m^2(X)) 
       \partial_p^\mu\partial_p^\nu \!+\!{\cal O} (\partial_X^3) \Big]i\Delta(X\!;\!p)=i
       \,,\quad
    \end{split}
     \label{eom in x}
\end{equation}
where the double product in the quadratic term vanishes as we apply a second mixed derivative.

We can now transform also the symmetric equation of motion \eqref{symmetric prop eq of motion} into Wigner space using equation \eqref{wigner transform with diamond} and, as before, keep only terms up to second order in the derivative expansion. Interestingly, we obtain a similar equation as before, but with the imaginary terms of the operator in the left-hand side of equation \eqref{eom in x} with opposite signs, implying that
(up to second order in gradients) the following equation must be satisfied,
\begin{equation}
 \Big[i p\! \cdot \!\partial_X -\frac{i}{2} (\partial_X m^2(X))\!\cdot\!\partial_p \Big]i\Delta(X;p)=0.
 \label{eom in y}
\end{equation}
In fact, the propagator must satisfy 
both the symmetric equations of motion~(\ref{integrated prop eq of motion}),
 in which the real parts of the operator 
remain unchanged, while the imaginary terms flip signs. 
The same conclusion can be reached from the transposition symmetry 
of the propagator~(\ref{propagator: symmetry requirement}), 
which in Wigner space reads, 
\begin{equation}
    i\Delta_{ac}[\phi_e](X;p)=i\Delta_{ca}[\phi_e](X;-p)
\,.
\label{propagator: symmetry requirement: Wigner}
\end{equation}

We can now exploit the inverse matrix expansion to obtain a derivative expansion of the propagator,
\begin{eqnarray}
    i\Delta(X,p)&\!=\!&
  i\Delta^{(0)} +  i\Delta^{(1)} +  i\Delta^{(2)} + \cdots 
\label{midpoint general propagator}\\    
&\!=\!& i\Delta^{(0)}-i(\Delta^{(0)}\mathcal{D}^{(1)} \Delta^{(0)} )+i\big[\!- \Delta^{(0)}\mathcal{D}^{(2)} \Delta^{(0)} +  
                \Delta^{(0)}\mathcal{D}^{(1)} [\Delta^{(0)} \mathcal{D}^{(1)}\Delta^{(0)} ]
                \big]+\cdots 
\,,
    \qquad
\label{derivative expansion: single field: b}
\end{eqnarray}
where we keep terms up to second order in derivatives.
At zeroth order, it is easy to see that,
\begin{equation}
\mathcal{D}^{(0)}i\Delta^{(0)}=i
    \;\Longrightarrow i\Delta^{(0)}(X,p) = \frac{i}{-p^2 -m^2(X)+i\epsilon}
  \,,\quad \left(\mathcal{D}^{(0)} = -p^2 -m^2(X)\right) 
\,,
\label{propagator expansion terms1}
\end{equation}
where we introduced the standard $i\epsilon$ prescription corresponding to the 
(initial value) boundary prescription for the Feynman (time-ordered) propagator.
For notational simplicity, in what follows we use a short-hand notation, $p_\epsilon^2 \equiv p^2-i\epsilon$.
The linear correction is given by:
\begin{equation}
    \mathcal{D}^{(0)}i\Delta^{(1)}=-\mathcal{D}^{(1)} i\Delta^{(0)}
    \,, \qquad   
    \mathcal{D}^{(1)} = -i p \cdot \partial_X +\frac{i}{2} (\partial_X m^2(X))\cdot\partial_p 
\,.
\end{equation}
Thus, the first order correction to the propagator 
in~(\ref{midpoint general propagator}--\ref{derivative expansion: single field: b})
can be obtained from,
\begin{equation}
\begin{split}
    \mathcal{D}^{(1)}i\Delta^{(0)}=&
    \Big[-i p \cdot \partial_X +\frac{i}{2} (\partial_X m^2(X))\!\cdot\!\partial_p\Big] \; 
    \frac{i}{-p_\epsilon^2-m^2(X)}
    =0,
\end{split}
\end{equation}
such that $i\Delta^{(1)}=0$. This is consistent with the fact that the transposition 
property~(\ref{propagator: symmetry requirement: Wigner}) implies that the propagator in the one-field case can depend on
even powers of $p^\mu$, i.e. only on the combination 
$p^2 \equiv p_\mu p^\mu$, which is a Lorentz scalar.
This property generalizes to higher order in gradients, in that all
odd powers in gradients give zero when acting on the propagator. 

As an example, the third order contribution to the propagator is given by:
\begin{equation}
\begin{split}
    \Delta^{(3)}=&-\Delta^{(0)}\mathcal{D}^{(3)}\Delta^{(0)}+\Delta^{(0)}\mathcal{D}^{(1)}\Delta^{(0)}\mathcal{D}^{(2)}\Delta^{(0)}\\
    &+\Delta^{(0)}\mathcal{D}^{(2)}\Delta^{(0)}\mathcal{D}^{(1)}\Delta^{(0)}-\Delta^{(0)}\mathcal{D}^{(1)}\Delta^{(0)}\mathcal{D}^{(1)}\Delta^{(0)}\mathcal{D}^{(1)}\Delta^{(0)}
    \,,
\end{split}
\end{equation}
where the last two terms vanish and the third order contribution to the inverse propagator is:
\begin{equation}
    \mathcal{D}^{(3)}=-\frac{i}{48}(\partial^X_{\mu} \partial^X_{\nu} \partial^X_{\sigma}m^2(X))\partial_p^\mu\partial_p^\nu \partial_p^\sigma\; .
\end{equation}
We find that:
\begin{equation}
\begin{split}
    -\Delta^{(0)}\mathcal{D}^{(3)}\Delta^{(0)}=&\frac{i(\partial^X_{\mu} \partial^X_{\nu} \partial^X_{\sigma}m^2(X)) p^{\mu} p^{\nu}p^{\sigma}}{(-p_\epsilon^2-m^2(X))^5}\\
    &+\frac{i}{6} \frac{(\partial^X_{\mu} \partial^X_{\nu} \partial^X_{\sigma}m^2(X))(\eta^{\mu \nu}p^{\sigma}+\eta^{\mu \sigma}p^{\nu}+\eta^{\nu \sigma}p^{\mu})}{(-p_\epsilon^2-m^2(X))^4}\; ,
\end{split}
\end{equation}
\begin{equation}
    \Delta^{(0)}\mathcal{D}^{(1)}\Delta^{(0)}\mathcal{D}^{(2)}\Delta^{(0)}=-\frac{i(\partial^X_{\mu} \partial^X_{\nu} \partial^X_{\sigma}m^2(X)) p^{\mu} p^{\nu}p^{\sigma}}{(-p_\epsilon^2-m^2(X))^5}-\frac{i}{2}\frac{(\partial_{\nu} \partial_{\mu}\partial^{\mu}m^2(X))p^{\nu}}{(-p_\epsilon^2-m^2(X))^4}\; .
\end{equation}
Summing the two terms, it is easy to check that $i\Delta^{(3)}=0$, as expected.

Therefore, the expansion of the propagator is simply given by $i \Delta= i \Delta^{(0)}+i \Delta^{(2)}+\cdots$, where only terms with an even number of derivatives contribute.

For a later use,  we also compute the quadratic contribution to the propagator which is given by:
\begin{equation}
    \mathcal{D}^{(0)}i\Delta^{(2)}=-i\mathcal{D}^{(2)} \Delta^{(0)},
\end{equation}
and knowing that $\mathcal{D}^{(2)}$ has a unique expression in the two symmetric equations of motion of the propagator. In Wigner space, we find that the second order contribution to the propagator has the following form:
\begin{equation}
    i\Delta^{(2)}(X,p)=\frac{i}{2}(\partial_{\mu}\partial_{\nu}m^2)\bigg[\frac{\eta^{\mu \nu}}{(p_{\epsilon}^2+m^2)^3}-2\frac{p^{\mu}p^{\nu}}{(p_{\epsilon}^2+m^2)^4}\bigg] -\frac{i}{2}\frac{(\partial_X m^2)^2}{(p_{\epsilon}^2+m^2)^4} \; .
    \label{propagator expansion terms2}
\end{equation}
%

{\bf The position space propagator.}
In what follows we construct the off-coincident propagator in position space
by integrating its momentum space expression~\eqref{propagator expansion terms1} and \eqref{propagator expansion terms2}.

We define the following integral,
\begin{eqnarray}
    J_n(x;x')=\int \frac{{\rm d}^D p}{(2\pi)^D}\frac{-i e^{i p \cdot\Delta x}}{(p_{\epsilon}^2+m^2)^n}\; .
    \label{J_n}
\end{eqnarray}
At zeroth order in gradients, the off-coincident propagator is found by setting $n=1$ in the previous integral. The solution is:
\begin{eqnarray}
    i\Delta^{(0)}(x;x')\equiv J_1(x;x')=\frac{m^{2\nu}}{(2\pi)^{\nu+1}}\frac{K_{\nu}(z)}{z^{\nu}}\; ,
    \label{prop0off}
\end{eqnarray}
where $\nu = \frac{D-2}{2}$ and
$K_{\nu}(z)$ is the modified Bessel function of the second kind, which can be conveniently 
expressed as the sum of two series,
\begin{eqnarray}
    \frac{K_{\nu}(z)}{z^{\nu}}=\frac{1}{2^{1+\nu}}\sum_{n=0}^{\infty}\bigg[\frac{\Gamma(-\nu )\big(\frac{z}{2}\big)^{2n}}{n!(\nu+1)_n}+\frac{\Gamma(\nu)\big(\frac{z}{2}\big)^{2n-2\nu}}{n!(1-\nu)_n}\bigg]\; ,
\end{eqnarray}
where $z=m\sqrt{\Delta x^2}$ and $\Delta x^2=-(|t-t'|-i\epsilon)^2+|| \Vec{x}-\vec{x}'||^2$.

We can then write the full off-coincident propagator as,
\begin{equation}
\begin{split}
    i\Delta(x;x')=\int \frac{{\rm d}^D p}{(2\pi)^D} e^{i p \cdot \Delta x}\bigg[&\frac{-i}{(
    p^2_{\epsilon}+m^2)}+\frac{i}{2}(\partial_{\mu}\partial_{\nu}m^2)\bigg(\frac{\eta^{\mu \nu}}{(p^2_{\epsilon}+m^2)^3}-\frac{2p^{\mu}p^{\nu}}{(p^2_{\epsilon}+m^2)^4}\bigg)\\
    &-\frac{i}{2}\frac{(\partial_X m^2)^2}{(p^2_{\epsilon}+m^2)^3}\bigg]\;,
\end{split}
\end{equation}
and using the replacement $p^{\mu} p^{\nu} \rightarrow -\frac{\partial}{\partial \Delta x_{\mu}} \frac{\partial}{\partial \Delta x_{\nu}}$, 
\begin{equation}
\begin{split}
    i\Delta(x;x')=\int \frac{{\rm d}^D p}{(2\pi)^D} \bigg[&\frac{-i e^{i p \cdot \Delta x}}{(p^2_{\epsilon}+m^2)}+\frac{i}{2}(\partial_{\mu}\partial_{\nu}m^2)\bigg(\frac{\eta^{\mu \nu} e^{i p \cdot \Delta x}}{(p^2_{\epsilon}+m^2)^3}\\
    &+ \frac{\partial}{\partial \Delta x_{\mu}} \frac{\partial}{\partial \Delta x_{\nu}} \frac{2 e^{i p \cdot \Delta x}} {(p^2_{\epsilon}+m^2)^4}\bigg)-\frac{i}{2}e^{i p \cdot \Delta x}\frac{(\partial_X m^2)^2}{(p^2_{\epsilon}+m^2)^3}\bigg]\;.
\end{split}
\end{equation}

Noting that $J_3=\frac{1}{2}(\partial_{m^2})^2J_1$ and $J_4=-\frac{1}{6}(\partial_{m^2})^3J_1$, we can rewrite the full off-coincident propagator as:
\begin{eqnarray}
\begin{split}
    i\Delta(x;x')=&\bigg\{1-\frac{1}{2} (\partial_{\mu} \partial_{\nu}m^2)\bigg[\eta^{\mu \nu} \frac{1}{2}\bigg(\frac{\partial}{\partial m^2}\bigg)^2-2  \frac{\partial}{\partial \Delta x_{\mu}} \frac{\partial}{\partial \Delta x_{\nu}} \frac{1}{6} \bigg(\frac{\partial}{\partial m^2}\bigg)^3 \bigg]\\
    &-\frac{1}{2} (\partial_X m^2)^2 \frac{1}{6} \bigg(\frac{\partial}{\partial m^2}\bigg)^3 \bigg\}J_1 \;.
\end{split}
\end{eqnarray}

Using the following relations,
\begin{eqnarray}
    \bigg(\frac{\partial}{\partial m^2}\bigg) m^{2\nu}&\!\!=\!\!&\frac{m^{2\nu}}{2 m^2}\bigg[z \frac{\partial}{\partial z} +2 \nu \bigg]\; ,
\\
    \bigg(\frac{\partial}{\partial m^2}\bigg)^2m^{2\nu}&\!\!=\!\!&\frac{m^{2\nu}}{4 m^4}\bigg[ z \frac{\partial}{\partial z} +2(\nu -1) \bigg]\bigg[ z \frac{\partial}{\partial z} +2 \nu \bigg]\; ,
\\
    \bigg(\frac{\partial}{\partial m^2}\bigg)^3m^{2\nu}&\!\!=\!\!&\frac{m^{2\nu}}{8 m^6}\bigg[ z \frac{\partial}{\partial z} +2(\nu -2) \bigg]\bigg[ z \frac{\partial}{\partial z} +2(\nu -1) \bigg]\bigg[ z \frac{\partial}{\partial z} +2 \nu \bigg]\; ,
\end{eqnarray}
and the recursive relation for the modified Bessel function of the second kind \cite{Gradshteyn:2007},
\begin{eqnarray}
    \bigg(z \frac{\partial}{\partial z}+\nu \bigg) K_{\nu}(z)=-z K_{\nu-1}(z)\; ,
\end{eqnarray}
we can rewrite the off-coincident propagator as the sum of three terms:
\begin{eqnarray}
\begin{split}
i\Delta(x;x')=i\Delta^{(0)}(x;x')+i\Delta^{(2)}_{\nu-2}(x;x')+i\Delta^{(2)}_{\nu-3}(x;x')\; .
\end{split}
\label{propoffcoinc}
\end{eqnarray}
The first term is the zeroth order propagator in equation \eqref{prop0off}, while the terms at second order in gradients are:
\begin{eqnarray}
    i\Delta^{(2)}_{\nu-2}(x;x')=-\frac{m^{2\nu-4}}{(2\pi)^{\nu+1}}\frac{1}{16}(\partial^2_X m^2)\frac{K_{\nu-2}(z)}{z^{\nu-2}}\; ,
\end{eqnarray}
\begin{eqnarray}
    i\Delta^{(2)}_{\nu-3}(x;x')=\frac{m^{2\nu-6}}{(2\pi)^{\nu+1}}\frac{1}{48}\bigg[\frac{1}{2}(\partial_X m^2)^2 -(\partial_{\mu}\partial_{\nu} m^2) \frac{\partial}{\partial \Delta x^{\mu}} \frac{\partial}{\partial \Delta x^{\nu}} \bigg] \frac{K_{\nu-3}(z)}{z^{\nu-3}}\; .
\end{eqnarray} 

Taking the coincident limit, $x= x'=X$, of the previous equation, we have that 
$\big[K_{\nu}(z)/z^{\nu}\big]_{z\rightarrow 0}=2^{-\nu-1}\Gamma(-\nu)$, and we can verify that the propagator at coincidence is,
\begin{equation}
i\Delta(X;X)=i\Delta^{(0)}(X;X)+i\Delta^{(2)}(X;X)
\,,
\label{coincident propagator: 0+2}
\end{equation}
where,
\begin{equation}
    i\Delta^{(0)}(X;X)=\frac{m^{D-2}}{(4 \pi)^{D/2}}\Gamma\bigg(\!1\!-\!\frac{D}{2}\bigg)\; ,
    \label{prop0coin}
\end{equation}
and
\begin{equation}
\begin{split}
    i\Delta^{(2)}(X;X)=&\frac{m^{D-8}}{12 (4\pi)^{D/2}}\Gamma\bigg(\!4\!-\!\frac{D}{2}\bigg)(\partial_X m^2)^2\\
    &-\frac{m^{D-6}}{6 (4\pi)^{D/2}}\Gamma\bigg(\!3\!-\!\frac{D}{2}\bigg) (\partial^2_X m^2) \; .
\end{split}
    \label{prop2coin}
\end{equation}
If expanded around $D \rightarrow 4$, the coincident propagator becomes,
\begin{eqnarray}
      i\Delta^{(0)}(X;X)&\!\!=\!\!& \frac{m^2 }{(4 \pi)^2} \bigg(\frac{2 \mu^{D-4}}{D-4}\!+\!\ln\bigg(\frac{m^2}{4\pi \mu^2}\bigg)\!-\!1\!+\!\gamma_E\bigg) \!+\!{\cal O}\big(D\!-\!4\big)
      \; ,
\\
    i\Delta^{(2)}(X;X)&\!\!=\!\!& \frac{ 1}{(4 \pi)^2}  \bigg(\frac{(\partial_X m^2)^2}{12 m^4}-\frac{(\partial^2_X m^2)}{6 m^2} \bigg)
    \!+\!{\cal O}\big(D\!-\!4\big)
\; .\qquad
\end{eqnarray}
As expected, the same expressions can be obtained performing the momentum integrals \eqref{J_n} in the limit $\Delta x\rightarrow 0$ of equations \eqref{propagator expansion terms1} and \eqref{propagator expansion terms2}.

We are now ready to compute the gradient corrections to the one-loop effective action.

\subsection{Direct integration of the logarithm}

The quantum corrections at one-loop order are given by the functional determinant
of the inverse propagator, which can be computed by performing a direct integration of the logarithm of the propagator. Considering equation~\eqref{1 loop effective action 2}, our starting point is:
\begin{equation}
    \Gamma^{(1)}=-\frac{i}{2}\int {\rm d}^Dx \int \frac{{\rm d}^D p}{(2\pi)^D}\log(\Delta)\; ,
    \label{one-loop_ea}
\end{equation}
where $\Delta$ is the propagator given in equations \eqref{propagator expansion terms1} and \eqref{propagator expansion terms2}. In the following, we adopt the notation: 
\begin{equation}    \Gamma^{(1)}=\Gamma^{(1,0)}+\Gamma^{(1,2)}\; ,
\end{equation}
with $\Gamma^{(1,0)}$ the zeroth order in gradients contribution to the one-loop effective action, while $\Gamma^{(1,2)}$ the second order in gradients one. 

Using the substitution $p^{\mu}p^{\nu}\rightarrow \eta^{\mu \nu}\frac{p^2}{D}$, valid up to second order in gradients expansion \footnote{It is possible to estimate the error we make using this substitution by rewriting the momentum integral of equation \eqref{one-loop_ea} as:
\begin{equation}
    \int_p \log[A(p)+B(p)(\partial_{\mu}\partial_{\nu}m^2)p^{\mu} p^{\nu}]=\int_p\bigg\{\log[A(p)]+\sum_{n=0}^{\infty}\frac{(-1)^n}{n+1}\bigg[\frac{B(p)}{A(p)}(\partial_{\mu}\partial_{\nu}m^2)p^{\mu} p^{\nu}\bigg]^{n+1}\bigg\}
    \label{log_expansion_k}
\end{equation} where $A(p)$ and $B(p)$ are momentum-dependent expressions that can be inferred from the explicit expression for the propagator in equations~\eqref{propagator expansion terms1} and \eqref{propagator expansion terms2}. \\
If $n=0$, the substitution $p^{\mu}p^{\nu}\rightarrow \eta^{\mu \nu}\frac{p^2}{D}$ is valid. On the other hand, for the next order term we have to use the substitution,
\begin{equation}
p^{\mu}p^{\nu}p^{\rho}p^{\sigma}\rightarrow \frac{p^4}{D(D+2)}(\eta^{\mu \nu}\eta^{\rho \sigma}+\eta^{\mu \rho}\eta^{\nu \sigma}+\eta^{\mu \sigma}\eta^{\rho \nu})
\end{equation}
such that the $n=1$ term in equation \eqref{log_expansion_k} is:
\begin{equation}
    \frac{p^4}{D(D+2)}[(\partial^2m^2)^2+2(\partial_{\mu}\partial_{\nu}m^2)(\partial^{\mu}\partial^{\nu}m^2)]\; .
\end{equation}
In the main text we use the substitution $p^{\mu}p^{\nu}\rightarrow \eta^{\mu \nu}\frac{p^2}{D}$ inside the logarithm and so have an error that at leading order is proportional to $(\partial_{\mu}\partial_{\nu}m^2)(\partial^{\mu}\partial^{\nu}m^2)-D(\partial^2m^2)^2$. This and all the higher order terms can be calculated exactly and included in the calculations. Since our purpose is to compute second-order corrections in number of derivatives, we can neglect these higher order corrections for simplicity.}, we find the following integrand:
\begin{equation}
   \log\Bigg[-\frac{(p_{\epsilon}^2+m^2)^3-\Big(\frac{D-2}{2D}\Big)(\partial^2m^2)(p_{\epsilon}^2+m^2)-\frac{m^2}{D}(\partial^2m^2)+\frac{1}{2}(\partial m^2)^2}{(p_{\epsilon}^2+m^2)^4}\Bigg]\; .
\end{equation}
The minus sign inside the logarithm is regulated to zero, and so we can rewrite the previous expression as,
\begin{equation}
\begin{split}
    &\log\bigg[(p_{\epsilon}^2)^3+3(p_{\epsilon}^2)^2m^2+\bigg(3m^4-\frac{(D-2)}{2D}(\partial^2m^2)\bigg)p_{\epsilon}^2\\
    &+\frac{1}{2}(\partial m^2)^2-\frac{m^2}{2}(\partial^2m^2)+m^6\bigg]-4\log(p_{\epsilon}^2+m^2)\; .
\end{split}
\end{equation}
Decomposing the cubic polynomial inside the first logarithm, we rewrite the previous expression as:
\begin{equation}
    \sum_{i=1}^3\log(p_{\epsilon}^2-x_i m^2)-4\log(p_{\epsilon}^2+m^2)\; ,
\end{equation}
with the cubic roots being:
\begin{equation}
    x_1=\frac{p_1^2}{m^2}=-1-\frac{2^{1/3}\delta}{\Big(z+\sqrt{4\delta^3+z^2}\Big)^{1/3}}+\frac{\Big(z+\sqrt{4\delta^3+z^2}\Big)^{1/3}}{2^{1/3}}\; , 
    \label{croot1}
\end{equation}
\begin{equation}
    x_2=\frac{p_2^2}{m^2}=-1+\frac{2^{1/3}e^{i\frac{\pi}{3}}\delta}{\Big(z+\sqrt{4\delta^3+z^2}\Big)^{1/3}}-\frac{e^{-i\frac{\pi}{3}}\Big(z+\sqrt{4\delta^3+z^2}\Big)^{1/3}}{2^{1/3}}\; ,
    \label{croot2}
\end{equation}
\begin{equation}
    x_3=\frac{p_3^2}{m^2}=-1+\frac{2^{1/3}e^{-i\frac{\pi}{3}}\delta}{\Big(z+\sqrt{4\delta^3+z^2}\Big)^{1/3}}-\frac{e^{i\frac{\pi}{3}}\Big(z+\sqrt{4\delta^3+z^2}\Big)^{1/3}}{2^{1/3}}\; ,
        \label{croot3}
\end{equation}
and with:
\begin{equation}
    \delta=\frac{\alpha}{3}-1 \; , \;\;\;\;\;\; z=\alpha-\beta-2\; ,
    \label{delta-z-factors}
\end{equation}
\begin{equation}
    \alpha=3-\frac{(D-2)}{2D}\frac{(\partial^2m^2)}{m^4}\; , \;\;\;\;\;\; \beta=\frac{(\partial m^2)^2}{2m^6}-\frac{(\partial^2 m^2)}{2m^4}+1\; .
    \label{alpha-beta-factor}
\end{equation}
We define the expressions in \eqref{delta-z-factors} such that both are second order in gradients. Moreover, it is worth noticing that the cubic roots transform one into each other by flipping the sign in front of the square root, {\it i.e.} by sending $+\sqrt{4\delta^3+z^2}\rightarrow -\sqrt{4\delta^3+z^2}$ in all three cubic roots, we still find a valid set of solutions for the cubic equation. As a consequence, symmetric combinations of the three roots have to be invariant under this transformation. 

Moreover, we notice that in the limit $\delta \rightarrow 0$ and $z<0$ the above solutions \eqref{croot1}, \eqref{croot2} and \eqref{croot3} are indeterminate. In this regime, we can consider the set of solutions obtained with the transformation $+\sqrt{4\delta^3+z^2}\rightarrow -\sqrt{4\delta^3+z^2}$.

To proceed we need to perform the following integral:
\begin{equation}
        \int \frac{{\rm d}^D p}{(2\pi)^D}\log(p_{\epsilon}^2+m^2)\; ,
\end{equation}
so we rotate to Euclidean space ($p^0 \rightarrow ip^0_E$) and use the spherical coordinates to write:
\begin{equation}
    \int \frac{{\rm d}^D p}{(2\pi)^D}\log(p_{\epsilon}^2+m^2)=\frac{2i}{(4\pi)^{D/2}\Gamma(\frac{D}{2})} \int_0^{\infty}dp_E \, p_E^{D-1}\log(p_E^2+m^2)\; ,
\end{equation}
and we notice that the integral of the modulus of the momentum is divergent. We thus introduce a cut-off $\Lambda$ and split the domain of integration:
\begin{equation}
\begin{split}
    &\frac{2i}{(4\pi)^{D/2}\Gamma(\frac{D}{2})}\bigg(\int_0^{\Lambda}dp_E \, p_E^{D-1}\log(p_E^2+m^2)+\int_{\Lambda}^{\infty}dp_E \, p_E^{D-1}\log(p_E^2+m^2)\bigg)\\
     =&\frac{2i}{(4\pi)^{D/2}\Gamma(\frac{D}{2})}\bigg[\frac{(-m)^D}{D}B\bigg(-\frac{\Lambda^2}{m^2},1+\frac{D}{2},0\bigg)\\
     &-\frac{2m^2\Lambda^{D-2}}{D(D-2)}\hypgeo{2}{1} \bigg(1,1-\frac{D}{2},2-\frac{D}{2},-\frac{m^2}{\Lambda^2}\bigg)+2\frac{\Lambda^D}{D^2}\bigg]\; ,
\end{split}
\end{equation}
where $B$ is the incomplete beta function and ${}_2F_1$ is Gauss' hypergeometric function. This result is found assuming that $\text{Re}(D)<0$ in the ultraviolet regime while $\text{Re}(D)>0$ in the infrared.
Analytically extending to all $D$ and expanding around $\Lambda \rightarrow \infty$, we obtain the desired result:
\begin{equation}
    \int \frac{{\rm d}^D p}{(2\pi)^D}\log(p_{\epsilon}^2+m^2)=-\frac{i}{(4 \pi)^{D/2}} m^D \Gamma\bigg(\!\!-\frac{D}{2}\bigg)\; .
\end{equation}

\begin{figure}
\centering    
\includegraphics[width=0.45\textwidth]{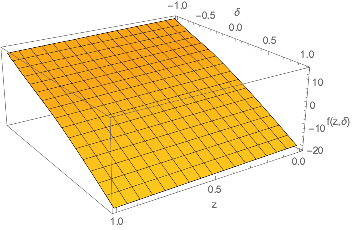}
\includegraphics[width=0.45\textwidth]{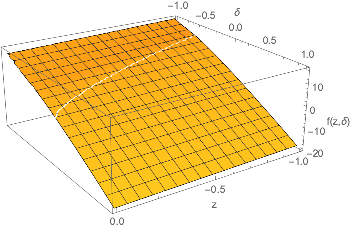}
\caption{$f(z,\delta)$ surface for $z>0$ on the left and for $z<0$ on the right. The line of white points in the right plot corresponds to a vanishing square root $\sqrt{4\delta^3+z^2}$. We verified that the function is continuous in this limit. }
\label{fig:f_zd}
\end{figure}

\begin{figure}
\centering    
\includegraphics[width=0.48\textwidth]{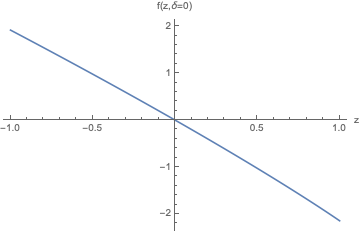}
\hskip 0.2cm
\includegraphics[width=0.48\textwidth]{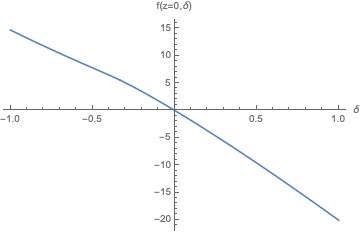}
\caption{On the left, $f(z,\delta=0)$ as given in equation \eqref{f_d0}, and on the right, $f(z=0,\delta)$ as given in equation \eqref{f_z0}.}
\label{fig:f_zd_limits}
\end{figure}

Using the previous expression, we can rewrite equation \eqref{one-loop_ea} as:
\begin{equation}
    \Gamma^{(1)}=\frac{1}{2(4 \pi)^{D/2}}\int {\rm d}^Dx \, m^D
\Gamma\bigg(\!-\frac{D}{2}\bigg)\bigg[4-\sum_{i=1}^3(-x_i)^{D/2}\bigg]\; .
\label{one-loop_general}
\end{equation}
Expanding the sum over the cubic roots for $D\rightarrow 4$, the previous expression can be rewritten as:
\begin{equation}
    \Gamma^{(1)}=\frac{1}{2(4 \pi)^{D/2}}\int {\rm d}^Dx \, m^D
\Gamma\bigg(\!\!-\frac{D}{2}\bigg) \bigg[1-\sum_{i=1}^3\bigg((-x_i)^2+\frac{D\!-\!4}{4}(-x_i)^2\log(-x_i)^2-1\bigg)\bigg]\; ,
\label{effetive_actionD4}
\end{equation}
the first term in the square brackets corresponds to the zeroth order in gradients contribution to the one-loop effective action. In Figure \ref{fig:f_zd}, we plot $f(z,\delta)=\sum_{i=1}^3 (-x_i)^2\log(-x_i)^2$. We can study the limit of this function when $\delta = 0$, and find:
\begin{equation}
\begin{split}
    f(z,\delta = 0)=&(1-z^{1/3})^2\log[(1-z^{1/3})^2]+(1+e^{i\pi/3}z^{1/3})^2\log[(1+e^{i\pi/3}z^{1/3})^2]\\
    &+(1+e^{-i\pi/3}z^{1/3})^2\log[(1+e^{-i\pi/3}z^{1/3})^2]\; ,
    \label{f_d0}
\end{split}
\end{equation}
where the solution is valid both for $z<0$ and $z>0$, as previously explained. Similarly, we can expand around $z = 0$, 
\begin{equation}
\begin{split}
    f(z = 0,\delta)=&(1-i\sqrt{3\delta})^2\log[(1-i\sqrt{3\delta})^2]+(1+i\sqrt{3\delta})^2\log[(1+i\sqrt{3\delta})^2]\\
    =&2(1-3\delta) \log(1+3\delta)-24 \delta \hypgeo{2}{1} \bigg(1,\frac{1}{2};\frac{3}{2};-3\delta\bigg)\; .
\end{split}
\label{f_z0}
\end{equation}
We notice that the expressions we obtained are well-defined at the origin where $f(0,0)=0$, as can be seen from Figure \ref{fig:f_zd_limits}.

Coming back to equation \eqref{one-loop_general}, we expand the sum over the cubic roots up to second order in number of derivatives, and we find:
\begin{equation}
\begin{split}
    \Gamma^{(1)}=\frac{1}{2(4 \pi)^{D/2}}\int & {\rm d}^Dx \, m^D
\Gamma\bigg(\!\!-\frac{D}{2}\bigg)\bigg[4-3+3\bigg(\frac{D}{2}-1\bigg)\frac{D}{2}\delta\\
&+\bigg(\frac{D}{2}-2\bigg)\bigg(\frac{D}{2}-1\bigg)\frac{D}{4}w^3\bigg]\; ,
\label{effective_action_gradients}
\end{split}
\end{equation}
where $w=\Big(\frac{z+\sqrt{4\delta^3+z^2}}{2}\Big)^{1/3}$ if $z>0$, while $w=\Big(\frac{z-\sqrt{4\delta^3+z^2}}{2}\Big)^{1/3}$ if $z<0$ .
At zeroth order, the result is:
\begin{equation}
    \Gamma^{(1,0)}=\frac{1}{2(4 \pi)^{D/2}}\int {\rm d}^Dx\, m^D
\Gamma\bigg(\!\!-\frac{D}{2}\bigg) \; ,
\end{equation}
which contains the primitive one-loop effective potential. At second order and for $z>0$ :
\begin{equation}
\begin{split}
    \Gamma^{(1,2)}=&\frac{1}{2(4 \pi)^{D/2}}\int {\rm d}^Dx\, m^D
\Gamma\bigg(\!\!-\frac{D}{2}\bigg)\bigg\{\bigg[\!-\frac{1}{2}\bigg(\frac{D}{2}-1\bigg)^2\\
&+\frac{1}{8}\bigg(\frac{D}{2}-1\bigg)\bigg(\frac{D}{2}-2\bigg)\bigg]\frac{(\partial^2m^2)}{m^4}-\frac{D}{16}\bigg(\frac{D}{2}-1\bigg)\bigg(\frac{D}{2}-2\bigg)\frac{(\partial m^2)^2}{m^6}\\
&+\frac{D}{8}\bigg(\frac{D}{2}-1\bigg)\bigg(\frac{D}{2}-2\bigg)\bigg[\frac{1}{D}\frac{(\partial^2m^2)}{m^4}-\frac{(\partial m^2)^2}{2m^6}\bigg]\bigg\}\; ,
\end{split}
\end{equation}
The same expression can be obtained in the regime $z<0$, so the effective action is continuous in $z=0$. We can now sum the terms and integrate by parts:
\begin{equation}
    \Gamma^{(1,2)}=\frac{1}{(4 \pi)^{D/2}}\int {\rm d}^Dx\, \bigg(\frac{1}{8}-\frac{1}{8}\bigg)m^{D-6}\Gamma\bigg(3-\frac{D}{2}\bigg)(\partial m^2)^2=0\; .
    \label{final result - direct integration}
\end{equation}
Upon neglecting a boundary term, one finds that the two bulk terms cancel each other exactly, giving a vanishing result. 

\begin{figure}
\centering    
\includegraphics[width=0.49\textwidth]{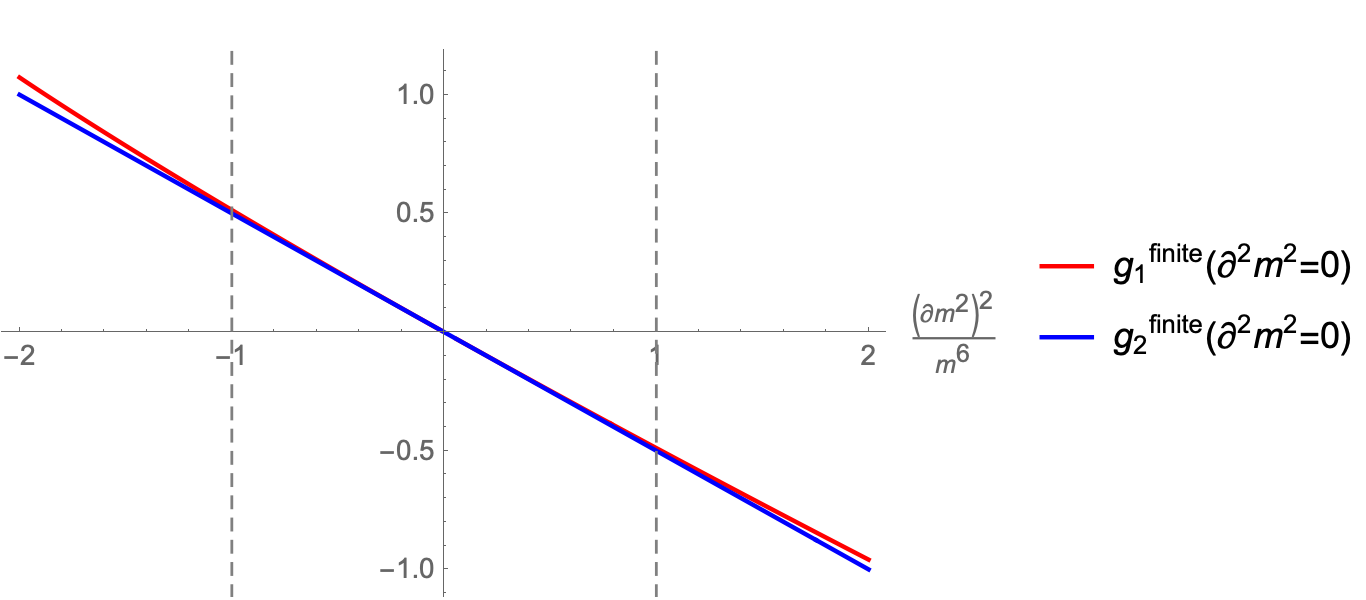}
\hskip 0.1cm
\includegraphics[width=0.49\textwidth]{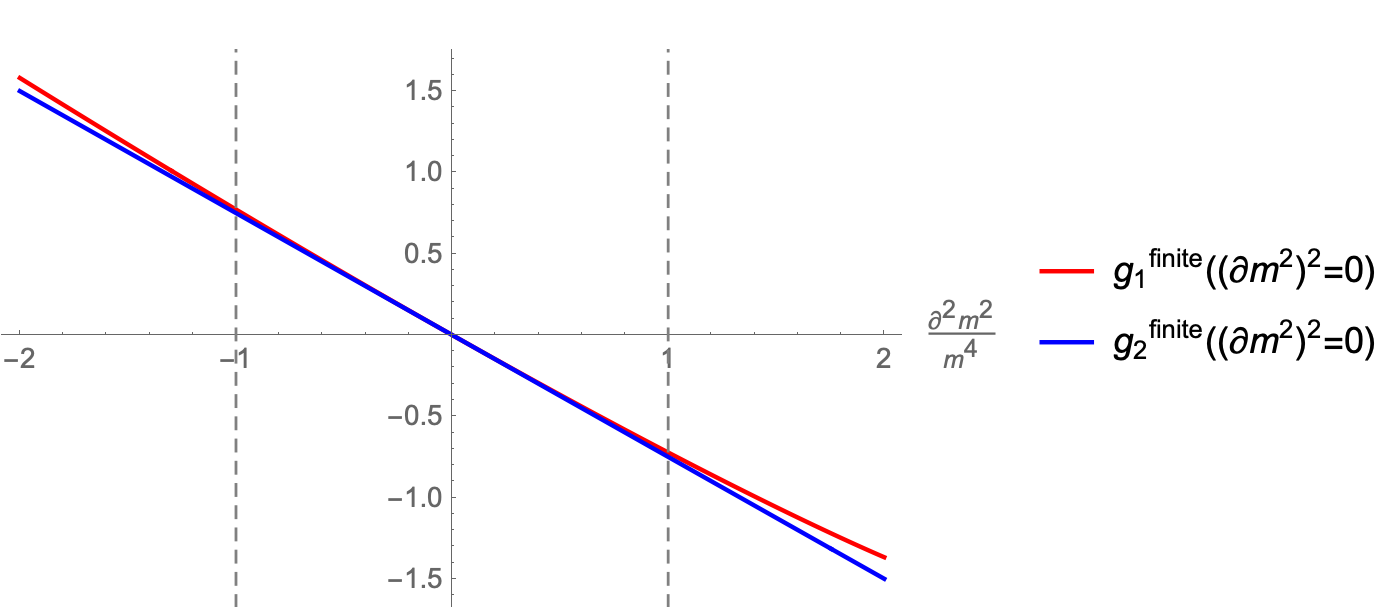}

\caption{The finite parts of the functions $g_1$ and $g_2$. In the right panel, we plot them as a function of $\frac{\partial^2m^2}{m^4}$ and setting $(\partial m^2)^2=0$, while in the left panel we choose $\partial^2m^2=0$ and plot as a function of $\frac{(\partial m^2)^2}{m^6}$. As expected, the agreement is good
for small ($<1$) values of the arguments.}
\label{fig:full_expanded}
\end{figure}

As a sanity check of the result, we consider the expression in the square brackets in equation \eqref{effetive_actionD4}:
\begin{equation}
    g_1(z,\delta)=\sum_{i=1}^3 -\frac{D-4}{4} (-x_i)^2\log(-x_i)^2-(-x_i)^2+3\; ,
    \label{g1full}
\end{equation}
where we have subtracted the contribution to the zeroth order in gradients one-loop effective action. Notice that to complete the expansion in $D\rightarrow 4$, we also have to account for the $D-$dependence of the roots. We can compare the expression in \eqref{g1full} with the one obtained expanding in gradients and keeping only the second order terms given in equation \eqref{effective_action_gradients}:
\begin{equation}
    g_2(z,\delta)=\frac{D-4}{2} \bigg(\frac{D}{2}-1\bigg)\frac{D}{4}w^3+3\bigg(\frac{D}{2}-1\bigg)\frac{D}{2}\delta \; .
    \label{g2}
\end{equation}
Taking the limit $D\rightarrow 4$ we obtain,
\begin{equation}
\begin{split}
    g_2(z,\delta)=&\,\frac{D\!-\!4}{2} (w^3+9\delta)+6\delta\\
    =&\,\frac{D\!-\!4}{2} \bigg(\!-\frac{(\partial m^2)^2}{2m^6}-\frac{3}{4}\frac{\partial^2m^2}{m^4}\bigg)-\frac{1}{2} \frac{\partial^2m^2}{m^4}\; .
\end{split}
\end{equation}
While the divergent parts are mathematically equal, in Figure~\ref{fig:full_expanded} we numerically compare the finite parts of $g_1$, which is expanded only around $D\rightarrow4$, and $g_2$, which is expanded both around $D\rightarrow4$ and up to second order in gradients. As can be seen, the two functions show perfect agreement for small values of the derivatives.

\subsection{Expanding the logarithm}
\label{Expanding the logarithm}

We also present an alternative, and quicker, way to compute the second-order corrections to the one-loop effective action. This second method is based on an expansion of the logarithm of the propagator. In the absence of linear correction to the propagator
(see Eq.~(\ref{eom in y})), we can write,
\begin{equation}
\begin{split}
    \Gamma^{(1)}=\frac{i}{2} \text{Tr}\log(\Delta^{-1})
    =&-\frac{i}{2} \text{Tr}\log(\Delta^{(0)}+\Delta^{(2)}+...)\\
    =&-\frac{i}{2} \big[\text{Tr}\log(\Delta^{(0)}) + \text{Tr}(\mathcal{D}^{(0)}\Delta^{(2)} )+\cdots\big]\; .
\label{Gamma1: expansion in gradients}
\end{split}
\end{equation}
The first term simply corresponds to the effective potential while the latter one contains the one-loop correction to the kinetic term.
After performing a Wick rotation to Euclidean space, the second-derivative term can be integrated in momentum space. Making use of the Wigner space expression for the propagator~(\ref{propagator expansion terms1}) and~(\ref{propagator expansion terms2}) one obtains,
\begin{eqnarray}
    \Gamma^{(1,2)}=-\frac{i}{2} \text{Tr}(\mathcal{D}^{(0)}\Delta^{(2)} )
    &\!\!=\!\!&\frac{i}{2} \int {\rm d}^Dx \int \frac{{\rm d}^Dp}{(2\pi)^D} \bigg[
    \frac{\frac{m^2}{D}(\partial_X^2m^2)-\frac{1}{2}(\partial_X m^2)^2}{(p_\epsilon^2+m^2(X))^3}
    \nonumber\\
  &&\hskip 5.1cm
  +\,\frac{\frac{(D-2)}{2D}(\partial^2_X m^2)}{(p_\epsilon^2+m^2(X))^2}
    \bigg]
 \label{final result: single field 0}\\
    &\!\!=\!\!&\frac{1}{(4\pi)^{D/2}} \int {\rm d}^Dx \bigg[
    \frac{1}{8}(\partial_{\mu}m^2)(\partial^{\mu}m^2) (m^2)^{-3+\frac{D}{2}} \Gamma\Big(3\!-\!\frac{D}{2}\Big) 
   \nonumber
    \\
    &&\hskip 3.cm
    - \frac{1}{8} (\partial_{\mu}\partial^{\mu}m^2)(m^2)^{-2+\frac{D}{2}} \Gamma\Big(2\!-\!\frac{D}{2}\Big)\bigg] 
 \label{final result: single field 1} \\
&\!\!=\!\!& 0 
\,,
\label{final result: single field}
\end{eqnarray}
where, to get the second equality, we made use of the integral~(\ref{appendix D: easy integral: In}), and in the last step we integrated by parts the second term in~(\ref{final result: single field 1}). We also used the substitution $p^{\mu}p^{\nu}\rightarrow \eta^{\mu \nu}\frac{p^2}{D}$, which is legitimate since we expanded the logarithm at second order in number of derivatives. This result agrees with equation \eqref{final result - direct integration}, obtained without expanding the logarithm of the propagator.
Moreover, it agrees with the recent result~\cite{Glavan:2023lvw}, obtained by using the propagator constructed 
in the special case of an accelerating cosmological background.

\subsection{Taking a parametric derivative}
\label{Taking a parametric derivative}

The third method we wish to present is based on taking 
a parametric derivative with respect to the mass parameter, and it is
the method of choice in the literature~\cite{Fraser,Aitchison-Fraser,Aitchison-Fraser2,Chan}
\footnote{The method used in~\cite{Iliopoulos-Itzykson-Martin} is based on a direct evaluation of the effective 
action by using the Schwinger-DeWitt proper time method.
The details of the evaluation are discussed in Appendix~\ref{Appendix: literature}.}.
Let us begin by rewriting the one-loop effective action~(\ref{1 loop effective action})
 in the simple case of one real scalar field with canonical 
kinetic term as, 
\begin{equation}
\Gamma_\alpha^{(1)} = \frac{i}{2}{\rm Tr}\log\left[\big(\partial_x^2 - \alpha m^2(x)\big)\delta^D(x\!-\!x')\right]
\,,\quad
\label{1 loop effective action: parametric}
\end{equation}
where $\alpha$ is a parameter to be taken equal to unity at the end of calculation
and $m^2(x)$ is generally a field dependent mass squared.
Taking derivative with respect to $\alpha$ yields,
\begin{equation}
\frac{\partial}{\partial \alpha}
\Gamma_\alpha^{(1)} = - \frac{1}{2}{\rm Tr}\left[m^2i\Delta_\alpha(x;x')\right]
 =  - \frac{1}{2}\int {\rm d}^Dx\, m^2i\Delta_\alpha(x;x)
\,,\quad
\label{1 loop effective action: parametric derivative}
\end{equation}
where $i\Delta_\alpha(x;x')$ denotes the scalar propagator with the mass parameter $\alpha m^2(x)$,
and $i\Delta_\alpha(x;x)$ is the corresponding coincident propagator. The effective action
$\Gamma^{(1)} \equiv\Gamma_{\alpha=1}^{(1)} $ is then obtained 
by integrating~(\ref{1 loop effective action: parametric derivative}) over $\alpha$,
which equals up to a field-independent integration constant,
\begin{equation}
\Gamma^{(1)} =  - \frac{1}{2}\int_0^1 {\rm d}\alpha\int {\rm d}^Dx\, m^2(x)i\Delta_\alpha(x;x)
\,,\quad
\label{1 loop effective action: alpha=1}
\end{equation}
The effective action~(\ref{1 loop effective action: alpha=1}) is evaluated by 
inserting the coincident propagator $i\Delta_\alpha(x;x)$ 
which is obtained from~(\ref{coincident propagator: 0+2})--(\ref{prop2coin})
by substituting $m^2(x)\rightarrow \alpha m^2(x)$. The result is, 
\begin{eqnarray}
\Gamma^{(1)} &\!\!=\!\!&  - \frac{1}{2 (4\pi)^{D/2}}\int_0^1 {\rm d}\alpha\int {\rm d}^Dx\, m^2(x)\bigg\{ 
(\alpha m^2)^{\frac{D}{2}-1}\Gamma\bigg(\!1\!-\!\frac{D}{2}\bigg)
\nonumber\\
 &\!\!&\hskip 1cm
+\,\frac{(\alpha m^2)^{\frac{D}{2}-4}}{12}\Gamma\bigg(\!4\!-\!\frac{D}{2}\bigg)
 (\partial_x \alpha m^2)^2
    -\frac{(\alpha m^2)^{\frac{D}{2}-3}}{6}\Gamma\bigg(\!3\!-\!\frac{D}{2}\bigg) (\partial^2_x\alpha m^2) 
\bigg\}
\\
 &\!\!=\!\!&  \frac{1}{2 (4\pi)^{D/2}}\int {\rm d}^Dx\bigg\{ 
(m^2)^{\frac{D}{2}}\Gamma\bigg(\!\!-\!\frac{D}{2}\bigg)
\nonumber\\
 &\!\!&\hskip 1cm
-\,\frac{(m^2)^{\frac{D}{2}-3}}{6(D\!-\!2)}\Gamma\bigg(\!4\!-\!\frac{D}{2}\bigg)
 (\partial_x m^2)^2
    +\frac{(m^2)^{\frac{D}{2}-2}}{3(D\!-\!2)}\Gamma\bigg(\!3\!-\!\frac{D}{2}\bigg) (\partial^2_x m^2) 
\bigg\}
\,,\quad
\label{1 loop effective action: alpha=1 2}
\end{eqnarray}
where we exchanged the order of integration, 
$\int_0^1 {\rm d}\alpha\int {\rm d}^Dx\rightarrow\int {\rm d}^Dx\int_0^1 {\rm d}\alpha$.
Upon  integrating by parts the last term in~(\ref{1 loop effective action: alpha=1 2})
and dropping a boundary term one obtains~\footnote{
Instead of using $\alpha$, 
we could have used the tree level mass $M^2$ as the parameter, defined as $m^2(x) = M^2 + \frac{\lambda}{2}\phi^2(x)$, as done in the literature~\cite{Fraser,Aitchison-Fraser,Aitchison-Fraser2,Chan}.
This would result in an ambiguity produced by 
${\rm d} M^2 ={\rm d}\left[M^2 + \frac{\lambda}{2}\phi^2(x)\right]$. 
This ambiguity would produce an off-shell effective action that differs from 
the action~(\ref{1 loop effective action: alpha=1 3}).}, 
\begin{eqnarray}
\Gamma^{(1)}  &\!\!=\!\!&   \frac{1}{2 (4\pi)^{D/2}}\int {\rm d}^Dx\bigg\{ 
(m^2)^\frac{D}{2}\Gamma\bigg(\!\!-\!\frac{D}{2}\bigg)
\!-\!\frac{(m^2)^{\frac{D}{2}-3}}{12}\Gamma\bigg(\!3\!-\!\frac{D}{2}\bigg)
 (\partial_x m^2)^2
\bigg\}
\,,\quad
\label{1 loop effective action: alpha=1 3}
\end{eqnarray}
which upon renormalization yields, 
\begin{eqnarray}
\Gamma^{(1)}_{\rm ren} &\!\!=\!\!&   \frac{1}{32\pi^2}\int {\rm d}^Dx\bigg\{ \!\!
-\frac{m^4}{2}\left[\log\left(\frac{m^2}{4\pi\mu^2}\right)\!+\!\gamma_E\!-\!\frac32\right]
\!-\!\frac{1}{12m^2}(\partial_x m^2)^2
\bigg\}
\,,\quad
\label{1 loop effective action: alpha=1 4}
\end{eqnarray}
The results obtained by the last 
method~(\ref{1 loop effective action: alpha=1 3}--\ref{1 loop effective action: alpha=1 4})
 agree with those found in the 
literature~\cite{Fraser,Aitchison-Fraser,Aitchison-Fraser2,Chan,Iliopoulos-Itzykson-Martin},
however the results in~\eqref{final result - direct integration} and \eqref{final result: single field}  
obtained by the first two methods disagree. 

To summarize, direct integration of the logarithm in the one-loop effective action
and a derivative expansion of the logarithm give a vanishing result for the second derivative correction to the 
 one-loop effective action,
while the method of parametric derivative/integration and  the method of proper time
give a different, non-vanishing result. Even though one of the two results cannot be correct, 
we were unable to find a convincing argument based on which one could reject one of the two answers~\footnote{
 One potential source of error might be in exchanging the order of integration, which is illegitimate
for divergent integrals, which is the case in our problem.}. A more detailed discussion regarding
the disagreement with previous works in the literature can be found in Appendix~\ref{Appendix: literature}
and in Section~\ref{Sec.V}.

In what follows, we generalise the single-field results of this section to the case of many fields, 
and we calculate the effective action by expanding the logarithm.

\section{Generalisation to multiple fields with mass mixing}
\label{Sec.III}

We now aim at generalising the result obtained with the expansion of the logarithm of the propagator to the case of multiple scalar fields with canonical kinetic terms and mixing mass terms,  which can be relevant for multifield inflationary models~\cite{Dvali,Bernardeau:2002jy,Rigopoulos:2005us,Senatore:2010wk,Kaiser:2012ak,Liu-Prokopec,Barnaveli}  and 
for elecroweak symmetry breaking in scaling symmetric extensions of the standard model~\cite{Chataignier:2018aud,Chataignier:2018kay,Rezacek,Carone, Englert,Mohamadnejad}.
In this case, the two symmetric equations of motion of the propagator are:
\begin{equation}
    \begin{split}
        (\mathbb{1} \partial_x^2 -\mathbb{M}^2(x))_{ab} i\Delta(x,y)_{bc}&=i\delta_{ac}\delta^D(x-y)\; ,\\
        i\Delta(x,y)_{cb}(\mathbb{1} \overleftarrow{\partial}_y^2 -\mathbb{M}^2(y))_{ba} &=i\delta_{ac}\delta^D(x-y)\; ,
    \end{split}
\label{two symmetric equations}
\end{equation}
where $\mathbb{M}^2$ is the off-diagonal mass matrix, which is taken to be real and transpose invariant. It can be obtained from the tree-level potential as  $\mathbb{M}^2_{ab}=\frac{\partial^2 V^{(0)}}{\partial \phi_b \partial \phi_a}$ and it can contain the tree-level mass but also field-dependent contributions coming from higher order interactions of the fields. 

In case of translation-invariant background fields, the off-diagonal mass matrix can be easily diagonalized and the effective potential is obtained by summing over all the diagonal terms. For spacetime dependent fields the situation is more involved however, and the result of the previous section does not carry over to the case of mixing flavours.

Similarly to the single field case, we can transform the two equations of motion into Wigner space using~\eqref{wigner transform with diamond}:
\begin{eqnarray}
    \Big[\!-\!\mathbb{1}p^2\!-\!\mathbb{M}^2(X)\!-\!i\mathbb{1} p \!\cdot\! \partial_X 
    \!+\!\frac{i}{2} (\partial_X \mathbb{M}^2)\!\cdot\! \partial_p 
    \!+\!\frac{1}{4}\mathbb{1} \partial^2_X
     \!+\!\frac{1}{8} (\partial_{\mu} \partial_{\nu}\mathbb{M}^2)\partial_{p_{\mu}}\partial_{p_{\nu}}
     \Big]\!\cdot\!
i\Delta(X;p)&\!=\!& i\mathbb{1}
\,,\qquad
\label{multifield propagator equation: Wigner 1}
\\ 
i\Delta(X;p)\!\cdot\!\Big[\!-\!\mathbb{1}p^2\!-\!\mathbb{M}^2(X)
       \!+\!i\mathbb{1} p \!\cdot\! \overleftarrow{\partial_X} 
    \!-\!\frac{i}{2} (\partial_X \mathbb{M}^2)\!\cdot\! \overleftarrow{\partial_p} 
    \!+\!\frac{1}{4}\mathbb{1} \overleftarrow{\partial}^2_X 
\!+\!\frac{1}{8} (\partial_{\mu} \partial_{\nu}\mathbb{M}^2)\overleftarrow{\partial}_{p_{\mu}}\overleftarrow{\partial}_{p_{\nu}}\Big]&\!=\!& i\mathbb{1}
\,. \qquad
\label{multifield propagator equation: Wigner 2}
\end{eqnarray}
Upon transposing the second equation and making use of 
Eq.~(\ref{propagator: symmetry requirement: Wigner}),
one sees that~(\ref{multifield propagator equation: Wigner 1})
can be obtained from~(\ref{multifield propagator equation: Wigner 2}), making them consistent.
Another way of saying the same is that one can replace 
equation~(\ref{multifield propagator equation: Wigner 2}) 
by the symmetry condition on the propagator~(\ref{propagator: symmetry requirement: Wigner}).

With this in mind, we can focus on solving the first propagator equation 
in~(\ref{two symmetric equations}).
Expanding the propagator in gradients as in the one-field case~(\ref{midpoint general propagator}), at zeroth order one gets:
\begin{equation}
    (-\mathbb{1}p^2-\mathbb{M}^2)\!\cdot\!i\Delta^{(0)} =i
     \; \Longrightarrow \;  i\Delta^{(0)}(X;p)=i\big[-\mathbb{1}p^2-\mathbb{M}^2\big]^{-1}
     \,,
\end{equation}
which is clearly transpose invariant (symmetric in flavour $ab$), as it ought to be.

Similarly, Eq.~(\ref{multifield propagator equation: Wigner 1}) implies
for the first-order correction,
\begin{equation}
        \Delta^{(1)}(X;p)=\Delta^{(0)}\Big[i\mathbb{1}p \!\cdot\! \partial_X -\frac{i}{2}(\partial_X \mathbb{M}^2)\! \cdot\!  \partial_p\Big] \Delta^{(0)}
        \,,
\label{first order correction: matrix case}
\end{equation}
which can be solved by exploiting the relation, ${\rm d}A=-A ({\rm d}A^{-1}) A$, 
with $A$ being a matrix and $A^{-1}$ its inverse. 
The matrices $\partial_X \mathbb{M}^2$
and $\Delta^{(0)}$  in~(\ref{first order correction: matrix case}) 
in general do not commute resulting in,
\begin{equation}
    \Delta^{(1)}(X;p)
     = - \Delta^{(0)} \mathcal{D}^{(1)}\Delta^{(0)}
     =\Delta^{(0)}[\Delta^{(0)},ip \!\cdot\! (\partial_X \mathbb{M}^2)] \Delta^{(0)}
    \,,
\label{first order correction: matrix case 2}
\end{equation}
which is clearly transpose invariant in the sense 
of Eq.~(\ref{propagator: symmetry requirement: Wigner}).
Therefore, the first-order propagator correction vanishes only if the two matrices commute. 

Proceeding to the second-order correction, from~(\ref{multifield propagator equation: Wigner 1})
 one obtains,
\begin{equation}
    \Delta^{(2)}(X;p)=-\Delta^{(0)} \mathcal{D}^{(2)}\Delta^{(0)}-\Delta^{(0)}\mathcal{D}^{(1)} \Delta^{(1)}
    \,.
    %
\label{2order propagator0}
\end{equation}
This can be evaluated by exploiting the matrix relation ${\rm d}^2A=-A ({\rm d}^2A^{-1}) A+2A ({\rm d}A^{-1})A({\rm d}A^{-1})A$ 
and keeping in mind that $\partial_{\mu} \partial_{\nu} \mathbb{M}^2$ and $\Delta^{(0)}$ do not commute,
\begin{eqnarray}
    \Delta^{(2)}&\!\!=\!\!&-\frac{1}{4} \Delta^{(0)} (\partial^2_X \mathbb{M}^2)(\Delta^{(0)})^2-\frac{1}{4} (\Delta^{(0)})^2 (\partial^2_X \mathbb{M}^2)\Delta^{(0)}+\frac{1}{2} \Delta^{(0)} (\partial_X \mathbb{M}^2) (\Delta^{(0)})^2(\partial_X \mathbb{M}^2)\Delta^{(0)}
\nonumber\\
    &&-\Delta^{(0)} \big[ (p\!\cdot\!\partial)^2\mathbb{M}^2\big](\Delta^{(0)})^3
 -(\Delta^{(0)})^3 \big[ (p\!\cdot\!\partial)^2\mathbb{M}^2\big]\Delta^{(0)}
 +(\Delta^{(0)})^2\big[ (p\!\cdot\!\partial)^2\mathbb{M}^2\big](\Delta^{(0)})^2
\nonumber\\
    &&    -\frac{1}{2} (\Delta^{(0)})^2 (\partial_X \mathbb{M}^2)\Delta^{(0)}(\partial_X \mathbb{M}^2)\Delta^{(0)} -\frac{1}{2} \Delta^{(0)} (\partial_X \mathbb{M}^2)\Delta^{(0)}(\partial_X \mathbb{M}^2) (\Delta^{(0)})^2 
\nonumber\\ 
    &&  +2\Delta^{(0)}\big(\Delta^{(0)}p\!\cdot\!\partial\,\mathbb{M}^2\big)\Delta^{(0)}
      \big(p\!\cdot\!\partial\,\mathbb{M}^2\Delta^{(0)}\big)\Delta^{(0)}
    +2\Delta^{(0)}\big( p\!\cdot\!\partial\,\mathbb{M}^2\Delta^{(0)}\big)\Delta^{(0)}
      \big(\Delta^{(0)}p\!\cdot\!\partial\,\mathbb{M}^2\big)\Delta^{(0)}
 \nonumber     \\
 && -2\left(\Delta^{(0)}\right)^{\!3}\big(p\!\cdot\!\partial\,\mathbb{M}^2\big)\Delta^{(0)}
      \big(p\!\cdot\!\partial\,\mathbb{M}^2\big)\Delta^{(0)}
      -2\Delta^{(0)}\big(p\!\cdot\!\partial\,\mathbb{M}^2\big)\Delta^{(0)}
      \big(p\!\cdot\!\partial\,\mathbb{M}^2\big)\left(\Delta^{(0)}\right)^{\!3}
\,,\qquad
\label{2order propagator}
\end{eqnarray}
which is symmetric in flavour and quadratic in $p^\mu$, and therefore 
satisfies~(\ref{propagator: symmetry requirement: Wigner}). 

As before, we can compute the relative corrections to the one-loop effective action starting from the general expression for real scalar fields and expanding the propagator in number of derivatives. Unlike in the single field case, the first-order corrections do not vanish, thus the complete expression for the one-loop effective action is:
\begin{equation}
    \Gamma^{(1)}=-\frac{i}{2} \Big[\text{Tr}\log(\Delta^{(0)})+\text{Tr}(\mathcal{D}^{(0)}\Delta^{(1)} ) + \text{Tr}(\mathcal{D}^{(0)}\Delta^{(2)} )
    -\frac{1}{2}  \text{Tr}(\mathcal{D}^{(0)}\Delta^{(1)} )^2+...
    \Big]
    \,.
    \label{gradient expansion effective action}
\end{equation}
The first order correction vanishes on two accounts: (1) 
any contribution to a trace containing a commutator must vanish;
 (2) an integral over the momenta of a function $\propto p^\mu$ vanishes.
Next, using~(\ref{2order propagator})
the second-order correction in~(\ref{gradient expansion effective action}) becomes, 
\begin{eqnarray}
    \text{Tr}(\mathcal{D}^{(0)}\Delta^{(2)} )-\frac{1}{2}\text{Tr}(\mathcal{D}^{(0)}\Delta^{(1)} )^2&\!\!=\!\!& \text{Tr}\bigg(\!-\!\frac{1}{2} (\partial^2_X \mathbb{M}^2)(\Delta^{(0)})^2
\!-\!\frac{1}{2}(\partial_X \mathbb{M}^2)\Delta^{(0)}(\partial_X\mathbb{M}^2)(\Delta^{(0)})^2
\nonumber\\
&& \!-\! \big[ (p\!\cdot\!\partial)^2\mathbb{M}^2\big](\Delta^{(0)})^3 
\!\!+\! \big[p\!\cdot\!\partial\mathbb{M}^2\big](\Delta^{(0)})^2\big[p\!\cdot\!\partial\mathbb{M}^2\big](\Delta^{(0)})^2 
\nonumber\\
&& \!-\! \big[p\!\cdot\!\partial\mathbb{M}^2\big]\Delta^{(0)}\big[p\!\cdot\!\partial\mathbb{M}^2\big](\Delta^{(0)})^3\!\bigg) \;,
\qquad
\label{2 order ea multifield}
\end{eqnarray}
where to simplify the expression we moved the terms cyclically, which is allowed under a trace.

We perform the computations for a generic number of scalar fields with mass mixing and we do not assume a particular form for the mass matrix $\mathbb{M}^2$. Firstly, we diagonalise the zeroth-order propagator using a suitable rotation matrix $\mathcal{R}$. Having a diagonal propagator, the momentum integral can be performed in the standard way. We find then convenient to separate the sum over the diagonal and off-diagonal terms and to transform all the remaining mass matrices to be diagonal. Depending on the terms in equation \eqref{2 order ea multifield}, we have to deal with first- or second-order spatial derivatives of $\mathbb{M}^2$. Consequently, the diagonalisation procedure introduces derivatives of the rotation matrices. It is then useful to define the matrix product $T=\mathcal{R}(\partial_X \mathcal{R}^T)$. The detailed calculations can be found in Appendix~\ref{Appendix: multifields}.

Using equation \eqref{second order appendix result} we can rearrange the different terms, such that we can write the final expression for the second-order corrections of the effective action as:
\begin{equation}
\begin{split}
    &\text{Tr}(\mathcal{D}^{(0)}\Delta^{(2)} )-\frac{1}{2} \left[ \text{Tr}(\mathcal{D}^{(0)}\Delta^{(1)} )^2\right]
    =\frac{-i}{(4\pi)^{D/2}} \Gamma\bigg( 1-\frac{D}{2}\bigg) \int {\rm d}^DX \sum_{i\neq k} T_{ik} T_{ki}\\
    &\hskip 3cm 
    \times \frac{2D(8-D)m_i^2 m_k^2+(48-14D+D^2)m_k^4+(D^2-2D)m_i^4}{m_i^2-m_k^2} \frac{(m_k^2)^{\frac{D}{2}-2}}{8D}
    \; ,\quad
    \label{second gradient general formula}
\end{split}
\end{equation}
where $m_i^2$ are the spacetime dependent eigenvalues of the diagonalised mass matrix.
Interestingly, we see that all the diagonal terms cancel and the second-order corrections to the effective action are given only by the summation of off-diagonal terms. This is consistent with the conclusions that we obtained in the single-field case.

It is important to remember that in four dimensions the Gamma function in equation \eqref{second gradient general formula} is divergent so we ought to investigate if this correction has to be renormalised with appropriate counterterms, similarly to what is usually done for the one-loop effective potential. As it can be seen from the analysis in Appendix~\ref{Appendix: scale-independent}, the divergences in equation \eqref{second gradient general formula} cancel exactly so we are left with finite second-order gradient corrections. Consequently, we can also write our result in a manifestly scale-independent way.
Considering equation \eqref{D second order ea} and taking the limit $D\rightarrow 4$ yields to,
\begin{equation}
    \Gamma^{(1,2)}=\frac{-1}{8(4\pi)^2} \int {\rm d}^4X \sum_{i\neq k} T_{ik} T_{ki} \bigg[3m_i^2+ \frac{m_i^4+2m_i^2m_k^2}{m_i^2-m_k^2}\log\bigg(\frac{m_k^2}{m_i^2} \bigg)\bigg]\; .
    \label{4dim second order ea}
\end{equation}
This is our result for the second-order gradient correction of the one-loop effective action. The derivatives of the fields are hidden in the matrix product $T=\mathcal{R}(\partial_X \mathcal{R}^T)$. 
Notice that if $m_i^2=m_k^2$ for some $i$ and $k$, then the expression in the square brackets reduces to $0$, which is consistent with our single field
calculation~(\ref{final result: single field}) in section~\ref{Sec.II}. 

\subsection*{Application to a simple toy model}

We are now interested in applying the previous result to a simple toy model of two real scalar fields with mass mixing. We consider the following action:
\begin{equation}
    \mathcal{S}=\int {\rm d}^Dx \bigg[ -\frac{1}{2}(\partial_{\mu}\phi)( \partial^{\mu}\phi)-\frac{1}{2}(\partial_{\mu}\psi) (\partial^{\mu}\psi)-V^{(0)}(\phi,\psi) \bigg]\; ,
    \label{L of 2 scalar fields with mass mixing}
\end{equation}
where $V^{(0)}$ is the tree-level potential of the theory. 

Without assuming a specific expression for the tree-level potential, we have a tree-level mass matrix with the following generic form:
\begin{equation}
    \mathbb{M}^2=\begin{pmatrix}
     m_{11}^2 & m_{12}^2\\
    m_{12}^2&m_{22}^2\\
    \end{pmatrix}\; ,
    \label{tree-level mass matrix}
\end{equation}
where $m^2_{ab}=\frac{\partial^2 V^{(0)}}{\partial \phi_a \partial \phi_b}$.
We can diagonalise the mass matrix using the following rotation matrix:
\begin{equation}
\mathcal{R}=
 \begin{pmatrix}
\cos \theta& -\sin \theta \\
\sin \theta  & \cos \theta\\
\end{pmatrix}\; ,
\end{equation}
with:
\begin{equation}
    \tan (2 \theta)=\frac{2m^2_{12}}{m^2_{22}-m^2_{11}}\; ,
    \label{tan 2theta}
\end{equation}
and we find the following mass eigenvalues:
\begin{equation}
  m_{\pm}^2=\frac{1}{2}\bigg(m_{11}^2+m_{22}^2 \pm \sqrt{\Delta}\bigg)\; ,
\end{equation}
with the discriminant $\Delta=m_{11}^4+4m_{12}^4-2m_{11}^2m_{22}^2+m_{22}^4$. We can now compute the second-order gradient corrections to the effective action of this theory. To do so, we apply equation \eqref{4dim second order ea}. We can start by noticing that the product:
\begin{equation}
\begin{split}
    \mathcal{R}(\partial_{\mu}\mathcal R^T)=&
    \begin{pmatrix}
\cos \theta& -\sin \theta \\
\sin \theta  & \cos \theta\\
\end{pmatrix} \cdot \begin{pmatrix}
\partial_{\mu}\cos \theta& \partial_{\mu}\sin \theta \\
-\partial_{\mu}\sin \theta  & \partial_{\mu}\cos \theta\\
\end{pmatrix}\\
=&\begin{pmatrix}
0& 1 \\
-1  & 0\\
\end{pmatrix}\cdot (\partial_{\mu}\theta)\; ,
\end{split}
\end{equation}
is an antisymmetric matrix and thus:
\begin{equation}
    \mathcal{R}(\partial_{\mu}\mathcal R^T)\mathcal{R}(\partial^{\mu}\mathcal R^T)=-\mathbb{1}(\partial_{\mu}
    \theta)(\partial^{\mu}
    \theta)\; ,
\end{equation}
such that the condition of summing only over off-diagonal terms in equation \eqref{4dim second order ea} is automatically satisfied. Exploiting equation \eqref{4dim second order ea} and taking the trace, we obtain for the effective action:
\begin{equation}
\begin{split}
    &\Gamma^{(1)}=\frac{1}{8(4\pi)^2} \int {\rm d}^4x (\partial_{\mu}
    \theta)(\partial^{\mu}
    \theta) \bigg[3m_+^2+3m_-^2+ \frac{m_+^4+m_-^4+4m_+^2m_-^2}{m_+^2-m_-^2}\log\bigg(\frac{m_-^2}{m_+^2} \bigg)\bigg]\\
    &=\frac{1}{32(4\pi)^2} \int {\rm d}^4x \frac{(\partial_{\mu}\tan 2\theta)^2}{(1+(\tan2\theta)^2)^2} \bigg[3m_+^2+3m_-^2+ \frac{m_+^4+m_-^4+4m_+^2m_-^2}{m_+^2-m_-^2}\log\bigg(\frac{m_-^2}{m_+^2} \bigg)\bigg] ,
\end{split}
\label{two fields 2gradient primitive}
\end{equation}
which, using \eqref{tan 2theta}, can be rewritten as,
\begin{equation}
\begin{split}
    \Gamma^{(1)}=\frac{1}{8(4\pi)^2} & \int {\rm d}^4x  \frac{1}{\Delta^{5/2}}[(m_{11}^4+m_{22}^4+4m_{11}^2m_{22}^2-2m_{12}^4)\log\bigg(\frac{m_-^2}{m_+^2} \bigg)\\
    & +3(m_{11}^2+m_{22}^2)\sqrt{\Delta}][(m_{11}^2-m_{22}^2)\partial_{\mu}m_{12}^2+m_{12}^2(\partial_{\mu}m_{22}^2-\partial_{\mu}m_{11}^2)]^2\; .
\end{split}
\label{two fields 2gradient final}
\end{equation}
Clearly, expression \eqref{two fields 2gradient primitive} is invariant under the exchange $m_+^2\leftrightarrow m_-^2$. Of course, this still holds in \eqref{two fields 2gradient final} since if $m_+^2\leftrightarrow m_-^2$, the sign change of the logarithm is compensated by $\sqrt{\Delta}\rightarrow -\sqrt{\Delta}$, such that the effective action remains invariant. 
Moreover, it is worth noting that, if the mass eigenvalue $m_-^2$ turns negative, {\it i.e.} if
\begin{equation}
m_{11}^2m_{22}^2<m_{12}^4\; ,
\label{m_- neg}
\end{equation}
we obtain the logarithm of a negative quantity which gives complex one-loop corrections to the kinetic term. This is not surprising, since in this case there is at least one tachyonic mass, and for actions with tachyonic masses 
there is no stable equilibrium and thus also no well-defined {\it in-out} effective action. The suitable formalism to use in these 
situations is the {\it in-in} formalism.
Similarly, complex results are obtained in case when $\Delta<0$. 
Finally, vanishing second-order kinetic corrections are found in the limit $m_{11}^2=m_{22}^2$. In this limit the tangent in equation \eqref{tan 2theta} is undefined. We notice that if $m_{11}^2=m_{22}^2$, we have that the action is invariant under exchange of the two scalar fields $\phi \leftrightarrows \psi$.


\section{Two-loop gradient expansion}
\label{Sec.IV}

So far, we have performed the gradient expansion of the effective action up to one-loop order. As already said, the gradient expansion can be seen as an expansion in $\hbar  \partial_x $. Similarly, in the loop expansion, each loop brings a factor of $\hbar$.
It is therefore important to highlight that in our framework we are treating the loop and the gradient expansion as different and independent expansions in $\hbar$. 

Coming back to the single-field case, in this section we report the results for the two-loop effective action expanded up to second order in gradients. 
\begin{figure}
    \centering    \includegraphics[width=0.6\textwidth]{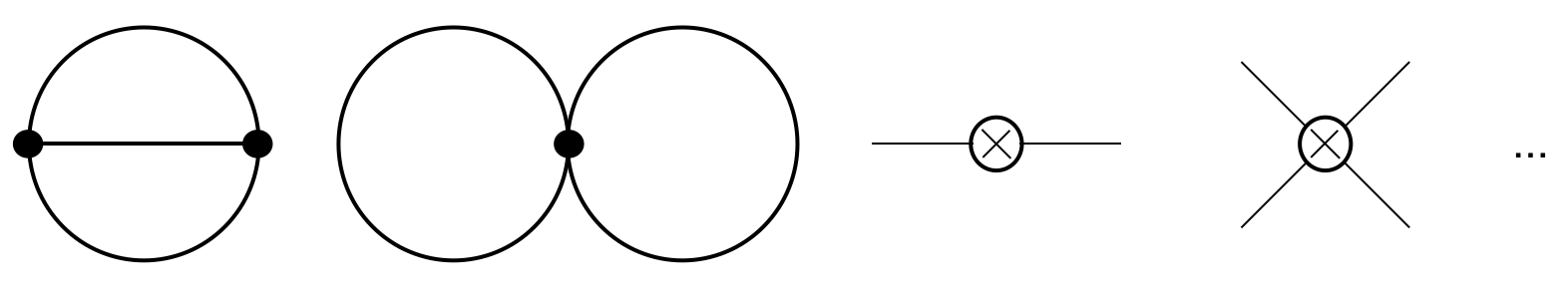}
    \caption{Diagrams contributing to the two-loop 1PI effective action and counterterms. The dots represents the infinite series of counterterms diagrams with 2{\it n} external legs.
    }
    \label{fig:vacuumdiagramsl}
\end{figure}
In Figure~\ref{fig:vacuumdiagramsl}, the two one-particle irreducible diagrams that contribute to the two-loop effective action are shown. The corresponding contributions are:
\begin{equation}
    \Gamma^{(2)}=\frac{i}{12} \int {\rm d}^Dx\, {\rm d}^Dx'\, \frac{\delta^3 S}{\delta \phi^3(x)}\frac{\delta^3 S}{\delta \phi^3(x')}[i\Delta(x,x')]^3-\frac{\lambda}{8} \int {\rm d}^Dx \, [i \Delta(x,x)]^2 + S^{(2)}_{\rm ct}.
\end{equation}
For the counterterm action, we assume the following form,
\begin{equation}
\begin{split}
    S_{\rm ct}^{(2)}=\int {\rm d}^Dx \; \bigg(&\!\!-\!\frac{\Lambda}{8 \pi G}
    \!-\!\frac{Z_0}{2}(\partial_{X}\phi)^2\!-\!\frac{m_0^2}{2} \phi^2\!-\! \frac{\lambda_0}{4!} \phi^4
    \\
    &\! -\!\sum_{n=3}^{\infty} \frac{\lambda_{(n)}}{(2n)!}\phi^{2n}
    -\!\sum_{n=1}^{\infty}\frac{ Z_{(2n)}}{(2n)!}\phi^{2n}(\partial_{X}\phi)^2
    \bigg)
\,,
\label{ct action}
\end{split}
\end{equation}
where the counterterm action is fixed by imposing that $\Gamma^{(2)}$ remains finite in the limit when 
$D\rightarrow 4$. As we will see, surprisingly, an infinite series of counterterms is needed to subtract the divergent parts
in $\Gamma^{(2)}$, rendering the theory nonrenormalisable.

It is convenient to transform to midpoint and relative coordinates and rewrite the previous expression as:
\begin{equation}
\begin{split}
    \Gamma^{(2)}=&\frac{i}{12} \int {\rm d}^DX {\rm d}^Dr \frac{{\rm d}^Dp}{(2\pi)^D} \frac{{\rm d}^Dp'}{(2\pi)^D} \frac{{\rm d}^Dp''}{(2\pi)^D} \lambda^2 \phi\bigg(X+\frac{r}{2}\bigg) \phi\bigg(X-\frac{r}{2}\bigg)e^{i(p+p'+p'')r}\\
    &\times i\Delta(X,p)i \Delta(X,p')i\Delta(X,p'')
    -\frac{\lambda}{8} \int d^DX \int \frac{{\rm d}^D p}{(2 \pi)^D} \frac{{\rm d}^D p'}{(2 \pi)^D} i\Delta(X,p) i\Delta(X,p')\\
    &+ S^{(2)}_{ct}.
\end{split}
\label{two loop effective action}
\end{equation}
Exploiting the results obtained for the one-loop effective action, we can now perform a midpoint gradient expansion of the propagator as $i\Delta=i\Delta^{(0)}+i\Delta^{(2)}+...$ as in equation \eqref{midpoint general propagator}. Moreover, we can also expand the field around the midpoint coordinate as $\phi(X+r/2)=\phi(X)+\partial_{\mu}\phi \frac{r^{\mu}}{2}+\frac{1}{2}\partial_{\mu}\partial_{\nu}\phi \frac{r^{\mu}r^{\nu}}{4}$.

In the following, we adopt the notation: 
\begin{equation}
    \Gamma^{(2)}=\Gamma^{(2,0)}+\Gamma^{(2,2)}.
\end{equation}
$\Gamma^{(2,0)}$ is the zeroth order in gradients contribution to the two-loop effective action, while $\Gamma^{(2,2)}$ is the second order in gradients one. It is straightforward to check that the contribution of first order in gradients vanishes, as expected.

\subsection{$\Gamma^{(2,0)}$}

To find the contribution of zeroth order in gradients, we use the zeroth order propagator:
\begin{equation}
   i \Delta^{(0)}(X,p)=-\frac{i}{p_\epsilon^2+m^2(X)}\; .
    \label{prop0}
\end{equation}
and from now on, we assume: 
\begin{equation}
m^2(X)=M^2+\frac{1}{2}\lambda \phi^2(X)\; ,
\label{space dependent mass}
\end{equation} 
where $M$ denotes the tree-level mass. The zeroth order contribution in gradients is then given by,
\begin{equation}
\Gamma^{(2,0)}=\Gamma^{(2,0)}_a+\Gamma^{(2,0)}_b
\,,
\label{Gamma 20 split}
\end{equation}
where the momentum integrals that we ought to solve are:
\begin{equation}
\begin{split}
   \Gamma^{(2,0)}_a=&-\frac{\lambda^2}{12} \int_{X,p,p'} \phi^2(X) \,
    \frac{1}{p_\epsilon^2+m^2(X)}\frac{1}{p_\epsilon'^2+m^2(X)} \frac{1}{(p+p')_\epsilon^2+m^2(X)}\; ,
\end{split}
\label{I1a: two loop integral}
\end{equation}
\begin{equation}
   \Gamma^{(2,0)}_b=\frac{\lambda}{8} \int_{X,p,p'}  \frac{1}{p_\epsilon^2+m^2(X)}\frac{1}{p_\epsilon'^2+m^2(X)}
    \,,
\label{I1b: simple integral}
\end{equation}
where we used the shorthand notation $\int_{X,p,p'}=\int {\rm d}^DX \int \frac{{\rm d}^D p}{(2 \pi)^D} \frac{{\rm d}^D p'}{(2 \pi)^D} $. 

These contributions are evaluated in Appendix~\ref{Two-loop calculations}, 
and the results are given in Eqs.~(\ref{appendix D: Gamma 20a final}) 
and~(\ref{appendix D: Gamma 20b final}),
 \begin{eqnarray}
     \Gamma^{(2,0)}_a&\!\!=\!\!&-\frac{\lambda}{2(4 \pi)^4}\int {\rm d}^DX 
     \lambda\phi^2(X)m^2
     \Bigg\{\frac{\mu^{2(D-4)}}{(D-4)^2}
     +\frac{\mu^{D-4}}{D-4}\left[\ln\left(\frac{m^2}{4\pi\mu^2}\right)\!+\!\gamma_E 
            \!-\!\frac32\right]
  \nonumber\\
  &\!\!\!\!& \hskip -.6cm
    \!+\,\frac12\left[\ln^2\!\left(\frac{m^2}{4\pi\mu^2}\right)
   \! -\!(3\!-\!2\gamma_E)\ln\left(\frac{m^2}{4\pi\mu^2}\right)
 \!+\!\gamma_E^2 \!-\!3\gamma_E\!+\!\frac{\pi^2}{12}\!+\!\frac{13}{6}\!+\!\frac{2}{3}c_1^*
            \right]
             \!\Bigg\}
\!\!+\!{\cal O}\big(D\!-\!4\big)
,\qquad\;
\label{Gamma 20a final} 
\end{eqnarray}
and
\begin{eqnarray}
\Gamma^{(2,0)}_b 
&\!\!=\!\!& -\frac{\lambda}{2(4\pi)^4}\int {\rm d}^DX m^4(X)
\Bigg\{\frac{\mu^{2(D-4)}}{(D-4)^2}
+\frac{\mu^{D-4}}{D-4}\bigg[\ln\left(\frac{m^2}{4\pi\mu^2}\right)\!+\!\gamma_E 
            \!-\!1\bigg]
\label{Gamma 20b final} \\
&\!\!\!\!&
+ \frac12\bigg[\ln^2\left(\frac{m^2}{4\pi\mu^2}\right)
 \!-\!2(1-\gamma_E)\ln\left(\frac{m^2}{4\pi\mu^2}\right)
 +\gamma_E^2-2\gamma_E+\frac{\pi^2}{12}+\frac{3}{2}\bigg]            
\Bigg\}
\nonumber
\!+\!{\cal O}\big(D\!-\!4\big)
\,,\qquad
\end{eqnarray} 
where $c_1^*\simeq 0.242\cdots$ is defined in~(\ref{appendix D: c1*}).

It is important to notice that the divergent parts of these results are non-renormalisable due to spacetime dependence of the logarithm terms. As anticipated, we need a counterterm action with an infinite series of terms.
Using the counterterm action in equation \eqref{ct action}, we find the following coefficients:
\begin{equation}
    \frac{\Lambda}{8 \pi G}=-\frac{ \lambda M^4}{2 (4 \pi)^4}  \bigg[ \frac{ \mu^{2(D-4)}}{(D-4)^2}+\frac{\mu^{D-4}}{(D-4)}\bigg(\ln\bigg(\frac{M^2}{4 \pi \mu^2}\bigg)+\gamma_E-1\bigg)\bigg]\;,
\end{equation}
\begin{equation}
    \frac{m_0^2}{2}=-\frac{\lambda^2 M^2}{(4 \pi)^4}\bigg[ \frac{ \mu^{2(D-4)}}{(D-4)^2}+\frac{\mu^{D-4}}{(D-4)}\bigg(\ln\bigg(\frac{M^2}{4 \pi \mu^2}\bigg)+\gamma_E-1\bigg)\bigg] \;,
\end{equation}
\begin{equation}
    \frac{\lambda_0}{4!}=-\frac{3 \lambda^3}{8(4 \pi)^4}\bigg[ \frac{ \mu^{2(D-4)}}{(D-4)^2}+\frac{\mu^{D-4}}{(D-4)}\bigg(\ln\bigg(\frac{M^2}{4 \pi \mu^2}\bigg)+\gamma_E-\frac{1}{6}\bigg)\bigg]\;,
\end{equation}
and for $n\geq3$:
\begin{equation}
    \frac{\lambda_{(n)}}{(2n)!}=\bigg(-\frac{\lambda}{2}\bigg)^{n+1} \frac{M^{4-2n}}{(4\pi)^4 } \frac{ \mu^{D-4}}{(D-4)} \bigg(-\frac{1}{n}+\frac{4}{n-1}-\frac{3}{n-2}\bigg)\;.
\end{equation}  

At leading order, the renormalised effective action has the form of an effective potential,
\begin{equation}
\Gamma^{(2,0)}_{\rm ren}\big[\phi\big] = - \int {\rm d}^4 X \; V^{(2)}(\phi)\; ,
\label{renormalized Gamma 20: Veff}
\end{equation}
where the renormalised two-loop effective potential is given by, 
\begin{eqnarray}
V^{(2)}(\phi)&\!\!=\!\!& \frac{\lambda}{4(4\pi)^4}\Bigg\{
 m^2\big(m^2 \!+\!\lambda\phi^2\big)
        \ln^2\!\left(\frac{m^2}{4\pi\mu^2}\right)
   \! -\!m^2\Big(2(1\!-\!\gamma_E)m^2 \!+\!(3\!-\!2\gamma_E)\lambda\phi^2\Big)
             \ln\left(\frac{m^2}{4\pi\mu^2}\right)
\nonumber\\
 &\!\!\!\!&\hskip 1.5cm
 +\,\Big(\gamma_E^2\!-\!2\gamma_E\!+\!\frac{\pi^2}{12}\!+\!\frac{3}{2}\Big)m^4
+\lambda\phi^2m^2\Big(\gamma_E^2 \!-\!3\gamma_E\!+\!\frac{\pi^2}{12}
       \!+\!\frac{13}{6}\!+\!\frac{2}{3}c_1^*\Big)
            \bigg]\Bigg\}
\, .\qquad\;
\label{renormalized Gamma 20: Veff 2}
\end{eqnarray}

\subsection{$\Gamma^{(2,2)}$}
We can proceed and calculate the second-order gradient corrections of the two-loop effective action. To do so, we need to consider the second-order propagator,
\begin{eqnarray}
i\Delta^{(2)} = -\Delta^{(0)} {\cal D}^{(2)}i\Delta^{(0)}
    =\frac12(\partial_X^2 m^2 )\big[i\Delta^{(0)}\big]^3
    - i\Big[\big(p\cdot\partial_X\big)^2 m^2 + \frac12 \big(\partial_X m^2 \big)^2\Big]
    \big[i\Delta^{(0)}\big]^4
    \,,\qquad
\label{prop2m}
\end{eqnarray}
where ${\cal D}^{(2)} = \frac14\partial_X^2 
 + \frac18\big(\partial_\mu \partial_\nu m^2(X)\big)\partial_{p_\mu}\partial_{p_\nu}$.
Aided by $m^2(X) = M^2 + \frac{\lambda}{2}\phi^2(X)$, this can be rewritten as, 
\begin{eqnarray}
 \Delta^{(2)}(X,p)=\frac{\frac{\lambda m^2}{D}((\partial_{X}\phi)^2 +\phi (\partial_{X}^2\phi))-\frac{\lambda^2 \phi^2}{2}(\partial_{X}\phi)^2}{(p_\epsilon^2+m^2(X))^4}
 +\frac{\frac{\lambda(D-2)}{2D}((\partial_{X}\phi)^2+\phi (\partial_{X}^2\phi))}{(p_\epsilon^2+m^2(X))^3}
 \,,\quad
    \label{prop2}
\end{eqnarray}
where we made use of the fact that, under the integral over the momentum, 
the following replacement,
$p^\mu p^\nu \rightarrow p^2\eta^{\mu\nu}/D$,  which is legitimate in the coincident limit.

The gradient suppressed contributions to the two-loop effective 
action~(\ref{two loop effective action}) can be 
neatly split into three contributions. The first comes from 
expanding the vertex around the midpoint,  the second (third) comes from the second-order 
contribution to the propagator~(\ref{prop2})
 in the part of the diagram coming 
from the cubic (quartic) vertex,
\begin{equation}
\Gamma^{(2,2)}=\Gamma^{(2,2)}_a+\Gamma^{(2,2)}_b+\Gamma^{(2,2)}_c
\,.
\label{splitting Gamma 22}
\end{equation}
By expanding the vertex in~(\ref{two loop effective action}) one obtains,
\begin{eqnarray}
\Gamma^{(2,2)}_a &\!\!=\!\!& \frac{\lambda^2}{24 D} \int_{X,p,p'} \Big[(\partial_{X}\phi)^2 -\phi( \partial_{X}^2 \phi)\Big]
\bigg[\frac{D\!-\!4}{(p_\epsilon^2+m^2)^2}+\frac{4m^2 }{(p_\epsilon^2+m^2)^3}\bigg]
\nonumber\\
    &\!\!\!\!&\hskip 6.5cm
    \times\frac{1}{(p_\epsilon'^2+m^2)[(p+p')_\epsilon^2+m^2]}
\nonumber\\
&\!\!=\!\!&\frac{\lambda^2}{24 D} \int_X
\Big[(\partial_{X}\phi)^2 -\phi( \partial_{X}^2 \phi)\Big]
\bigl[(D\!-\!4)I^*_2+4m^2 I^*_3\bigr]
    \,.\qquad\;
\label{Gamma 22a: vertex}
\end{eqnarray}
Expanding the propagators results in,
\begin{eqnarray}
    \Gamma^{(2,2)}_b &\!\!=\!\!& -\frac{\lambda}{4} \int_{X,p,p'}
    \!\bigg[
 \frac{\frac{\lambda(D-2)}{2D}\big[(\partial_{X}\phi)^2\!+\!\phi (\partial_{X}^2 \phi)\big]}
 {(p_\epsilon^2+m^2)^3}
 +\frac{\frac{\lambda m^2}{D}\big[(\partial_{X}\phi)^2\!+\!\phi (\partial_{X}^2 \phi)\big]-\frac{\lambda^2 \phi^2}{2}(\partial_{X}\phi)^2}{(p_\epsilon^2+m^2)^4}\bigg]
 \nonumber\\
&&\hskip 1.99cm
 \times\,
 \frac{1}{p'^2\!+\!m^2}
  \nonumber\\
  &\!\!=\!\!& -\frac{\lambda}{4} \int_{X}
    \!\bigg\{
\frac{\lambda(D-2)}{2D}\big[(\partial_{X}\phi)^2\!+\!\phi (\partial_{X}^2 \phi)\big]I_3
 +\bigg[\frac{\lambda m^2}{D}\big[(\partial_{X}\phi)^2\!+\!\phi (\partial_{X}^2 \phi)\big]
  \nonumber\\
&&\hskip 1.99cm
 -\frac{\lambda^2 \phi^2}{2}(\partial_{X}\phi)^2\bigg]I_4\bigg\}  I_1
 \,,
 \label{Gamma 22b: prop}
\end{eqnarray}
and
\begin{eqnarray}
   \Gamma^{(2,2)}_c &\!\!=\!\!& \frac{\lambda^2}{4}\int_{X,p,p'}\!\phi^2 
   \bigg[\frac{\frac{\lambda(D-2)}{2D}\big[(\partial_{X}\phi)^2
   \!+\!\phi (\partial_{X}^2 \phi)\big]}{(p_\epsilon^2+m^2)^3}
   +\frac{\frac{\lambda m^2}{D}\big[(\partial_{X}\phi)^2\!+\!\phi (\partial_{X}^2 \phi)\big]
   -\frac{\lambda^2 \phi^2}{2}(\partial_{X}\phi)^2}{(p_\epsilon^2+m^2)^4}
 \bigg] 
  \nonumber\\
 &&\hskip 2.2cm
 \times\,  \frac{1}{p_\epsilon'^2+m^2}\frac{1}{(p+p')_\epsilon^2+m^2}
 \label{Gamma 22c: prop}\\
&\!\!\!\!&\hskip -1.cm 
 =\frac{\lambda^2}{4}\!\int_{X}\!\phi^2\bigg\{\! 
   \frac{\lambda(D\!-\!2)}{2D}\big[(\partial_{X}\phi)^2\!+\!\phi (\partial_{X}^2 \phi)\big]I^*_3
   \!+\!\bigg[\frac{\lambda m^2}{D}\big[(\partial_{X}\phi)^2\!+\!\phi (\partial_{X}^2 \phi)\big]
   \!-\!\frac{\lambda^2 \phi^2}{2} (\partial_{X}\phi)^2\bigg]  I^*_4
 \! \bigg\}
\,,
\nonumber
\end{eqnarray}
where $I_n\;(n=1,3,4)$ and $I^*_n\;(n=2,3,4)$ denote the integrals, respectively, in equations \eqref{appendix D: easy integral: In} and \eqref{appendix D: hard integral: In*}, which are evaluated in Appendix~\ref{Two-loop calculations}. 

Let us first evaluate the easiest contribution, namely $\Gamma^{(2,2)}_b$. Making  use of equation \eqref{appendix D: easy integral: In}, one immediately obtains, 
\begin{eqnarray}
   \Gamma^{(2,2)}_b &\!\!=\!\!& \frac{\lambda^2}{24(4\pi)^D}\big(m^2\big)^{D-4}
   \Gamma\left(\!1\!-\!\frac{D}{2}\right)\Gamma\left(\!3\!-\!\frac{D}{2}\right)
\nonumber\\
       &\!\!\times\!\!&       
       \int {\rm d}^DX\bigg\{\Big[(\partial_X\phi)^2 \!+\!\phi(\partial_X^2\phi)\Big]
                -\frac{(6\!-\!D)}{2}\frac{\lambda\phi^2}{2m^2}(\partial_X\phi)^2
               \bigg\}
\,.
\label{Gamma 22b: primitive}
\end{eqnarray}
This is linearly divergent in $(D\!-\!4)$~\footnote{This is a general feature of perturbation theory.
Namely, at 
 $n$ loops, the leading order divergence in the effective potential is 
of the type 
$\sim 1/(D-4)^n$, 
such that the corresponding leading order logarithm is
$\ln^n\big(m^2/(4\pi\mu^2)\big)$. 
On the other hand, the leading 
divergence to the two-derivative (kinetic) term is  
$\sim 1/(D\!\!-4)^{n-1}$,
implying that the leading order logarithm is 
$\ln^{n-1}\big(m^2/(4\pi\mu^2)\big)$.
We suspect that this pattern continues for higher order derivative contributions to the effective
action.
}, 
such that the expansion around the pole is straightforward,
\begin{eqnarray}
   \Gamma^{(2,2)}_b &\!\!=\!\!& \frac{\lambda^2\mu^{D-4}}{12(4\pi)^4}\int {\rm d}^DX
\Bigg\{\Bigg[\frac{\mu^{D-4}}{D\!-\!4}+\!\ln\left(\!\frac{m^2}{4\pi\mu^2}\!\right)\!-\!\psi(1)\!-\!\frac12\Bigg]
\label{Gamma 22b: expanded}\\ 
&\!\!\!\!& \hskip 2.5cm
   \times \left[\Big[(\partial_X\phi)^2 \!+\!\phi(\partial_X^2\phi)\Big]
     \!-\!\frac{\lambda\phi^2}{2m^2}(\partial_X\phi)^2\right]
\!+\!\frac{\lambda\phi^2}{4m^2}(\partial_X\phi)^2
    \!\Bigg\}
.
\nonumber
\end{eqnarray}

Summing all three contributions, given in Appendix~\ref{Two-loop calculations} in equations \eqref{appendix D: Gamma 22a: final}, \eqref{appendix D: Gamma 22b: final} and \eqref{appendix D: Gamma 22c: final}, yields equation \eqref{appendix D: Gamma 22: final}: 
\begin{eqnarray}
   \Gamma^{(2,2)} &\!\!=\!\!& 
 \frac{\lambda^2\mu^{D-4}}{12(4 \pi)^4}\!\int\!\!{\rm d}^D{X}\!
\Bigg\{\! \Bigg[
   \bigg[(\partial_{X}\phi)^2\!+\!\phi (\partial_{X}^2 \phi)
    \!-\!\frac{\lambda \phi^2}{2m^2} (\partial_{X}\phi)^2\bigg] 
    \bigg(\!1\!+\!\frac{\lambda\phi^2}{m^2}\bigg)
    \!+\! \frac14\bigg[(\partial_{X}\phi)^2\!-\!\phi (\partial_{X}^2 \phi)\bigg] \Bigg]
\nonumber\\    
 &\!\!\!\!& \hskip 2.95cm    
  \times \bigg[\frac{\mu^{D-4}}{D\!-\!4} 
    \!+\!\ln\Big(\frac{m^2}{4\pi\mu^2}\Big)\!+\!\gamma_E\bigg]
\nonumber\\    
 &\!\!\!\!& \hskip .7cm
 +\,\big[(\partial_{X}\phi)^2\!+\!\phi (\partial_{X}^2 \phi)\big]
    \bigg[\!\!-\!\frac{1}{2}\!+\!\frac{\lambda \phi^2}{m^2}
    \Big(\frac1{24}\!+\!\frac{3c_3^*\!+\!c_4^*}{4}\Big)\!\bigg]
    \!+\!\big[(\partial_{X}\phi)^2\!-\!\phi (\partial_{X}^2 \phi)\big]
              \bigg[\!\frac1{16}\!+\!\frac{c_3^*}{2}\bigg]
\nonumber\\    
 &\!\!\!\!& \hskip .7cm
    \!+\, (\partial_{X}\phi)^2  \bigg(\frac{\lambda \phi^2}{m^2} \bigg)
    \bigg[\frac{1}{4}\!+\!\frac{\lambda \phi^2}{m^2}\Big(\frac16\!-\!\frac{c_4^*}{2}\Big)\bigg]
 \! \Bigg\}
.\qquad\;
 \label{Gamma 22: final}
\end{eqnarray}

Similarly to the zeroth order in gradient case contribution, we find that the divergent part is non-renormalisable. In the general case, an infinite series of counterterms is needed to subtract the divergent part. Using equation \eqref{ct action}, we find the following coefficients:
\begin{equation}
    Z_0=\frac{\lambda^2}{12 (4\pi)^4}\frac{\mu^{2(D-4)}}{D-4}\; , 
\end{equation}
and for $n\geq 1$,
\begin{equation}
    \frac{Z_{(2n)}}{(2n)!}= \bigg(\frac{-\lambda}{2 M^2}\bigg)^{n} \frac{\lambda^2}{12 (4\pi)^4}\frac{\mu^{2(D-4)}}{D\!-\!4}  (2n\!+\!3)\; .
\end{equation}  

We are then left with the renormalised result:
\begin{eqnarray}
    \Gamma^{(2,2)}_{\rm ren}&\!\!=\!\!& 
 \frac{\lambda^2}{12(4 \pi)^4}\!\int\!\!{\rm d}^4{X}\!
\Bigg\{\! \Bigg[
   \bigg[(\partial_{X}\phi)^2\!+\!\phi (\partial_{X}^2 \phi)
    \!-\!\frac{\lambda \phi^2}{2m^2} (\partial_{X}\phi)^2\bigg] 
    \bigg(\!1\!+\!\frac{\lambda\phi^2}{m^2}\bigg)
    \!+\! \frac14\bigg[(\partial_{X}\phi)^2\!-\!\phi (\partial_{X}^2 \phi)\bigg] \Bigg]
\nonumber\\    
 &\!\!\!\!& \hskip 2.95cm    
  \times \bigg[\ln\Big(\frac{m^2}{4\pi\mu^2}\Big)\!+\!\gamma_E\bigg]+\,\big[(\partial_{X}\phi)^2\!+\!\phi (\partial_{X}^2 \phi)\big]
    \bigg[\!\!-\!\frac{1}{2}\!+\!\frac{\lambda \phi^2}{m^2}
    \Big(\frac1{24}\!+\!\frac{3c_3^*\!+\!c_4^*}{4}\Big)\!\bigg]
\nonumber\\    
 &\!\!\!\!& \hskip .7cm 
    \!+\,\big[(\partial_{X}\phi)^2\!-\!\phi (\partial_{X}^2 \phi)\big]
              \bigg[\frac1{16}\!+\!\frac{c_3^*}{2}\bigg]    \!+\, (\partial_{X}\phi)^2  \bigg(\frac{\lambda \phi^2}{m^2} \bigg)
    \bigg[\frac{1}{4}\!+\!\frac{\lambda \phi^2}{m^2}\Big(\frac16\!-\!\frac{c_4^*}{2}\Big)\bigg]
 \! \Bigg\}
,\qquad\;
    \label{two-loop result R}
\end{eqnarray}
which, after integrating by parts, becomes~\footnote{As mentioned, Ref.~\cite{Iliopoulos-Itzykson-Martin} performs the two-loop calculations for the effective action at zeroth and second order in gradients. Therefore, it is possible to compare the results, in particular focusing on the logarithmic terms. Reference \cite{Iliopoulos-Itzykson-Martin} finds:
\begin{equation}
    V^{(2)}_{\rm IIM}=\frac{\lambda}{4(4\pi)^4}\bigg[\frac{1}{2}m^2(m^2+\lambda\phi^2)\log^2\bigg(\frac{m^2}{4\pi\mu^2}\bigg)-\frac{3}{2}m^2\lambda\phi^2\log\bigg(\frac{m^2}{4\pi\mu^2}\bigg)\bigg]\; ,
\end{equation}
\begin{equation}
    \Gamma^{(2,2)}_{\rm IIM}=\frac{\lambda^2}{12(4\pi)^4}\int {\rm d}^4x \,\frac{(\partial \phi)^2}{2}\log\bigg(\frac{m^2}{4\pi\mu^2}\bigg)\bigg(1-\frac{5}{2}\frac{\lambda \phi^2}{m^2}+\frac{1}{2}\frac{\lambda^2\phi^4}{m^4}\bigg)\; .
    \label{gamma22_iim}
\end{equation}
Comparing the previous expressions with the logarithmic terms in equations \eqref{renormalized Gamma 20: Veff 2} and \eqref{gamma22_ren}, we notice that the leading logarithms differ by a factor of $\frac{1}{2}$ with respect to the ones we found (except for the first term in Eq.~\eqref{gamma22_iim}, on which we find agreement). On the other hand, some of the differences in the subleading logarithmic terms 
and in the field-dependent terms containing no logarithms can be explained by a different choice of the finite counterterms. In our case, the minimal subtraction scheme is adopted, where only divergent terms are subtracted.
},
\begin{equation}
\begin{split}
    \Gamma^{(2,2)}_{\rm ren}=
 \frac{\lambda^2}{12(4 \pi)^4}\!\int\!\!{\rm d}^4{X} \frac{(\partial_X \phi)^2}{2} \bigg[& \bigg(1-5\frac{\lambda \phi^2}{m^2}+\frac{\lambda^2\phi^4}{m^4}\bigg)\bigg(\ln\Big(\frac{m^2}{4\pi\mu^2}\Big)\!+\!\gamma_E\bigg)+\frac{1}{4}+2c_3^*\\
 &-\frac{\lambda \phi^2}{m^2}\bigg(\frac{7}{6}+3c_3^*+c_4^*\bigg)-\frac{\lambda^2\phi^4}{m^4}\bigg(\frac{19}{12}-\frac{3c_3^*-c_4^*}{2}\bigg)
 \bigg]\; .
 \label{gamma22_ren}
\end{split}
\end{equation}  

\subsection{2PI formalism}

As of now, we considered renormalisation in the context of the 1PI formalism. As mentioned in Section~\ref{Intro}, the 1PI effective action is a functional of the background field or one-point function only. Conversely, the 2PI effective action is usually expressed as a functional of two independent objects: the background field and the propagator, or two-point function \cite{Berges:2004,Carrington:2015}. In our calculations, we utilise the resummed propagator as defined by Coleman-Weinberg \cite{Coleman-Weinberg}. However, differently from previous literature, we lift the assumption of constant background fields and work with a propagator in which the spacetime dependent mass term $\frac{\lambda}{2}\phi^2(x)$ is resummed. In this context, the 2PI formalism provides a suitable framework to perform renormalisation of our theory. 

\subsubsection{2PI renormalisation}

Our previous results suggest that this theory is not renormalisable. Indeed, we found that an infinite series of counterterms is needed to subtract the divergences. However, we notice that the 1PI renormalisation framework is not suitable for the resummed propagator we are using. As noted, we are working with a resummed two-point function
(obtained by resumming mass insertions) and with an off-shell one-point function. In this context, the 2PI formalism provides a suitable framework to renormalise the theory. 

To show this we consider the 2PI counterterm action:
\begin{equation}
    S^{\rm (2PI)}_{\rm ct}\!=\!\int_x\!  \bigg(\!\! -\frac{ \Tilde{Z}_0}{2}[\partial_{\mu}\partial_{\nu}'i\Delta(x;x')]_{x=x'}\eta^{\mu \nu}\! - \frac{\Tilde{m}_0^2 }{2}i\Delta(x;x) - \frac{\Tilde{\lambda}_0\phi^2}{4}  i\Delta(x;x)- \frac{\Tilde{\lambda}_1}{4}[i\Delta(x;x)]^2\bigg).
    \label{2PI ct action}
\end{equation}
As proven in the following, only the second and third terms in the previous expression are needed in our case. Their explicit form is,
\begin{eqnarray}
\begin{split}
    -\bigg[\frac{ \Tilde{m}^2_0}{2}+\frac{\Tilde{\lambda}_0}{4}\phi^2\bigg]i\Delta(X;X)=&-\bigg[\frac{\Tilde{m}^2_0}{2}+\frac{\Tilde{\lambda}_0}{4}\phi^2\bigg]\bigg\{ \frac{m^2}{(4\pi)^2}\bigg[ \frac{2\mu^{D-4}}{D\!-\!4}\!+\!\log\bigg(\frac{m^2}{4\pi \mu^2}\bigg)\!+\!\gamma_E\!-\!1\bigg]\qquad\\
    &-\frac{\lambda}{6m^2(4\pi)^2}\bigg[(\partial_{X}\phi)^2\!+\!\phi (\partial_{X}^2 \phi)
    \!-\!\frac{\lambda \phi^2}{2m^2} (\partial_{X}\phi)^2\bigg] \bigg\}\; .
    \label{2PI ct}
\end{split}
\end{eqnarray}
 The first counterterm in equation \eqref{2PI ct action} can be calculated by taking the second derivative of the off-coincident propagator in equation \eqref{propoffcoinc} and then considering the coincidence limit of the result. The explicit expression of this first counterterm is,
\begin{equation}
\begin{split}
    -\int_X  \frac{\Tilde{Z}_0}{2}[\partial_{\mu}\partial_{\nu}'i\Delta(x,x')]_{x=x'=X}\eta^{\mu \nu}=&-\int_X \frac{\Tilde{Z}_0}{2}\bigg[\!-\frac{m^D}{(4 \pi)^{D/2}}\Gamma\bigg(1-\frac{D}{2}\bigg)\\
    \hskip -1.7cm
    &-\frac{m^{D-4}}{(4 \pi)^{D/2}}\lambda\big((\partial_{X}\phi)^2\!+\!\phi (\partial_{X}^2 \phi)\big) \frac{D+2}{12}
    \Gamma\bigg(2-\frac{D}{2}\bigg)\\
    &+\!\frac{m^{D-6}}{(4 \pi)^{D/2}}\frac{\lambda^2 \phi^2}{2}(\partial_{X}\phi)^2 \frac{D\!+\!6}{12}\Gamma\bigg(3-\frac{D}{2}\bigg)\bigg]
\,,     
\label{footnote 8}
\end{split}
\end{equation}
which can be calculated by recalling that:
$ \partial_{\mu}\partial'_{\sigma}\frac{K_{\nu}}{z^{\nu}}\Big|_{x'=x}=-\frac{m^2 \eta_{\mu \sigma}}{2^{\nu+2}}\frac{\Gamma(-\nu)}{(1+\nu)}$. Since $\tilde Z_0$ does not depend on the fields, one can integrate by parts the second term in~(\ref{footnote 8}) and expand around $D\rightarrow 4$ to
obtain,
\begin{equation}
\begin{split}
    -\!\int_X\!  \frac{\Tilde{Z}_0}{2}[\partial_{\mu}\partial_{\nu}'i\Delta(x,x')]_{x=x'=X}\eta^{\mu \nu}=&\frac{\Tilde{Z}_0}{2}\bigg\{ \frac{m^4}{(4\pi)^2}\bigg[ \frac{2\mu^{D-4}}{D\!-\!4}\!+\!\log\bigg(\frac{m^2}{4\pi \mu^2}\bigg)\!+\!\gamma_E\!-\!1\bigg]\\
    &+\frac{\lambda^2\phi^2}{12m^2(4\pi)^2}(\partial_X \phi)^2\bigg\}
    \,.
\end{split}
\label{footnote 8b}
\end{equation}
Clearly, the structure of~(\ref{footnote 8b}) differs from the 2PI counterterms in~(\ref{2PI ct}), which is precisely the structure of the divergence to be renormalised. Similar considerations are valid for the fourth counterterm in equation \eqref{2PI ct action}. In fact, this counterterm is quadratically divergent for $D\rightarrow 4$, so it is needed to renormalise the action at three and higher loops. Moreover, when $\Tilde{\lambda}_1$ is finite, the fourth counterterm provides no new divergent structure from that already given by the second and third counterterms in~\eqref{2PI ct action}. Thus 
we can already choose $ \Tilde{\lambda}_1=0$. Considering the first three counterterms in \eqref{2PI ct action}, we can fix their coefficients by imposing that $\Gamma_{\rm div}+S_{\rm ct}=0$. Focusing only on the nonrenormalisable terms within the 1PI formalism, we have three equations:
\begin{equation}
    \frac{\Tilde{m}^2_0}{2}-\frac{\Tilde{Z}_0}{2}M^2=-\frac{\lambda M^2}{2 (4\pi)^2}\frac{\mu^{D-4}}{D-4}\; ,
\end{equation}
\begin{equation}
    \frac{\Tilde{\lambda}_0}{4}-\frac{\Tilde{Z}_0}{4}\lambda=-\frac{3\lambda^2}{4 (4\pi)^2}\frac{\mu^{D-4}}{D-4} \; ,
\end{equation}
\begin{equation}
\begin{split}
    &\bigg[\frac{\Tilde{m}^2_0}{2}+\frac{\Tilde{\lambda}_0}{4}\phi^2\bigg] \frac{\lambda}{6m^2(4\pi)^2}\bigg[(\partial_{X}\phi)^2\!+\!\phi (\partial_{X}^2 \phi)
    \!-\!\frac{\lambda \phi^2}{2m^2} (\partial_{X}\phi)^2\bigg] +\frac{\Tilde{Z}_0}{2}\frac{\lambda^2\phi^2}{12m^2(4\pi)^2}(\partial_X \phi)^2\\
    &=-\frac{\lambda^2}{12(4\pi)^4}\frac{\mu^{2(D-4)}}{D-4}\bigg[(\partial_{X}\phi)^2\!+\!\phi (\partial_{X}^2 \phi)
    \!-\!\frac{\lambda \phi^2}{2m^2} (\partial_{X}\phi)^2\bigg]\bigg(1+\frac{\lambda\phi^2}{m^2}\bigg)\; .
\end{split}
\end{equation}
This is a system of three equations in three unknowns, whose unique solution is given by:
\begin{equation}
    \frac{\Tilde{m}^2_0}{2}=-\frac{\lambda M^2}{2 (4\pi)^2}\frac{\mu^{D-4}}{D-4}\;,\;\;\;\; \frac{\Tilde{\lambda}_0}{4}=-\frac{3\lambda^2}{4 (4\pi)^2}\frac{\mu^{D-4}}{D-4} \; , \;\;\;\; \Tilde{Z}_0=0\; .
    \label{2PI coefficients}
\end{equation}
Curiously, we find that $\Tilde{Z}_0$ vanishes and the two counterterms in \eqref{2PI ct} subtracts both the logarithmically divergent terms of $\Gamma^{(2,0)}$ in equations \eqref{Gamma 20a final} and \eqref{Gamma 20b final} and the `nonrenormalisable' divergent terms of $\Gamma^{(2,2)}$ in equation \eqref{Gamma 22: final}. The remaining divergences can be easily subtracted with the canonical 1PI 
counterterms with the following coefficients,
\begin{equation}
    \frac{\Lambda}{8\pi G}=\frac{\lambda M^4}{2(4 \pi)^4}\frac{\mu^{2(D-4)}}{(D-4)^2} \; ,
\end{equation}
\begin{equation}
    \frac{m_0^2}{2}=\frac{\lambda^2 M^2}{(4\pi)^4}\bigg(\frac{\mu^{2(D-4)}}{D-4}+\frac{1}{4}\frac{\mu^{D-4}}{D-4}\bigg) \; ,
\end{equation}
\begin{equation}
    \frac{\lambda_0^2}{4!}=\frac{\lambda^3}{8(4\pi)^4}\bigg(3\frac{\mu^{2(D-4)}}{D-4}+\frac{\mu^{D-4}}{D-4}\bigg) \; ,
\end{equation}
\begin{equation}
    \frac{Z_0}{2}=\frac{\lambda^2 M^4}{24(4 \pi)^4}\frac{\mu^{2(D-4)}}{D-4} \; .
\end{equation}
For simplicity here we have used a minimal subtraction scheme, but we note that renormalization 
can be easily generalized to a non-minimal subtraction scheme.

Summarising, the two 2PI counterterms in equation~\eqref{2PI ct}, aided by the canonical 1PI counterterm action selected from~(\ref{ct action}), can fully subtract the divergences of the two-loop effective action expanded up to second order in gradients. We conclude that the 2PI formalism is the correct framework to perform renormalisation when working with resummed propagators. 

\section{Discussion and conclusion}
\label{Sec.V}

In this paper we performed a gradient expansion of the scalar sector of the quantum effective action. To do so we considered the scalar propagator and we expanded its equation of motion in Wigner space. We made use of the fact that the Feynman propagator is non-trivially constrained by an additional equation of motion~(\ref{integrated prop eq of motion}), which can be obtained, for example, by noticing that the propagator $i\Delta_{ab}(x;y)$ should be symmetric~(\ref{propagator: symmetry requirement}) under the exchange of points $x \leftrightarrows y$, in the single field case, and under the exchange of points and transposition in flavour $a \leftrightarrows b$, in the multi-field case. An expansion in derivatives (also known as gradient expansion) is performed by using a midpoint expansion of the propagator.
In our calculations, we verified that both the symmetric equations of motion of the propagator remain satisfied after expanding in number of derivatives. The kinetic corrections to the one-loop effective action are then obtained from the functional determinant of the expanded inverse propagator. Firstly, we performed the calculations for the simple case of a single scalar field with generic tree-level potential. We proceeded both with an exact calculation, by a direct integration of the logarithm of the propagator, and with an approximated one, by expanding the logarithm in gradients. By both methods
we find that the second-order gradient corrections to the effective action vanish at one-loop~(\ref{final result: single field}). Moreover, by considering both equations of motion for the propagator, we have argued that, when performing a midpoint expansion of the propagator, all the terms with odd number of derivatives should be zero. We explicitly verified this for the first- and third-order corrections to the propagator. 
The one-loop effective action we thus obtained disagrees with previous works~\cite{Fraser,Chan,Iliopoulos-Itzykson-Martin},
which used parametric differentiation and subsequent integration~\cite{Fraser,Chan},
or a proper-time representation of the effective action~\cite{Iliopoulos-Itzykson-Martin} to obtain the result.
In Appendix~\ref{Appendix: literature} we elucidate some aspects of these works, in particular clarifying the differences between their methods and improving them in various aspects. We also show that at coincidence our gradient expanded propagator in equations~\eqref{prop0coin} and~\eqref{prop2coin} corresponds to the one of these previous works. 

Furthermore, we generalized the calculations to the case of multiple scalar fields with canonical kinetic terms, but with mass mixing in the tree-level potential. Considering again a generic tree-level mass matrix, we performed similar calculations as in the one-field case. Interestingly, we find that the previous result does not carry over to the multi-field case. Indeed, we obtain non-vanishing one-loop kinetic corrections for the scalar fields. Moreover, we verify that the second-order corrections to the propagator are manifestly transpose invariant, confirming that both the symmetric equations of motion are satisfied. We then exploited the propagator expanded up to second order in gradients to find consistent unique one-loop kinetic corrections for the effective action~(\ref{4dim second order ea}). 

From this general result, we can notice that, unlike the one-loop corrections of the potential, the one-loop second-order gradient corrections to the effective action are finite and do not need to be renormalised by adding appropriate counterterms. This is interesting to highlight since it signifies that these gradient corrections do not introduce any renormalization scale dependence, thus not breaking scaling symmetry, as demonstrated in Appendix~\ref{Appendix: scale-independent}. 

As a simple application, we also used our one-loop result for the case of two real scalar fields with mass mixing~(\ref{two fields 2gradient final}). 
From this example, it is worth highlighting that the kinetic terms cannot be written in the canonical form since the second-order one-loop gradient corrections introduce kinetic mixing terms between the two scalar fields
which has a non-vanishing configuration space curvature.  

Finally, we performed the two-loop computations for the single-field case only and by using the propagator obtained by the midpoint expansion in gradients. 
Similarly to \cite{Iliopoulos-Itzykson-Martin}, at second-order in gradients, we found a non-vanishing and divergent result, given in equations \eqref{Gamma 20a final} and \eqref{Gamma 20b final} for the zeroth order in gradients contribution and \eqref{Gamma 22: final} for the second order in gradients one. However, we found that the logarithmically divergent terms and the second-order in gradients divergent terms generate an infinite series of divergent terms, which cannot be renormalised by the standard 1PI counterterm action. As a solution, we considered the counterterm action extended within the 2PI formalism. By imposing that the 2PI counterterms in equation \eqref{2PI ct action} cancel the `nonrenormalisable' two-loop terms, we obtained the coefficients of the 2PI counterterms, given in equation \eqref{2PI coefficients}. We verified that within this framework, the theory becomes renormalisable. We conclude that, when working with the 1PI propagator, obtained by resumming spacetime dependent field insertions, the 2PI formalism is the correct framework to renormalise the theory. We highlight that this is one of the main results of this work. The fact that the perturbative 2PI counterterms renormalise the theory can be viewed as an independent support for the gradient expansion approach used in this work.

Our computations have a wide range of applicability for early Universe phenomena when quantum corrections cannot be neglected. For example, in Ref.~\cite{Canevarolo-Prokopec} we study the effect of gradient corrections for the case of bubble nucleation during the electroweak phase transition in a conformal extension of the standard model. Indeed, to obtain the tunneling path from the false to the true vacuum state of the theory the full effective action is needed. In turn, this brings the necessity to have a good estimate of the loop corrections for the kinetic terms besides the well-known ones for the potential. Similarly, we expect that many other instances in cosmology where spatial or time gradients and multiple scalar fields are present, such as multifield
inflationary models, and various baryogenesis and leptogenesis scenarios, will provide a manifold of applications for our results.

\section{Acknowledgements}

The authors would like to thank Tanja Hinderer and Elisa Chisari for comments and Yaseen Asad for independently checking significant parts of the calculations presented in this work \cite{Asad}. This work is part of the Delta ITP consortium, a program of the Netherlands Organisation for Scientific Research (NWO) that is funded by the Dutch Ministry of Education, Culture and Science (OCW) — NWO project number 24.001.027.

\appendix
\section{Second-order gradient corrections in the multi-fields case}
\label{Appendix: multifields}

In the following, we present the explicit computations for the second-order gradient corrections of the one-loop effective action. To do so, we will consider term by term in equation \eqref{2 order ea multifield}.
We start by considering the first term and we use index notation. We define the matrix $N=\mathcal{R} (\partial^2_X \mathbb{M}^2)\mathcal{R}^{T}$. We have to compute the following integral in momentum space:
\begin{equation}
\begin{split}
    \text{Tr}\Big(\!-\frac{1}{2} (\partial^2_X \mathbb{M}^2)(\Delta^{(0)})^2\Big)=&-\frac{1}{2}\int {\rm d}^DX \int \frac{{\rm d}^Dp}{(2\pi)^D} \text{tr} \big[(\partial^2_X \mathbb{M}^2)(\Delta^{(0)})^2\big]\\
    =&-\frac{1}{2}\int {\rm d}^DX \int \frac{{\rm d}^Dp}{(2\pi)^D} \delta^{ij} N_{ik} (\Delta^{(0)}_d)^2_{kj}\\
    =&-\frac{1}{2} \int {\rm d}^DX \; \delta^{ij} N_{ij}\int \frac{{\rm d}^Dp}{(2\pi)^D} \frac{1}{(-p_\epsilon^2-m_j^2)^2}\\
    =&-\frac{i}{2}\frac{\Gamma(2-D/2)}{(4\pi)^{D/2}}\sum_i \int {\rm d}^DX \; N_{ii} (m_i^2)^{\frac{D}{2}-2}\; ,
\end{split}
\label{1 term}
\end{equation}
where $m_i^2$ are the eigenvalues of the diagonalized mass matrix and $\Delta^{(0)}_d$ the diagonalized zeroth-order propagator. With a similar procedure we can calculate also the second term,
\begin{equation}
    -\int {\rm d}^DX \int \frac{{\rm d}^Dp}{(2\pi)^D} \text{tr} \big[p^{\mu} p^{\nu} (\partial_{\mu} \partial_{\nu}\mathbb{M}^2)(\Delta^{(0)})^3\big]\; ,
\end{equation}
which becomes:
\begin{equation}
\begin{split}
    &- \int {\rm d}^DX \; \delta^{ij} N_{ij}\int \frac{{\rm d}^Dp}{(2\pi)^D} \frac{p^2}{D(-p_\epsilon^2-m_j^2)^3}=\\
    =&-i \frac{(D-2)\Gamma(1-D/2)}{8(4\pi)^{D/2}} \sum_i \int {\rm d}^DX \;  N_{ii} (m_i^2)^{\frac{D}{2}-2}\; .
\end{split}
\label{2 term}
\end{equation}
Using the property of the Gamma function $\Gamma(z+1)=z \Gamma(z)$, we see that the sum of \eqref{1 term} and \eqref{2 term} is simply,
\begin{equation}
    -\frac{i}{4}\frac{\Gamma\big(2-\tfrac{D}{2}\big)}{(4\pi)^{D/2}}\int {\rm d}^DX \; \delta^{ij} N_{ij} \big(m_j^2\big)^{\frac{D}{2}-2}\; .
    \label{1 2 terms}
\end{equation}
By partial integration, we transform the remaining mass matrix to be diagonal and \eqref{1 2 terms} becomes the following:
\begin{equation}
\begin{split}
    &\frac{-i}{4(4\pi)^{D/2}} \Gamma\bigg(2-\frac{D}{2}\bigg) \int {\rm d}^DX \; \sum_{i,k} \bigg( 2 T_{ik}T_{ki} (m_{i}^2)^{D/2-1}-2T_{ik}T_{ki} (m_{k}^2)(m_{i}^2)^{D/2-2}\\
    &-\bigg(\frac{D}{2}-2\bigg) (\partial_X m_{i}^2)^2(m_{i}^2)^{D/2-3}\bigg)\; ,
\end{split}
\label{1 2 term partial int}
\end{equation}
where we have defined the matrix product $T=\mathcal{R}(\partial_X \mathcal{R}^T)$. It is now convenient to separate the sum over the diagonal terms from the off-diagonal ones, such that equation \eqref{1 2 term partial int} becomes,
\begin{equation}
\begin{split}
    &-\frac{i}{4(4\pi)^{D/2}} \Gamma\bigg(2-\frac{D}{2}\bigg) \frac{4-D}{2} \int {\rm d}^DX \;  \sum_{i=k} \bigg( (\partial_X m_{i}^2)^2(m_{i}^2)^{D/2-3}\bigg)\\
    &+\frac{i}{(4\pi)^{D/2}} \Gamma\bigg(1-\frac{D}{2}\bigg)\frac{D-2}{4} \int {\rm d}^DX \;  \sum_{i\neq k} T_{ik}T_{ki}\bigg((m_{i}^2)^{D/2-1}- (m_{k}^2)(m_{i}^2)^{D/2-2}\bigg)\; ,
    \label{1 off-diagonal}
\end{split}
\end{equation}
and we see that the rotation matrices contribute only to off-diagonal terms.

Following the same procedure, we compute also the remaining terms containing first derivatives of the mass matrix. It is therefore useful to define $L=\mathcal{R}(\partial_X \mathbb{M}^2)\mathcal{R}^T$ and separate again the sum over diagonal and off-diagonal terms. We have,
\begin{equation}
\begin{split}
    &\text{Tr}\Big(-\frac{1}{2} \Delta^{(0)} (\partial_X \mathbb{M}^2)\Delta^{(0)}(\partial_X \mathbb{M}^2)\Delta^{(0)}\Big)\\
    &=-\frac{1}{2} \int {\rm d}^DX \; \sum_{i,k} L_{ik} L_{ki} \int \frac{{\rm d}^Dp}{(2\pi)^D} \frac{1}{(-p_\epsilon^2-m_k^2)}\frac{1}{(-p_\epsilon^2-m_i^2)^2}\\
    &=-\frac{1}{2} \int {\rm d}^DX \; \bigg( \sum_{i} (L_{ii})^2  \int \frac{{\rm d}^Dp}{(2\pi)^D} \frac{1}{(-p_\epsilon^2-m_i^2)^3}\\
    &+\sum_{i\neq k} L_{ik} L_{ki} \int \frac{{\rm d}^Dp}{(2\pi)^D} \bigg(\frac{(m_i^2-m_k^2)^{-2}}{(-p_\epsilon^2-m_k^2)}-\frac{(m_i^2-m_k^2)^{-2}}{(-p_\epsilon^2-m_i^2)}+\frac{(m_i^2-m_k^2)^{-1}}{(-p_\epsilon^2-m_i^2)^2}\bigg) \bigg)\; ,\\
    \end{split}
    \label{3 term}
\end{equation}
where in the last step we have also decomposed in partial fractions. Let us now focus on the summation over the diagonal part. As before, we can transform the mass matrix in the $L$ product to be diagonal. We find that:
\begin{equation}
     \sum_{i} (L_{ii})^2 =  \sum_{i} \big([\mathcal{R}(\partial_X \mathbb{M}^2)\mathcal{R}^T]_{ii}\big)^2
     =\sum_{i}\big([\mathcal{R}(\partial_X (\mathcal{R}^T \mathbb{M}_d^2 \mathcal{R})
     \mathcal{R}^T]_{ii}\big)^2= \sum_{i} \big(\partial_X m_{i}^2\big)^2\; .
\end{equation}
Performing the momentum integral of the diagonal term, we obtain the final expression:
\begin{equation}
    \frac{i}{4(4\pi)^{D/2}} \Gamma\bigg(3-\frac{D}{2}\bigg) \int {\rm d}^DX \;  \sum_i \bigg( (\partial_X m_{i}^2)^2(m_{i}^2)^{D/2-3} \bigg)\; .
    \label{2 diagonal}
\end{equation}
Exploiting again the properties of the Gamma function, we can sum the diagonal terms that we have found so far in equations \eqref{1 off-diagonal} and \eqref{2 diagonal}. This yields,
\begin{equation}
\begin{split}
    &\frac{i}{4(4\pi)^{D/2}} \Gamma\bigg(3-\frac{D}{2}\bigg) \int {\rm d}^DX \;  \sum_i \bigg( (\partial_X m_{i}^2)^2(m_{i}^2)^{D/2-3} \bigg)\\
    &+\frac{i}{4(4\pi)^{D/2}} \Gamma\bigg(2-\frac{D}{2}\bigg) \frac{D-4}{2} \int {\rm d}^DX \;  \sum_i \bigg( (\partial_X m_{i}^2)^2(m_{i}^2)^{D/2-3}\bigg) =0.
\end{split}
\end{equation}
Therefore, we see that the diagonal terms of the first three terms in equation \eqref{2 order ea multifield} do not contribute to the second-order corrections of the effective action. 

On the other hand, we can perform similar computations for the off-diagonal part of the term in equation \eqref{3 term}. Performing the momentum integral yields,
\begin{equation}
\begin{split}
        \frac{i}{(4\pi)^{D/2}} \Gamma\bigg(1-\frac{D}{2}\bigg) &\int {\rm d}^DX \;  \sum_{i \neq k}  T_{ik} T_{ki} \bigg( \frac{(m_i^2)^{\frac{D}{2}-1}-(m_k^2)^{\frac{D}{2}-1}}{2}\\
        &+\frac{D-2}{4}(m_k^2-m_i^2)(m_i^2)^{\frac{D}{2}-2} \bigg)\; .
        \label{2 off-diagonal}
\end{split}
\end{equation}

Finally, we can consider the last term in equation \eqref{2 order ea multifield}. In this case we have,
\begin{equation}
\begin{split}
    &\text{Tr}\bigg(\frac{1}{2}p^{\mu} p^{\nu} [\partial_{\mu}\mathbb{M}^2,\Delta^{(0)}]\Delta^{(0)}[\partial_{\nu}\mathbb{M}^2,\Delta^{(0)}]\Delta^{(0)}\bigg)\\
    &=\text{Tr}(+p^{\mu} p^{\nu}(\partial_{\mu}\mathbb{M}^2)(\Delta^{(0)})^2(\partial_{\nu}\mathbb{M}^2)(\Delta^{(0)})^2-p^{\mu} p^{\nu}(\partial_{\mu}\mathbb{M}^2)(\Delta^{(0)})^3(\partial_{\nu}\mathbb{M}^2)(\Delta^{(0)}))\\
    &= -\int {\rm d}^DX \; \sum_{i\neq k} L_{ik} L_{ki} \int \frac{{\rm d}^Dp}{(2\pi)^D} \frac{1}{D} \bigg(\frac{2m_i^2+m_k^2}{(m_k^2-m_i^2)^3(-p_\epsilon^2-m_i^2)}\\
    &-\frac{2m_i^2+m_k^2}{(m_k^2-m_i^2)^3(-p_\epsilon^2-m_k^2)}+\frac{m_i^2+m_k^2}{(m_k^2-m_i^2)^2(-p_\epsilon^2-m_k^2)^2}\\
    &-\frac{m_k^2}{(m_k^2-m_i^2)(-p_\epsilon^2-m_k^2)^3}+\frac{m_i^2}{(m_k^2-m_i^2)^2(-p_\epsilon^2-m_i^2)^2}\bigg)\; .
\end{split}
\label{last term}
\end{equation}
It is worth noticing that this last term does not contribute with any diagonal component. Performing the momentum integral of equation \eqref{last term}, we obtain,
\begin{equation}
\begin{split}
    &\frac{-i}{(4\pi)^{D/2}} \Gamma\bigg( 1-\frac{D}{2}\bigg) \int {\rm d}^DX \sum_{i\neq k} T_{ik} T_{ki} \bigg( \frac{2m_i^2+m_k^2}{m_k^2-m_i^2} \frac{(m_i^2)^{\frac{D}{2}-1}-(m_k^2)^{\frac{D}{2}-1}}{D}\\
    &-(m_k^2-m_i^2)\frac{(D-4)(D-2)}{8D}(m_k^2)^{\frac{D}{2}-2}+(m_i^2+m_k^2)\frac{D-2}{2D}(m_k^2)^{\frac{D}{2}-2}\\
    &+(m_i^2)\frac{D-2}{2D}(m_i^2)^{\frac{D}{2}-2}\bigg)\; .
    \label{3 off-diagonal}
\end{split}
\end{equation}
Finally, it is possible to sum all the off-diagonal terms contributing to the second-order corrections, which are given in equations \eqref{1 off-diagonal}, \eqref{2 off-diagonal} and \eqref{3 off-diagonal}. To do so, it is convenient to keep in mind that all the indices are summed, therefore in each term we can rename $i \leftrightarrow k$ and then exploit the fact that $T_{ik}T_{ki}$ is symmetric under the exchange of indices. Applying this, we can notice that the sum of \eqref{1 off-diagonal} and \eqref{2 off-diagonal} is zero:
\begin{equation}
\begin{split}
    &\frac{i}{(4\pi)^{D/2}}\Gamma\bigg( 1-\frac{D}{2}\bigg) \int {\rm d}^DX \sum_{i\neq k} T_{ik}T_{ki} \bigg( \frac{(m_i^2)^{\frac{D}{2}-1}-(m_k^2)^{\frac{D}{2}-1}}{2} \bigg)=0\; ,
\end{split}
\end{equation}
and we are left only with the contribution of \eqref{3 off-diagonal}:
\begin{equation}
\begin{split}
&\text{Tr}(\mathcal{D}^{(0)}\Delta^{(2)} )-\frac{1}{2}  \text{Tr}(\mathcal{D}^{(0)}\Delta^{(1)} )^2=\\
&=\frac{-i}{(4\pi)^{D/2}} \Gamma\bigg( 1-\frac{D}{2}\bigg) \int {\rm d}^DX \sum_{i\neq k} T_{ik} T_{ki} \bigg( \frac{2m_i^2+m_k^2}{m_k^2-m_i^2} \frac{(m_i^2)^{\frac{D}{2}-1}-(m_k^2)^{\frac{D}{2}-1}}{D}\\
    &+(m_i^2-m_k^2)\frac{(D-4)(D-2)}{8D}(m_k^2)^{\frac{D}{2}-2}+(2m_i^2+m_k^2)\frac{D-2}{2D}(m_i^2)^{\frac{D}{2}-2}\bigg)\; .
\end{split}
\label{second order appendix result}
\end{equation}
This result is used in equation \eqref{second gradient general formula} in the main text.

\section{Manifestly scale-independent second-order gradient corrections}
\label{Appendix: scale-independent}

In this appendix, we investigate whether the second-order corrections to the one-loop effective action for multiple scalar fields are finite or not. Considering equation \eqref{second gradient general formula}, we can write the argument of the integral as:
\begin{equation}
\begin{split}
&\frac{1}{2}\sum_{i\neq k} T_{ik} T_{ki} \frac{2D(8-D)m_i^2 m_k^2+(48-14D+D^2)m_k^4+(D^2-2D)m_i^4}{m_i^2-m_k^2} \frac{(m_k^2)^{\frac{D}{2}-2}}{8D}\\
&+\frac{1}{2}\sum_{i\neq k} T_{ik} T_{ki} \frac{2D(8-D)m_i^2 m_k^2+(48-14D+D^2)m_i^4+(D^2-2D)m_k^4}{m_k^2-m_i^2} \frac{(m_i^2)^{\frac{D}{2}-2}}{8D}\; ,
\end{split}
\end{equation}
we can then expand the mass terms as $(m^2)^{\frac{D}{2}-2} \approx \mu^{D-4}\big[1+\frac{D-4}{2}\log(\frac{m^2}{\mu^2})\big]$ and sum the first two terms, such that the previous expression becomes:
\begin{equation}
\begin{split}
&\sum_{i\neq k} T_{ik} T_{ki}\bigg[ \frac{3\mu^{D-4}}{4D} (D-4)(m_k^2+m_i^2)\\
&+\frac{2D(8-D)m_i^2 m_k^2+(48-14D+D^2)m_k^4+(D^2-2D)m_i^4}{m_i^2-m_k^2} \frac{\mu^{D-4}}{16D} \frac{D-4}{2} \log \bigg(\frac{m_k^2}{\mu^2} \bigg)\\
&- \frac{2D(8-D)m_i^2 m_k^2+(48-14D+D^2)m_i^4+(D^2-2D)m_k^4}{m_i^2-m_k^2}\frac{\mu^{D-4}}{16D}\frac{D-4}{2} \log \bigg(\frac{m_i^2}{\mu^2} \bigg)\bigg]\; .
\end{split}
\end{equation}
We have now the sum of three terms proportional to a $(D-4)$ factor which cancels the divergence hidden in the Gamma function of equation \eqref{second gradient general formula}. From this result we can conclude that the second-order corrections to the effective action are finite and there is no need of subtracting divergences with counterterms. Furthermore, this assures us that our final result has to be independent from the renormalisation scale $\mu$.
Focussing on the last two terms in the previous expression, we can try to massage them to make the $\mu$-independence manifest:
\begin{equation}
\begin{split}
    &\sum_{i\neq k} T_{ik} T_{ki} \frac{\mu^{D-4}}{16D}\frac{D-4}{2}\bigg[\frac{2D(8-D)m_i^2m_k^2}{m_i^2-m_k^2}\log \bigg(\frac{m_k^2}{m_i^2}\bigg)+D^2\frac{m_k^4+m_i^4}{m_i^2-m_k^2}\log \bigg(\frac{m_k^2}{m_i^2}\bigg)\\
    &+\frac{48-14D}{m_i^2-m_k^2}\bigg(m_k^4 \log \bigg(\frac{m_k^2}{\mu^2}\bigg)-m_i^4 \log \bigg(\frac{m_i^2}{\mu^2}\bigg) \bigg)\\
    &+\frac{2D}{m_i^2-m_k^2}\bigg(m_k^4 \log \bigg(\frac{m_i^2}{\mu^2}\bigg)-m_i^4 \log \bigg(\frac{m_k^2}{\mu^2}\bigg) \bigg)\bigg]\; .
\end{split}
\label{mu-ind manifest}
\end{equation}
We now use a trick to handle the logarithms. We introduce in our calculations a second arbitrary scale with mass dimension, called $\nu$. In particular we will use it to rewrite the last two terms of the previous expression \eqref{mu-ind manifest} as: 
\begin{equation}
\begin{split}
&\frac{48-14D}{m_i^2-m_k^2}\nu^4\bigg(\frac{m_k^4}{\nu^4} \log \bigg(\frac{m_k^2}{\mu^2}\bigg)-\frac{m_i^4}{\nu^4} \log \bigg(\frac{m_i^2}{\mu^2}\bigg) \bigg)\\
&+\frac{2D}{m_i^2-m_k^2}\nu^4\bigg(\frac{m_k^4}{\nu^4} \log \bigg(\frac{m_i^2}{\mu^2}\bigg)-\frac{m_i^4}{\nu^4} \log \bigg(\frac{m_k^2}{\mu^2}\bigg) \bigg)\\
&=\frac{12(4-D)}{m_i^2-m_k^2}\nu^4 \log \bigg[ \bigg(\frac{m_k^2}{\mu^2}\bigg)^{\frac{m_k^4}{\nu^4}}   \bigg(\frac{m_i^2}{\mu^2}\bigg)^{-\frac{m_i^4}{\nu^4}} \bigg]-\frac{2D(m_i^4+m_k^4)}{m_i^2-m_k^2}\log  \bigg(\frac{m_k^2}{m_i^2}\bigg) \; .
\end{split}
\end{equation}
It is interesting to notice that only the first term is still dependent on the two arbitrary mass scales $\mu$ and $\nu$. Putting together all the terms, we find the following complete result:
\begin{equation}
\begin{split}
&\sum_{i\neq k} T_{ik} T_{ki}\mu^{D-4} (D-4)\bigg[ \frac{3}{4D} (m_k^2+m_i^2)+\frac{1}{32D}\bigg(\frac{2D(8-D)m_i^2 m_k^2}{m_i^2-m_k^2} \log\bigg(\frac{m_k^2}{m_i^2} \bigg) \\
&-2D \frac{m_i^4+m_k^4}{m_i^2-m_k^2}\log\bigg(\frac{m_k^2}{m_i^2} \bigg) +D^2 \frac{m_i^4+m_k^4}{m_i^2-m_k^2} \log\bigg(\frac{m_k^2}{m_i^2} \bigg)\\ &-\frac{12D(D-4)}{m_i^2-m_k^2}\nu^4 \log\bigg(\frac{(m_k^2)^{\frac{m_k^4}{\nu^4}}}{(m_i^2)^{\frac{m_i^4}{\nu^4}}} (\mu^2)^{\frac{m_i^4-m_k^4}{\nu^4}} \bigg)  \bigg)  \bigg]\; .
\end{split}
\end{equation}
It is worth highlighting that we obtained a result in which the $\nu$- and $\mu$-dependent term can be discarded in the limit $D \rightarrow 4$, such that, as expected, we are left with a result independent from any arbitrary renormalisation scale:
\begin{equation}
\begin{split}
    \Gamma^{(1)}=&\frac{-1}{2(4\pi)^{D/2}} \Gamma\bigg( 1-\frac{D}{2}\bigg) \int {\rm d}^DX \sum_{i\neq k} T_{ik} T_{ki} \frac{D-4}{4D} \bigg[3(m_k^2+m_i^2)\\
    &+\frac{1}{8} \frac{D(D-2)(m_i^4+m_k^4)-2D(D-8)m_i^2m_k^2}{m_i^2-m_k^2}\log\bigg(\frac{m_k^2}{m_i^2} \bigg)\bigg]\; ,
\end{split}
\end{equation}
which can be rewritten as:
\begin{equation}
\begin{split}
    \Gamma^{(1)}=&\frac{-1}{(4\pi)^{D/2}} \frac{\Gamma\bigg( 3-\frac{D}{2}\bigg)}{2D(D-2)} \int  {\rm d}^DX \sum_{i\neq k} T_{ik} T_{ki} \\
    &\times \bigg[6m_i^2+\frac{D}{4}\!\times\! \frac{(D-2)m_i^4-(D-8)m_i^2m_k^2}{m_i^2-m_k^2}\log\bigg(\frac{m_k^2}{m_i^2} \bigg)\bigg]\; .
    \label{D second order ea}
\end{split}
\end{equation}
This is the final result for the one-loop second-order gradient corrections of the effective action in $D$ dimensions. As claimed, 
it is finite and manifestly scale independent.

\section{Comparison with the literature}
\label{Appendix: literature}

Our vanishing result for the one-loop corrections to the kinetic term in the single-field case is in disagreement with works already present in the literature \cite{Fraser,Chan,Iliopoulos-Itzykson-Martin}. We present here a more detailed analysis of the result obtained by these previous works.

\subsubsection{Chan (1985)}

In the following we consider reference \cite{Chan}, whose result is in agreement with references \cite{Fraser,Iliopoulos-Itzykson-Martin}. The propagator $\Delta(x;y)$ given in \cite{Chan} is,
\begin{equation}
    \Delta(x;y)=\int \frac{{\rm d}^Dp}{(2\pi)^D}{\rm e}^{ip \cdot (x-y)}\big[-p^2-U(x-i\partial_p)\big]^{-1}\; ,
    \label{Chan propagator}
\end{equation}
where $U(\phi(x))=M^2+V''(\phi(x))$. We modified \eqref{Chan propagator} in order to be consistent with the sign conventions adopted in this paper. The form of \eqref{Chan propagator} can be obtained starting from,
\begin{equation}
    \Delta(x;y)=\bigg\langle x\bigg| \frac{1}{\partial^2-\hat{U}} \bigg|y\bigg\rangle \; ,
\end{equation}
and inserting the identity,
\begin{equation}
\begin{split}
    \Delta(x;y)=&\int \frac{{\rm d}^Dp}{(2\pi)^D}  \Big\langle x  \Big|\frac{1}{\partial^2-\hat{U}} \Big| p  \Big\rangle \langle p |y \rangle\\
    =&\int \frac{{\rm d}^Dp}{(2\pi)^D} \langle x | p \rangle \frac{-1}{p^2+U(-i \overleftarrow{\partial_p})} \langle p |y \rangle\\
    =&\int \frac{{\rm d}^Dp}{(2\pi)^D}  \frac{-1}{p^2+U(x-i \overleftarrow{\partial_p})}\langle x | p \rangle\langle p |y \rangle \; ,
    \label{prop_x_Chan}
\end{split}
\end{equation}
where we used $\langle x|p \rangle= e^{ip\cdot x}$. The last line in \eqref{prop_x_Chan} represents the propagator expanded around point $x$ in gradients, which can be contrasted with the midpoint expansion used in this work (see Section \ref{Sec.II}). Note that~(\ref{Chan propagator}) can be also written as~\footnote{In fact, we can insert the identity in a different position,
\begin{equation}
\begin{split}
    \Delta(x;y)=&\int \frac{{\rm d}^Dp}{(2\pi)^D} \langle x |p  \rangle  \Big\langle p \Big| \frac{1}{\partial^2-\hat{U}}  \Big|y  \Big\rangle\\
    =&\int \frac{{\rm d}^Dp}{(2\pi)^D} \langle x | p \rangle \frac{-1}{p^2+U(i \overrightarrow{\partial_p})} \langle p |y \rangle\\
    =&\int \frac{{\rm d}^Dp}{(2\pi)^D}  \langle x | p \rangle\langle p |y \rangle \frac{-1}{p^2+U(y+i \overrightarrow{\partial_p})}
\end{split}
\end{equation}
}, 
\begin{equation}
    \Delta(x;y)=\int \frac{{\rm d}^Dp}{(2\pi)^D}{\rm e}^{ip \cdot (x-y)} \big[-p^2-U(y+i\partial_p) \big]^{-1}
\; ,
\label{Chan propagator y}
\end{equation}
where the operator $\partial_p$ now acts to the right. Equation~\eqref{Chan propagator y} now represents the propagator expanded in gradients around point y. Notice that the propagator in equations \eqref{prop_x_Chan} and \eqref{Chan propagator y} is symmetric under the transformation,
\begin{equation}
    \Delta(x;p)=\Delta(y;-p) \; ,
\end{equation}
which can be compared with our symmetry requirement \eqref{propagator: symmetry requirement: Wigner}.

The equation of motion that \eqref{Chan propagator} should satisfy is:
\begin{equation}
    [\partial^2_x-U(\phi(x))]i\Delta(x;y)=i\delta^D(x-y)\; .
    \label{chan eom x}
\end{equation}
It is easy to verify that this equation is satisfied by writing its operator version,
\begin{equation}
    (\partial^2-\hat{U})(\partial^2-\hat{U})^{-1}=\hat{\mathbb{1}}\; .\quad
\label{eom operators}
\end{equation}
Indeed, using the fact that $\langle x| \hat{X}|p\rangle = -i\partial_p \langle x| p\rangle$ and similarly that $\langle p| \hat{X}|x\rangle = i\partial_p \langle p| x\rangle$, we can project \eqref{eom operators} on the states,
\begin{equation}
\begin{split}
    &\langle x | (\partial^2-\hat{U})(\partial^2-\hat{U})^{-1} |y \rangle = \langle x |y \rangle =\delta^D(x-y)\; ,\\
    \int_p & \langle x | (\partial^2-\hat{U})(\partial^2-\hat{U})^{-1}|p\rangle \langle p | y \rangle = \delta^D(x-y)\; ,\\
    \int_p &  \big[\big(-p^2-U(-i\partial_p)\big)\big(-p^2-U(-i\partial_p)\big)^{-1}\langle x |p\rangle \big]\langle p | y \rangle = \delta^D(x-y)\; ,\\
    \int_p & e^{ip\cdot(x-y)} \big(-p^2-U(x-i\partial_p)\big)\big(-p^2-U(x-i\partial_p)\big)^{-1} = \delta^D(x-y) \; ,\\
\end{split}
\end{equation}
such that the first equation of motion can be rewritten in momentum space as,
\begin{equation}
    \big(\!-p^2-U(x-i\partial_p)\big)\big(\!-p^2-U(x-i\partial_p)\big)^{-1}=1 \; .
    \label{Chan eom x momentum}
\end{equation}
One can now verify that equation \eqref{Chan eom x momentum} is satisfied order by order in the gradient expansion.

Let us now consider the second differential equation, which also must be satisfied by the propagator,
\begin{equation}
    [\partial^2_y-U(\phi(y))]i\Delta(x;y)=i\delta^D(x-y)\; .
    \label{chan eom y}
\end{equation}
Projecting again equation \eqref{eom operators} and inserting the identity in a different position,
\begin{equation}
\begin{split}
    &\langle x | (\partial^2-\hat{U})(\partial^2-\hat{U})^{-1} |y \rangle = \delta^D(x-y)\; ,\\
    \int_p &\langle x |p\rangle \langle p|(\partial^2-\hat{U})(\partial^2-\hat{U})^{-1} | y \rangle = \delta^D(x-y)\; ,\\
    \int_p & e^{ip\cdot(x-y)} \big(-p^2-U(y+i\partial_p)\big)\big(-p^2-U(y+i\partial_p)\big)^{-1} = \delta^D(x-y) \; ,\\
\end{split}
\end{equation}
such that the second equation of motion can be written as,
\begin{equation}
     \big(-p^2-U(y+i\partial_p)\big)\big(-p^2-U(y+i\partial_p)\big)^{-1} =1 \; ,
     \label{chan eom at y}
\end{equation}
and expanding in gradients, one can verify that both equations \eqref{chan eom x} and \eqref{chan eom y} are satisfied by the propagator of \cite{Chan}.

Using the momentum space expression for the propagator~\eqref{Chan propagator}, and expanding up to
the second order in derivatives, one obtains, 
\begin{equation}
    \begin{split}
        \Delta(x;p) =& -\frac{1}{p_{\epsilon}^2+U(x)} 
             +2i[\partial_\mu^x U(x)] \frac{p^\mu}{[p_{\epsilon}^2+U(x)]^3}\\
        &+[\partial_\mu^x\partial_\nu^x U(x)]\left[\frac{\eta^{\mu\nu}}{[p_{\epsilon}^2+U(x)]^3}
   -\frac{4p^\mu p^\nu}{[p_{\epsilon}^2+U(x)]^4} \right]\\
   & - 2[\partial_\mu^x U(x)][\partial_\nu^x U(x)]
                  \left[\frac{\eta^{\mu\nu}}{[p_{\epsilon}^2+U(x)]^4}-\frac{6p^\mu p^\nu}{[p_{\epsilon}^2+U(x)]^5} \right] \; ,
    \end{split}
\label{Chan propagator: momentum space}
\end{equation}
which differs from our propagator in equation (\ref{propagator expansion terms2}), both at first order, for which our correction vanishes, and at second order. Integrating over the momentum, we can obtain the off-coincident propagator following similar calculations as displayed in Subsection \ref{subsec:Midpoint expansion of the propagator}, i.e. we calculate:
\begin{equation}
\begin{split}
    i\Delta(x;y)=&\int \frac{{\rm d}^{\rm D}p}{(2\pi)^D}e^{ip\cdot(x-y)}i\Delta(x;p)\\
    =&i\Delta^{(0)}(x;y)+i\Delta_{\nu-2}^{(1)}(x;y)+i\Delta^{(2)}_{\nu-2}(x;y)+i\Delta^{(2)}_{\nu-3}(x;y)+i\Delta^{(2)}_{\nu-4}(x;y)
\end{split}
\label{off-coincident_prop_Chan}
\end{equation}
with $i\Delta^{(0)}(x;y)$ as given in equation \eqref{prop0off} and
\begin{equation}
    i\Delta_{\nu-2}^{(1)}= - \frac14\frac{U^{\nu-2}(x)}{(2\pi)^{\nu+1}}(\partial^x_{\mu} U(x)) \frac{\partial}{\partial \Delta x_{\mu}} \frac{K_{\nu-2}(z)}{z^{\nu-2}}
\end{equation}
\begin{equation}
    i\Delta^{(2)}_{\nu-2}=-\frac{U^{\nu-2}(x)}{(2\pi)^{\nu+1}}\frac{1}{8}(\partial^2_x U(x))\frac{K_{\nu-2}(z)}{z^{\nu-2}}
\end{equation}
\begin{equation}
    i\Delta^{(2)}_{\nu-3}=\frac{U^{\nu-3}(x)}{(2\pi)^{\nu+1}}\frac{1}{24}\bigg[(\partial_x U(x))^2-2(\partial_{\mu}^x\partial^x_{\sigma}U(x)) \frac{\partial}{\partial \Delta x_{\mu}}\frac{\partial}{\partial \Delta x_{\sigma}}\bigg] \frac{K_{\nu-3}(z)}{z^{\nu-3}}
\end{equation}
\begin{equation}
    i\Delta^{(2)}_{\nu-4}=\frac{U^{\nu-4}(x)}{(2\pi)^{\nu+1}}\frac{(\partial_{\mu}^xU(x))(\partial^x_{\sigma}U(x))}{32} \frac{\partial}{\partial \Delta x_{\mu}}\frac{\partial}{\partial \Delta x_{\sigma}}\frac{K_{\nu-4}(z)}{z^{\nu-4}}
\label{Chan propagator: nu - 4}
\,,\quad
\end{equation}
with $\nu=\frac{D-2}{2}$ and $z^2=U(x)\Delta x^2$. At first and second order, this clearly differs from our off-coincident propagator in equation \eqref{propoffcoinc}.

We now have two ways to obtain the coincident propagator: we can take the coincident limit of \eqref{off-coincident_prop_Chan}, or we can integrate \eqref{Chan propagator: momentum space} over the momenta (the first order term drops out and $p^\mu p^\nu \rightarrow \eta^{\mu\nu}p^2/D$). We verify that in both ways we obtain the following coincident propagator, 
\begin{equation}
\begin{split}
    i\Delta(x;x)=&\frac{1}{(4\pi)^{D/2}}\bigg[\Gamma\bigg(1-\frac{D}{2}\bigg)[U(x)]^{\frac{D}{2}-1}\\
    &-\frac{1}{6}\Gamma\bigg(3-\frac{D}{2}\bigg)[U(x)]^{\frac{D}{2}-3}(\partial_x^2 U)+\frac{1}{12}\Gamma\bigg(4-\frac{D}{2}\bigg)[U(x)]^{\frac{D}{2}-4}(\partial_x U)^2\bigg]
\,,
\end{split}
\label{coincident propagator Chan}
\end{equation}
which agrees with equation~(9) of~\cite{Chan}.
Curiously, upon identifying $U(x) \rightarrow m^2(x)$, the coincident propagator~(\ref{coincident propagator Chan})  
is identical to our coincident propagator given in equations~\eqref{prop0coin} and~\eqref{prop2coin}.

Up to this point, we have found that using Chan's procedure results in a different propagator~(\ref{Chan propagator: momentum space}), which at coincidence gives the same answer. Next we shall calculate the one-loop effective action.
In what follows we use two procedures, the first is the one advocated in this work, while the second is the one 
used in~\cite{Chan}.

We first compute the one-loop effective action by using our procedure and 
the propagator~\eqref{Chan propagator: momentum space}.
We recall that the action is first expanded as in~\eqref{gradient expansion effective action}, 
\begin{equation}
    \Gamma^{(1)}=-\frac{i}{2} \Big[\text{Tr}\ln[\Delta^{(0)}]+\text{Tr}(\mathcal{D}^{(0)}\Delta^{(2)})-\frac{1}{2}\text{Tr}(\mathcal{D}^{(0)}\Delta^{(1)})^2\Big]
\,,
\label{gamma1 chan}
\end{equation}
where we used the fact that $\text{Tr}(\mathcal{D}^{(0)}\Delta^{(1)})=0$ (since integrating over the momenta an odd function of $p^\mu$ vanishes). The first term in~(\ref{gamma1 chan}) is the zeroth order in gradient 
contribution and can be calculated by using standard methods. The other two terms can be calculated by making use 
of the propagator~(\ref{Chan propagator: momentum space}), resulting in
\begin{eqnarray}
    \Gamma^{(1)}&\!\!=\!\!&\frac{1}{2(4\pi)^{D/2}}\Gamma\left(-\frac{D}{2}\right)\int {\rm d}^D x [U(x)]^\frac{D}{2}
    \nonumber\\
      &&  -\,\frac{i}{2}\int {\rm d}^D x \int \frac{{\rm d}^Dp}{(2\pi)^D}\bigg( 
           \! \!-\![\partial_\mu^x\partial_\nu^x U(x)]\left[\frac{\eta^{\mu\nu}}{[p_{\epsilon}^2\!+\!U(x)]^2}
   \!-\!\frac{4p^\mu p^\nu}{[p_{\epsilon}^2\!+\!U(x)]^3} \right]
    \nonumber\\
       &&                \!+\,2[\partial_\mu^x U(x)][\partial_\nu^x U(x)]
\left[\frac{\eta^{\mu\nu}}{[p_{\epsilon}^2\!+\!U(x)]^3}
   \!-\!\frac{6p^\mu p^\nu}{[p_{\epsilon}^2\!+\!U(x)]^4} \right]
    \nonumber\\
        &&   -\,[\partial_\mu^x U(x)][\partial_\nu^x U(x)]
        \left[\frac{\eta^{\mu\nu}}{[p_{\epsilon}^2\!+\!U(x)]^3}
   \!-\!\frac{6p^\mu p^\nu}{[p_{\epsilon}^2\!+\!U(x)]^4} \right] \bigg)
        \,,\qquad
\label{Gamma 1: Chan}
\end{eqnarray}
where term in the last line, which contributes as $-1/2$ times the term in the penultimate line,
comes from the last term in~(\ref{gamma1 chan}).
This can be easily integrated (upon exacting the substitution, $p^\mu p^\nu \rightarrow \eta^{\mu\nu}p^2/D$
and making use of~(\ref{appendix D: easy integral: In})) to yield, 
\begin{eqnarray}
    \Gamma^{(1)}&\!\!=\!\!&\frac{\Gamma\left(\!-D/2\right)}{2(4\pi)^{D/2}}\!\int \!{\rm d}^D x [U(x)]^\frac{D}{2}
\nonumber\\
   &\!\!+\!\!& \frac{1}{2(4\pi)^{D/2}}\!\int\! {\rm d}^D x
   \Bigg\{\!-[U(x)]^{\frac{D}{2}-2}[\partial_x^2 U(x)]\Gamma\left(\!2\!-\!\frac{D}{2}\right)
       \bigg[1-\frac{4}{D}\bigg(1-\frac{2-D/2}{2}\bigg)\bigg]
\nonumber\\
    &&\hskip 3cm   +\,\frac12 [U(x)]^{\frac{D}{2}-3}[\partial_x U(x)]^2\Gamma\left(\!3\!-\!\frac{D}{2}\right)
           \bigg[1-\frac{6}{D}\bigg(1-\frac{3-D/2}{3}\bigg)\bigg]
   \Bigg\}\quad
\nonumber\\
    &\!\!=\!\!& 0 
        \,,\qquad
\label{Gamma 1: Chan 2}
\end{eqnarray}
which agrees with our result~(\ref{final result: single field}), even though here we did not even need to 
integrate by parts to get a vanishing result. This is probably due to the fact that the effective action is obtained taking the coincident limit, in which the propagators agree.

Chan in~\cite{Chan} calculates the effective action by a different procedure. He first 
 takes a (functional) derivative of the effective action with respect to 
$U(x)$, and then integrates over the momenta, resulting in (see equation (7) of \cite{Chan}), 
\begin{eqnarray}
\Gamma^{(1)} = - \frac{1}{2}\int {\rm d}^Dx \int {\rm d}U i\Delta(x;x)
\,,
\label{Gamma 1: Chan 3}
\end{eqnarray}
where $i\Delta(x;x)$ is the coincident propagator given in~(\ref{coincident propagator Chan}). Upon integrating over $U$ 
one obtains, 
\begin{eqnarray}
    \Gamma^{(1)}&\!\!=\!\!&\frac{1}{2(4\pi)^{D/2}}\!\int \!{\rm d}^D x\Bigg\{\Gamma\left(\!-\frac{D}{2}\right) [U(x)]^\frac{D}{2}
\nonumber\\
&\!\!\!\!&  -\,\frac{1}{6}\Gamma\left(2\!-\!\frac{D}{2}\right)[U(x)]^{\frac{D}{2}-2}[\partial_x^2 U(x)]
        \!+\!\frac{1}{12}\Gamma\left(3\!-\!\frac{D}{2}\right)[U(x)]^{\frac{D}{2}-3}[\partial_x U(x)]^2\Bigg\}
        \,.\qquad
\label{Gamma 1: Chan 4}
\end{eqnarray}
Upon integrating the second term by parts, one finally obtains, 
\begin{eqnarray}
    \Gamma^{(1)}&\!\!=\!\!&\frac{1}{2(4\pi)^{D/2}}\!\int \!{\rm d}^D x\Bigg\{\Gamma\left(\!-\frac{D}{2}\right) [U(x)]^\frac{D}{2}
        \!-\!\frac{1}{12}\Gamma\left(3\!-\!\frac{D}{2}\right)[U(x)]^{\frac{D}{2}-3}[\partial_x U(x)]^2\Bigg\}
        \,,\qquad
\label{Gamma 1: Chan 5}
\end{eqnarray}
which disagrees with~(\ref{Gamma 1: Chan 2}). This is exactly the result obtained in Ref.~\cite{Chan}, 
which also agrees with earlier literature~\cite{Iliopoulos-Itzykson-Martin,Fraser}, and with the result obtained in Section~\ref{Taking a parametric derivative}.

\subsubsection{Iliopoulos, Itzykson, Martin (1975)}

We now proceed with the paper \cite{Iliopoulos-Itzykson-Martin}, which expanded the (quantum) field in powers of $x^\mu$ (to quadratic order in $\delta x^\mu$)
around some fiducial point $x^\mu$. The idea is that integrating out over the (quantum) fluctuations of 
$\delta x^\mu$ will account for the spacetime dependence of the field in the one-loop effective action, 
at least approximately.

We start by considering equation (2.5.18) for the quantum mechanical kernel~\cite{Sakurai:2020}
 valid for the one-dimensional harmonic oscillator,
\begin{equation}
\begin{split}
    &K(x,x';t,t')=\big\langle x \big| \exp \bigg[-\frac{i}{\hbar}\bigg(\frac{\hat{P}^2}{2m}+\frac{m \omega^2}{2}\hat{Q}^2\bigg)(t-t')\bigg]\big |x'\big\rangle\\
    &=\bigg(\frac{m\omega}{2 \pi \hbar i \sin(\omega \Delta t)} \bigg)^{1/2}\exp\bigg[\! -\frac{i m \omega}{2 \hbar \sin( \omega \Delta t)} \big[(x^2+x'^2)\cos(\omega \Delta t)-2x x'\big] \bigg]\; , 
    \label{ho IIM}
\end{split}
\end{equation}
where $\Delta t=t-t'$ and $\hat{P}$ and $\hat{Q}$ are the quantum mechanical momentum and position
operators satisfying $[\hat Q,\hat P]=i$. 

We start from the propagator and working in Euclidean signature we calculate,
\begin{equation}
\Delta(x;y)=-i \int_0^{\infty}{\rm d}s \, \Big\langle x \Big|e^{s[\partial^2-M^2-\hat{U}]}\Big|y\Big\rangle
    \label{IIP 1}
\end{equation}
where $\hat{U}=\frac{\lambda \phi^2(\hat{X})}{2}$ and $M^2$ denotes the tree level mass term. As done in \cite{Iliopoulos-Itzykson-Martin}, we
expand $\phi^2(\hat{X})$  around a reference point $x_0$ by writing $\hat{X}=x_0+\Delta \hat{ x}$ being $\Delta\hat{x}$ a quantum fluctuation, such that we have:
\begin{equation}
    \hat{U}=U(x_0)+\Delta\hat{ x}^{\mu}(\partial_{\mu}U(x_0))+\frac{1}{2} \Delta\hat{ x}^{\mu} \Delta\hat{ x}^{\nu} (\partial_{\mu}\partial_{\nu} U(x_0))+\cdots
\,.
\end{equation}
In order to exploit the result for the one-dimensional harmonic oscillator~(\ref{ho IIM}), 
we consider the {\it Ansatz},
\begin{equation}
\begin{split}
    &\Delta\hat{x}^{\mu}(\partial_{\mu}U(x_0)) \!+\!\frac{1}{2} \Delta\hat{x}^{\mu} \Delta\hat{x}^{\nu} (\partial_{\mu}\partial_{\nu} U(x_0))\\
    &=\frac{1}{2} (\partial_{\mu}\partial_{\nu} U(x_0))
    ( \Delta\hat{x}^{\mu}\!+\!a^{\mu})( \Delta\hat{x}^{\nu}\!+\!a^{\nu})\!-\!\frac{1}{2}a^{\mu} a^{\nu} (\partial_{\mu} \partial_{\nu} U(x_0))
\,,
\end{split}
\end{equation}
where vector $a^\nu$ is defined by,
\begin{equation}
     (\partial_{\mu} \partial_{\nu} U)a^{\nu} =(\partial_{\mu}U)
    \,,
\end{equation}
and we call $\Omega_{\mu \nu}^2=\partial_{\mu} \partial_{\nu} U$. Differently from \cite{Iliopoulos-Itzykson-Martin}, we choose not to restrict the expansion only around one direction and we generalise the result from four to $D$ dimensions. We can then rewrite \eqref{IIP 1} as,
\begin{equation}
     -i \int {\rm d}s\,\Big\langle x \Big| \exp\bigg\{s\bigg[\Hat{P}^2-\frac{1}{2}(\Delta\hat{x}+a)\!\cdot \!\Omega^2\!\cdot \! (\Delta\hat{x}+a)\bigg]\bigg\}\Big|y\Big\rangle \times e^{-s\big[M^2-\frac{1}{2}a\cdot \Omega^2\cdot a+U(x_0)\big]}\; .
\label{Iliop intermediate}
\end{equation}
Going to the basis in which $\Omega^2$ is diagonal, and exacting the replacements $\Delta t \rightarrow -is$, $\hbar \rightarrow 1$, $m \rightarrow \frac{1}{2}$, $\omega \rightarrow \sqrt{2} \Omega$, we can exploit the result of the harmonic oscillator in \eqref{ho IIM}, to rewrite~(\ref{Iliop intermediate}) as, 
\begin{equation}
\begin{split}
     -i\int{\rm d}s &\,e^{-s\big[M^2-\frac{1}{2}a\cdot\Omega^2\cdot a+U(x_0)\big]}\prod_{i=1}^D \bigg(\frac{\Omega_i}{2\sqrt{2}\pi \sinh(\sqrt{2}\Omega_i s)}\bigg)^{1/2}\\
    &\times \exp\bigg\{-\frac{\Omega_i}{2\sqrt{2}\sinh(\sqrt{2}\Omega_i s)}\bigg[[(\Delta x +a)_i^2+(\Delta y+a)_i^2]\cosh(\sqrt{2}\Omega_i s)\\
    &-2(\Delta x+a)_i(\Delta y+a)_i\bigg]\bigg\}\; .
\end{split}
\end{equation}
Expanding up to second order in derivatives, we obtain the following off-coincident result,
\begin{equation}
\begin{split}
    -\frac{i}{(4\pi)^{D/2}}\int\frac{{\rm d}s}{s^{D/2}}&e^{-s\big[M^2-\frac{1}{2}a\cdot\Omega^2\cdot a+U(x_0)\big]}e^{-\frac{1}{4s}\big((\Delta x +a)-(\Delta y+a)\big)^2}e^{-\frac{s^2}{6}\sum_i(\Omega_i^2)}\\
    &\times e^{-\frac{s}{6}\sum_i\big[\Omega_i^2\big((\Delta x +a)_i^2+(\Delta y+a)_i^2+(\Delta x +a)_i(\Delta y+a)_i\big)\big]}\\
    &\times e^{\frac{s^3}{180}\sum_i\big\{\Omega_i^4\big[4\big((\Delta x +a)_i^2+(\Delta y+a)_i^2\big)+7(\Delta x +a)_i(\Delta y+a)_i\big]\big\}}\; ,\quad
\end{split}
\end{equation}
where here $x_0$ is an arbitrary point of expansion. It is instructive to study how the result depends on the choice of $x_0$. For the purpose of this paper, two choices are of interest: $x_0 \rightarrow X=\frac{x+y}{2}$ and $x_0 \rightarrow x$ (the third choice of interest $x_0 \rightarrow y$ can be determined from the symmetry requirement). Let us start with $x_0 \rightarrow X$,
keeping only terms up to the second order in the number of derivatives, we can rewrite the propagator as,
\begin{equation}
\begin{split}
    \Delta(x;y)=&-\frac{i}{(4\pi)^{D/2}}\int\frac{{\rm d}s}{s^{D/2}}e^{-s\big[M^2+U(X)\big]}e^{-\frac{(x-y)^2}{4s}}
\\
    &\times \bigg[1-\frac{s}{24}(x-y)^{\mu}(\partial_{\mu}\partial_{\nu}U)(x-y)^{\nu}\bigg]\bigg[1-\frac{s^2}{6}(\partial^2 U)\bigg]\bigg[1+\frac{s^3}{12}(\partial U)^2\bigg] \; .
\end{split}
\end{equation}
Using the substitution,
\begin{equation}
    (x-y)^{\mu}(x-y)^{\nu} e^{-\frac{(x-y)^2}{4s}} = 
     4s^2\frac{\partial}{\partial \Delta x_{\mu}}\frac{\partial}{\partial \Delta x_{\nu}}
     e^{-\frac{(x-y)^2}{4s}}+ 2s \, \eta^{\mu \nu} e^{-\frac{(x-y)^2}{4s}}
\label{appendix C: substitution}
\end{equation}
and performing the integral over $s$, we obtain the final expression for the propagator expanded around the midpoint coordinate,
\begin{equation}
\begin{split}
i\Delta(x;y)=&\frac{1}{(4\pi)^{D/2}}\Bigg[m^{D-2}\frac{K_{\frac{D-2}{2}}(z)}{z^{\frac{D-2}{2}}}
-\frac{ m^{D-8} }{48}(\partial_{\mu}\partial_{\nu} m^2)
 \frac{\partial}{\partial \Delta x_{\mu}}\frac{\partial}{\partial \Delta x_{\nu}}
\bigg(\frac{K_{\frac{D-8}{2}}(z)}{z^{\frac{D-8}{2}}}\bigg)\\
&-\frac{1}{16} m^{D-6} (\partial^2m^2)\frac{K_{\frac{D-6}{2}}(z)}{z^{\frac{D-6}{2}}}+\frac{1}{96}m^{D-8}(\partial m^2)^2\frac{K_{\frac{D-8}{2}}(z)}{z^{\frac{D-8}{2}}}\Bigg]\, ,
\end{split}
\end{equation}
with $z=m\sqrt{\Delta x^2}$ and  $m^2(x) \equiv M^2 + U(x)$. This exactly corresponds to our off-coincident propagator given in equation \eqref{propoffcoinc}.

Next we construct the propagator by expanding around point $x$. Starting from Eq.~(C34), 
 and taking account of
$U(x_0)\rightarrow U(x)$, $\Delta x\rightarrow 0$, $\Delta y\rightarrow y\!-\!x$, 
we can write the propagator (truncated to second order 
in derivatives) as, 
\begin{eqnarray}
i\Delta(x;y) &\!\!=\!\!& \frac{1}{(4\pi)^\frac{D}{2}}\int\frac{{\rm d}s}{s^\frac{D}{2}}
    e^{-s[M^2+U(x)]}e^{-\frac{(y-x)^2}{4s}} \bigg\{\!1\!-\!\frac{s}{6}\Big[3(y\!-\!x)^\mu \partial_\mu U(x)
      \!+\!(y\!-\!x)^\mu\partial_\mu\partial_\nu U(x)(y\!-\!x)^\nu\Big]
\nonumber\\
&\!\!&\hskip 0cm
      +\,\frac{s^2}{8}(\partial_\mu U)(\partial_\nu U)(y\!-\!x)^\mu(y\!-\!x)^\nu
      -\,\frac{s^2}{6}(\partial^2_xU)\!+\!\frac{s^3}{12}(\partial_xU)^2\bigg\}
   \; .
\end{eqnarray}
Next, inserting into this the substitution,
\begin{equation}
(y\!-\!x)^\mu e^{-\frac{(y-x)^2}{4s}} \equiv (y\!-\!x)^\mu e^{-\frac{(y-x)^2}{4s}}
 \longrightarrow -2s\frac{\partial}{\partial(y\!-\!x)_\mu}e^{-\frac{(y-x)^2}{4s}}
 \,,\quad
\end{equation}
and~(\ref{appendix C: substitution}) and integrating over $s$ one obtains, 
\begin{equation}
\begin{split}
i\Delta(x;y)=\,&\frac{1}{(2\pi)^\frac{D}{2}}\Bigg\{m^{D-2}\frac{K_{\frac{D-2}{2}}(z)}{z^{\frac{D-2}{2}}}
\!-\!\frac{m^{D-6}}{8}\left[\big(\partial^2m^2\big)
\!+\!2\big(\partial_\mu m^2(x)\big)\frac{\partial}{\partial\Delta x_\mu} \right] \frac{K_{\frac{D-6}{2}}(z)}{z^{\frac{D-6}{2}}}
\\
&\hskip 1.45cm
+\,\frac{m^{D-8}}{24}\left[\big(\partial m^2\big)^2
\!-\!2\big(\partial_\mu\partial_\nu m^2(x)\big)
 \frac{\partial}{\partial\Delta x_\mu}\frac{\partial}{\partial\Delta x_\nu}\right] 
  \frac{K_{\frac{D-8}{2}}(z)}{z^{\frac{D-8}{2}}}
  \\
&\hskip 1.5cm
  +\,
\frac{m^{D-10}}{32}(\partial_{\mu}m^2)(\partial_{\nu} m^2)
 \frac{\partial}{\partial \Delta x_{\mu}}\frac{\partial}{\partial \Delta x_{\nu}}
 \bigg(\frac{K_{\frac{D-10}{2}}(z)}{z^{\frac{D-10}{2}}}\bigg)
\Bigg\}\, ,
\end{split}
\end{equation}
where here $z^2 \equiv m^2(x)(x\!-\!y)^2\equiv m^2(x)\Delta x^2$. This expression completely agrees
with the Chan propagator~(\ref{off-coincident_prop_Chan})-(\ref{Chan propagator: nu - 4})
 obtained by integrating over the momenta Eq.~(\ref{Chan propagator: momentum space}).

With similar calculations we can now find the one-loop effective action, which is given by,
\begin{equation}
    \frac{i}{2}\text{Tr}\log\bigg(\frac{\partial^2-M^2-\hat{U}}{\mu^2}\bigg)=\lim_{\mu \rightarrow \infty} \frac{i}{2}\int {\rm d}^Dx\int_{\mu^{-2}}^{\infty}\frac{{\rm d}s}{s}\Big\langle x \Big|e^{s[\partial^2-M^2-\hat{U}]}\Big|y\Big\rangle_{x=y}\, ,
    \label{IIP 1}
\end{equation}
where the equality holds in the limit $\mu \rightarrow \infty$, up to an irrelevant field-independent constant. To calculate the effective action, we are interested in the coincident limit,
\begin{eqnarray}
&&    \frac{i}{2}\Big\langle x \Big| \log\bigg(\frac{\partial^2-M^2-\hat{U}}{\mu^2}\bigg)\Big|x\Big\rangle
\nonumber\\
&& \hskip 2cm
   =\frac{i}{2(4\pi)^{D/2}}\int_x\int\frac{{\rm d}s}{s^{1+D/2}}e^{-s[M^2+U(x)]}e^{-\frac{s^2}{6}\sum_i(\Omega_i^2)} e^{\frac{s^3}{12}\sum_i\big(\Omega_i^4(\Delta x +a)_i^2\big)}\; ,\qquad
\end{eqnarray}
as done before, we can now integrate over $s$ using the fact that $\Gamma(z)=\int_0^{\infty}{\rm d}t \, t^{z-1}e^{-t}$, and keep only terms up to second order in derivatives. We find the one-loop contribution to the effective action in Euclidean space,
\begin{equation}
\begin{split}
   &\Gamma^{(1)}=\frac{i}{2(4\pi)^{D/2}}\int {\rm d}^Dx\bigg[(m^2)^{\frac{D}{2}}\Gamma\bigg(\! \! -\frac{D}{2},\frac{m^2}{\mu^2}\bigg)\\
    &-\frac{1}{6}(m^2)^{\frac{D}{2}-2}\Gamma\bigg(2-\frac{D}{2}\bigg)(\partial^2_x m^2(x))+\frac{1}{12}(m^2)^{\frac{D}{2}-3}\Gamma\bigg(3-\frac{D}{2}\bigg)(\partial_x m^2)^2\bigg]
\; ,
\end{split}
\end{equation}
where $\Gamma(\nu,z)$ denotes the incomplete gamma function.
Rotating back to Lorentzian space (${\rm d}^D x \rightarrow - i {\rm d}^D x$)
 and integrating by parts, we find the following one-loop contribution to the effective action,
\begin{equation}
\begin{split}
\Gamma^{(1)}=\frac{1}{2(4\pi)^{\frac{D}{2}}}\int {\rm d}^Dx
\bigg[(m^2)^\frac{D}{2}\Gamma\bigg(\! \! -\frac{D}{2},\frac{m^2}{\mu^2}\bigg)
   -\frac{1}{12}(m^2)^{\frac{D}{2}-3}\Gamma\bigg(3-\frac{D}{2}\bigg)(\partial_x m^2)^2\bigg]
\; , 
\end{split}
\label{Gamma 1: Iliopoulos et al}
\end{equation}
which corresponds to the result obtained by \cite{Iliopoulos-Itzykson-Martin} generalised to $D$ dimensions.
 The result~(\ref{Gamma 1: Iliopoulos et al}) is divergent in $D=4$, and it can be 
  regularized and renormalised as follows.
First note that incomplete gamma function has the following expansion around 
$\mu\rightarrow \infty$,
\begin{equation}
\begin{split}
  \Gamma\bigg(\! \! -\frac{D}{2},\frac{m^2}{\mu^2}\bigg)
   =& \Gamma\bigg(\! \! -\frac{D}{2}\bigg) + \left(\frac{m^2}{\mu^2}\right)^{\!\!-\frac{D}{2}}\!
   \left[\frac{2}{D}\!-\!\frac{2}{D\!-\!2}\frac{m^2}{\mu^2}
     \!+\!\frac{1}{D\!-\!4}\left(\frac{m^2}{\mu^2}\right)^{\!\!2}\right] 
     \!\\
     &+{\cal O}\!\left(\!\left(\frac{m}{\mu}\right)^{\!\!6-D}\right)
\;.
\end{split}
\label{incomplete gamma function}
\end{equation}
The first two terms in the square brackets are quartically and quadratically divergent in powers of $\mu$ in $D=4$, and are automatically 
subtracted in dimensional regularization (by assuming $\Re[D]<0$ when taking the limit $\mu\rightarrow\infty$,
and then analytically extending to all $D$), leaving us with the last term which is logarithmically divergent
when $D\rightarrow 4$. Combining this with the expansion of $\Gamma\big( \! -\frac{D}{2}\big)$
around $D=4$ yields a finite result in $D=4$,
\begin{equation}
  \Gamma\bigg(\! \! -\frac{D}{2},\frac{m^2}{\mu^2}\bigg)
   \rightarrow -\frac12
   \left[\log\left(\frac{m^2}{\mu^2}\right)+\gamma_E-\frac{3}{2}\right] 
     \!+\!{\cal O}\!\left(D\!-\!4\right)
\,.
\label{incomplete gamma function 2}
\end{equation}
When this is inserted into~(\ref{Gamma 1: Iliopoulos et al}) one can take the $D\rightarrow 4$ limit 
to obtain,
\begin{equation}
\Gamma^{(1)}_{\rm ren}=\frac{1}{32\pi^2}\int {\rm d}^4x
\bigg\{\!-\frac{m^4}{2}
   \left[\log\left(\frac{m^2}{\mu^2}\right)\!+\!\gamma_E\!-\!\frac{3}{2}\right] 
   \!-\!\frac{(\partial_x m^2)^2}{12m^2}\bigg\}
\; ,
\label{Gamma 1: Iliopoulos et al D=4}
\end{equation}
which agrees with~\cite{Iliopoulos-Itzykson-Martin}.

\subsubsection{Fraser (1985)}

Finally, we consider~\cite{Fraser}. The method used in this reference is based on the idea that we can expand the field as $\hat{\phi}(x)=\phi_0+\hat{\psi}(x)$, where $\phi_0$ is a constant referent field and $\hat{\phi}(x)$ a spacetime dependent perturbation on top of $\phi_0$. Next, they expand the one-loop effective action 
to second order in powers of the field operator, and evaluate the contribution of fluctuations 
to quadratic order in fluctuations. 
As a first step, we write the propagator of \cite{Fraser} as,
\begin{equation}
        \Delta(x,y)=\bigg \langle x\bigg| \frac{1}{\partial^2-\mu^2-\frac{\lambda}{2}\hat{\phi}^2} \bigg|y\bigg \rangle \; .
\end{equation}
Next, calling $\hat{B}=\lambda(\phi_0\hat{\psi}+\frac{\hat{\psi}^2}{2})$ and $M^2=\mu^2+\frac{\lambda}{2}\phi_0^2$, expanding and inserting the identity $\int_k |k\rangle \langle k|$ one finds,
\begin{equation}
\begin{split}
  &\Delta(x,y)=\int_k e^{ik\cdot(x-y)}\frac{-1}{k_{\epsilon}^2+M^2}+\int_{k,p}e^{ik\cdot x-ip\cdot y} \int_z e^{i(p-k)\cdot z}B(z) \frac{-1}{p_{\epsilon}^2+M^2}\\
  &+\int_{p,q,k}e^{ik\cdot x-iq\cdot y}\frac{-1}{k_{\epsilon}^2+M^2}\int_z e^{i(p-k)\cdot z}B(z)\frac{-1}{p_{\epsilon}^2+M^2}\int_{z'} e^{i(q-p)\cdot z'}B(z')\frac{-1}{q_{\epsilon}^2+M^2}\; .
\end{split}
\end{equation}
Using the fact that $e^{i(p-k)\cdot z}B(z)=B(i\partial_k)e^{i(p-k)\cdot z}$ and integrating by parts, the previous expression can be written as,
\begin{equation}
\begin{split}
     \Delta(x,y)=&\int_k e^{ik\cdot (x-y)}\frac{-1}{k_{\epsilon}^2+M^2}+\int_{k}\frac{e^{-ik\cdot y}}{k_{\epsilon}^2+M^2}\bigg[B(-i\partial_k)\frac{e^{ik\cdot x}}{k_{\epsilon}^2+M^2}\bigg]\\
     &-\int_k \bigg[B(-i\partial_k)\frac{e^{ik\cdot x}}{k_{\epsilon}^2+M^2}\bigg]\frac{1}{k_{\epsilon}^2+M^2}\bigg[B(i\partial_k)\frac{e^{-ik\cdot y}}{k_{\epsilon}^2+M^2}\bigg]\; ,
\end{split}
\end{equation}
which can then be recast as,
\begin{equation}
    \begin{split}
     \Delta(x,y)=&\int_k e^{ik\cdot (x-y)}\bigg\{-\frac{1}{k_{\epsilon}^2+M^2}+\frac{1}{k_{\epsilon}^2+M^2}B(x-i\partial_k)\frac{1}{k_{\epsilon}^2+M^2}\\
     &-\frac{1}{k_{\epsilon}^2+M^2}B(x-i\partial_k)\bigg[\frac{1}{k_{\epsilon}^2+M^2}B(x-i\partial_x)\frac{1}{k_{\epsilon}^2+M^2}\bigg]\bigg\}\; .
\end{split}
\end{equation}
Finally, in Wigner space the previous expression reads,
\begin{eqnarray}
\Delta(x;p) &\!\!=\!\!& -\frac{1}{k_{\epsilon}^2\!+\!m^2(x)\!} 
             +2i[\partial_\mu^x B(x)]\frac{k^\mu}{[k_{\epsilon}^2\!+\!m^2(x)]^3}
\label{Fraser propagator: momentum space}\\
              && \hskip 0cm
             +\,[\partial_\mu^x\partial_\nu^x B(x)]\left[\frac{\eta^{\mu\nu}}{[k_{\epsilon}^2\!+\!m^2(x)]^3}
   \!-\!\frac{4k^\mu k^\nu}{[k_{\epsilon}^2\!+\!m^2(x)]^4} \right]
\nonumber\\
              && \hskip 0cm
                  -\, 2[\partial_\mu^x B(x)][\partial_\nu^x B(x)]
                  \left[\frac{\eta^{\mu\nu}}{[k_{\epsilon}^2\!+\!m^2(x)]^4}-\frac{6k^\mu k^\nu}{[k_{\epsilon}^2\!+\!m^2(x)]^5} \right]
\,,
\nonumber
\end{eqnarray}
where $m^2(x)=M^2+B(x)$ and the propagator corresponds to the one of \cite{Chan}, given in equation \eqref{Chan propagator: momentum space}. As a consequence, all the considerations in the previous subsections of Appendix~\ref{Appendix: literature} apply also for this method,  and therefore
 the method proposed in
Ref.~\cite{Fraser} (when resummed) does not bring anything new.

\subsubsection{Summary}

The starting point of these works is the one-loop effective action~(\ref{1 loop effective action}),
$\Gamma^{(1)}[\phi_{a}] \!=\!\frac{i}{2} \hbar \text{Tr} 
             \log\big[ \mathcal{D}_{ab}[\phi_c](x;x')\big]$, obtained by integrating out 
quantum fluctuations of the field to quadratic order around a fixed classical background field $\phi_c(x)$. 
Our approach is to evaluate the operator $\mathcal{D}_{ab}[\phi_c](x;x')$ by evaluating 
its inverse -- the propagator -- in some consistent approximation scheme, and use it to compute the one-loop effective action, both with an exact calculation and with an approximate one, obtaining the same result. The approximation scheme used in this work is an expansion of the propagator and subsequently the effective action in spacetime gradients, which is valid when the field varies sufficiently slowly in space and time. Computing the propagator with the methods of \cite{Chan,Iliopoulos-Itzykson-Martin,Fraser} we find that, although the off-coincident propagators differ due to different choice of expansion points, at coincidence our propagator in equations~\eqref{prop0coin} and~\eqref{prop2coin} is equal to the one of the previous literature. This is remarkable since the coincident propagator expanded up to second order in gradients is the starting point for the one-loop effective action calculation. The physics should not depend
on the arbitrary point chosen around which one expands, the effective 
actions calculated with these different propagators should be physically equivalent. Even though establishing independence of physical results on the choice of the referent expansion point
is beyond the scope of this work, an encouraging observation is that the coincident propagators, which can be used 
to calculate the one-loop effective action, are identical.

In this work we used the techniques of expanding the logarithm in the effective action in derivatives and of integrating the logarithm of the propagator exactly, avoiding its expansion.
Surprisingly, these methods produce the gradient corrections to the effective action that differ from the ones obtained
by the methods used in the literature. These methods involve taking a parametric derivative of the action
with respect to the mass parameter or using the Schwinger-DeWitt proper time method.
Up to this point, we were unable to resolve the origin of the discrepancy. For now we just point out that both techniques contain potential pitfalls. Namely, the one-loop effective action is a divergent quantity, so integration over 
the mass parameter involves changing the order of integration, which may not be legitimate.
On the other hand, a na\^{i}ve Taylor expansion of the logarithm function may produce a wrong answer if the original expression is a trans-series. Even though the preliminary results presented in this work are not completely satisfactory, we have 
decided to make them public, in the hope to promote a discussion on that question.

\section{Two-loop calculations}  
\label{Two-loop calculations}

We present here in more details the procedure we adopted to calculate the two-loop contributions to the effective action~(\ref{two loop effective action}).

The zeroth order contribution in gradients is obtained by inserting~(\ref{prop0})
into~(\ref{two loop effective action}) and expanding in gradients the vertex functions,
\begin{equation}
\Gamma^{(2,0)}=\Gamma^{(2,0)}_a+\Gamma^{(2,0)}_b
\,,
\label{appendix D: Gamma 20: 0th derivative}
\end{equation}
where the momentum integrals that ought to be solved are:
\begin{equation}
\begin{split}
   \Gamma^{(2,0)}_a=&-\frac{\lambda^2}{12} \int_{X,p,p'} \phi^2(X)
    \frac{1}{p_{\epsilon}^2+m^2(X)}\frac{1}{p_{\epsilon}'^2+m^2(X)} \frac{1}{(p+p')_{\epsilon}^2+m^2(X)},
\end{split}
\label{appendix D: I1a: two loop integral}
\end{equation}
\begin{equation}
   \Gamma^{(2,0)}_b=\frac{\lambda}{8} \int_{X,p,p'}  \frac{1}{p_{\epsilon}^2+m^2(X)}\frac{1}{p_{\epsilon}'^2+m^2(X)}
    \,.
\label{appendix D: I1b: simple integral}
\end{equation}

It is convenient to introduce the following integrals, 
\begin{equation}
I_n = \int \frac{{\rm d}^Dp}{(2\pi)^D}\frac{1}{(p_{\epsilon}^2+m^2)^n}
    = \frac{i}{(4\pi)^\frac{D}{2}}\frac{\Gamma\left(n-\frac{D}{2}\right)}{\Gamma(n)}
\big(m^2\big)^{\frac{D}{2}-n}
\,,\quad (n=1,2,3,\cdots)
\,,\qquad
\label{appendix D: easy integral: In}
\end{equation}
and
\begin{equation}
    I_n^{*}= \int_{p,p'} \frac{1}{(p_{\epsilon}^2+m^2(X))^n}\frac{1}{p_{\epsilon}'^2+m^2(X)} \frac{1}{(p+p')_{\epsilon}^2+m^2(X)}
    \,.
\label{appendix D: hard integral: In*}
\end{equation}
The zeroth-order contributions~(\ref{appendix D: Gamma 20: 0th derivative})
can be simply expressed in terms of these integrals, 
\begin{eqnarray}
    \Gamma^{(2,0)}_a&\!\!=\!\!&-\frac{\lambda^2}{12}\int {\rm d}^DX \phi^2(X)I_1^{*}
\label{appendix D: I1* to Sigma20} 
\\
\Gamma^{(2,0)}_b &\!\!=\!\!&\frac{\lambda}{8}\int {\rm d}^DX I_1^{2}
\,.
\label{appendix D: I1 to Sigma20}
\end{eqnarray}

The derivative suppressed contributions to the two-loop effective 
action~(\ref{two loop effective action}) can be 
neatly split into three contributions. The first comes from 
expanding the vertex around the midpoint,  the second (third) comes from the second order 
contribution to the propagator in the part of the diagram coming 
from the cubic (quartic) vertex,
\begin{equation}
\Gamma^{(2,2)}=\Gamma^{(2,2)}_a+\Gamma^{(2,2)}_b+\Gamma^{(2,2)}_c
\,.
\label{appendix D: splitting Gamma 22}
\end{equation}
By expanding the vertex in~(\ref{two loop effective action}) one obtains,
\begin{eqnarray}
\Gamma^{(2,2)}_a &\!\!=\!\!& \frac{\lambda^2}{24 D} \int_{X,p,p'} \Big[(\partial_X \phi)^2-\phi (\partial^2_X \phi)\Big]
\bigg[\frac{D\!-\!4}{(p_{\epsilon}^2+m^2)^2}+\frac{4m^2(X) }{(p_{\epsilon}^2+m^2)^3}\bigg]
\nonumber\\
    &\!\!\!\!&\hskip 6.5cm
    \times\frac{1}{(p_{\epsilon}'^2+m^2)((p+p')_{\epsilon}^2+m^2)}
\nonumber\\
&\!\!=\!\!&\frac{\lambda^2}{24 D} \int {\rm d}^DX 
\Big[(\partial_X \phi)^2-\phi (\partial^2_X \phi)\Big]
\bigl[(D\!-\!4)I^*_2+4m^2(X) I^*_3\bigr]
    \,.\qquad\;
\label{appendix D: Gamma 22a: vertex}
\end{eqnarray}
Expanding the propagators results in,
\begin{eqnarray}
    \Gamma^{(2,2)}_b &\!\!=\!\!& -\frac{\lambda}{4} \int_{X,p,p'}
    \!\bigg[
 \frac{\frac{\lambda(D-2)}{2D}\big[(\partial_{X}\phi)^2\!+\!\phi (\partial_{X}^2 \phi)\big]}{(p_{\epsilon}^2+m^2(X))^3}
 +\frac{\frac{\lambda m^2}{D}\big[(\partial_{X}\phi)^2\!+\!\phi (\partial_{X}^2 \phi)\big]-\frac{\lambda^2 \phi^2}{2}(\partial_{X}\phi)^2}{(p_{\epsilon}^2+m^2(X))^4}\bigg]
 \nonumber\\
&&\hskip 1.99cm
 \times\,
 \frac{1}{p_{\epsilon}'^2\!+\!m^2}
  \label{appendix D: Gamma 22b: prop} \\
  &\!\!\!\!&\hskip -1.cm
   =-\frac{\lambda}{4} \int_{X}
    \!\bigg\{
\frac{\lambda(D\!-\!2)}{2D}\big[(\partial_{X}\phi)^2\!+\!\phi (\partial_{X}^2 \phi)\big]I_3
 \!+\!\bigg[\frac{\lambda m^2}{D}\big[(\partial_{X}\phi)^2\!+\!\phi (\partial_{X}^2 \phi)\big]-\frac{\lambda^2 \phi^2}{2}(\partial_{X}\phi)^2\bigg]I_4\bigg\}I_1
,
 \nonumber
\end{eqnarray}
and
\begin{eqnarray}
   \Gamma^{(2,2)}_c &\!\!=\!\!& \frac{\lambda^2}{4}\int_{X,p,p'}\!\phi^2 
   \bigg[\frac{\frac{\lambda(D-2)}{2D}\big[(\partial_{X}\phi)^2\!+\!\phi (\partial_{X}^2 \phi)\big]}{(p_{\epsilon}^2+m^2(X))^3}
   +\frac{\frac{\lambda m^2}{D}\big[(\partial_{X}\phi)^2\!+\!\phi (\partial_{X}^2 \phi)\big]
   -\frac{\lambda^2 \phi^2}{2}(\partial_{X}\phi)^2}{(p_{\epsilon}^2+m^2(X))^4}
 \bigg] 
  \nonumber\\
 &&\hskip 2.2cm
 \times\,  \frac{1}{p_{\epsilon}'^2+m^2}\frac{1}{(p+p')_{\epsilon}^2+m^2}
 \label{appendix D: Gamma 22c: prop}\\
&\!\!\!\!&\hskip -1.cm 
 =\frac{\lambda^2}{4}\!\int_{X}\!\lambda\phi^2\bigg\{\! 
   \frac{(D\!-\!2)}{2D}\big[(\partial_{X}\phi)^2\!+\!\phi (\partial_{X}^2 \phi)\big]I^*_3
   \!+\!\bigg[\frac{ m^2}{D}\big[(\partial_{X}\phi)^2\!+\!\phi (\partial_{X}^2 \phi)\big]
   \!-\!\frac{\lambda \phi^2}{2} (\partial_{X}\phi)^2\bigg]  I^*_4
 \! \bigg\}
\,,
\nonumber
\end{eqnarray}
where $I_n$ and $I^*_n\;(n=1,2,3,4)$ 
denote the integrals in~(\ref{appendix D: easy integral: In})
and~(\ref{appendix D: hard integral: In*}), respectively.

\subsection{Evaluation of the integrals}
\label{Evaluation of the integrals}

Using the Feynman parametrisation we can rewrite~(\ref{appendix D: hard integral: In*}) as,
\begin{equation}
    I_n^{*}=\int_{p'}\frac{1}{(p_{\epsilon}'^2+m^2)^n}\int_p \int_0^1 \int_0^1\!{\rm d}u_1 {\rm d}u_2
    \frac{\delta(1\!-\!u_1\!-\!u_2)}{(u_1(p_{\epsilon}^2+m^2)+u_2((p+p')_{\epsilon}^2+m^2))^2}
    \,,
\end{equation}
where we have introduced parameter $n$ for a later convenience.
The integral we are interested in here is obtained simply by setting $n=1$.

Performing the shift $l=p+u_2 p'$ and exploiting the delta function, the previous expression becomes:
\begin{equation}
    \int_{p'}\frac{1}{(p_{\epsilon}'^2+m^2)^n}\int_l \int_0^1  {\rm d}u 
     \frac{1}{\big(l_{\epsilon}^2+u(1-u)p_{\epsilon}'^2+m^2\big)^2}
    \,,
\end{equation}
we can now perform the integral over the momentum $l$:
\begin{equation}
  I_n^{*}=   \frac{\Gamma\big(2-\frac{D}{2}\big)}{(4 \pi)^{D/2}} i \int_{p'} \int_0^1 {\rm d}u 
  \big[u(1\!-\!u)\big]^{\frac{D}{2}-2}  \frac{\big(p'^2\!+\!\frac{m^2}{u(1\!-\!u)}\big)^{\frac{D}{2}-2}}{(p_{\epsilon}'^2\!+\!m^2)^n}
    \,.
\end{equation}
The integral over $p'$ is now a standard momentum integral and one can solve it using spherical coordinates, and recalling that the volume of the $(D\!-\!1)$-dimensional unit sphere is $\Omega_{D-1}=2\pi^{D/2}/\Gamma(D/2)$. For the radial part, we use the following expression,
\begin{equation}
    \int_0^{\infty}\!\! {\rm d}x \; x^{\nu-1}(\beta+x)^{-\mu}(x+\gamma)^{-\rho}
    =\beta^{-\mu}\gamma^{\nu-\rho}
    \frac{\Gamma(\nu)\Gamma(\mu\!-\!\nu\!+\!\rho)}{\Gamma(\mu\!+\!\rho)}
     \times\hypgeo{2}{1}\Big(\mu,\nu;\mu+\rho;1-\frac{\gamma}{\beta}\Big)
.\quad
\end{equation}
Performing the $p'$ integral in this way, one finds,
 \begin{equation}
 \begin{split}
     I_n^{*}=&-\frac{\Gamma\big(2-\frac{D}{2}\big)\Gamma(n+2-D)}
         {(4 \pi)^D\Gamma\big(n+2-\frac{D}{2}\big)}(m^2)^{D-2-n}\\
     &\times \int_0^1 {\rm d}u \hypgeo{2}{1}\bigg(\frac{D}{2},2-\frac{D}{2},n+2-\frac{D}{2},1-u(1\!-\!u)\bigg)
     \; .\quad
\label{appendix D: I1*} 
\end{split}
 \end{equation}
By making use of Eq.~(9.131.2) in~\cite{Gradshteyn:2007} this can be transformed into,
\begin{eqnarray}
     I_n^{*}&\!=\!&-\frac{(m^2)^{D-2-n}}{(4 \pi)^D}
     \int_0^1 {\rm d}u 
     \Bigg[\frac{\Gamma\Big(2-\frac{D}{2}\Big)\Gamma\Big(n-\frac{D}{2}\Big)}{\Gamma(n)}
      \times\hypgeo{2}{1}\bigg(\frac{D}{2},2\!-\!\frac{D}{2};\frac{D}{2}+1\!-\!n;u(1\!-\!u)\bigg)
     \nonumber\\
     &\!+\!&\frac{\Gamma(n+2-D)\Gamma\big(\frac{D}{2}\!-\!n\big)}{\Gamma\big(\frac{D}{2}\big)}
     \big[u(1\!-\!u)\big]^{n-\frac{D}{2}}
     \times\hypgeo{2}{1}\bigg(n,n\!+\!2\!-\!D;n\!+\!1\!-\!\frac{D}{2};u(1\!-\!u)\bigg)
     \Bigg].\qquad\;
\label{appendix D: I1*: 2} 
\end{eqnarray}
The hypergeometric functions can be expanded in powers of $u(1\!-\!u)$, after which 
the integral over $u$ can be performed, resulting in two sums, both of which 
can be summed to (generalized) hypergeometric functions, 
\begin{eqnarray}
     I_n^{*}&\!=\!&-\frac{(m^2)^{D-2-n}}{(4 \pi)^D} \Bigg\{
\frac{\Gamma\big(2-\frac{D}{2}\big) \Gamma\big(n-\frac{D}{2}\big)}{\Gamma(n)}
   \times\hypgeo{3}{2}\bigg(1,\frac{D}{2},2-\frac{D}{2};\frac{D}{2}+1-n,\frac32;\frac14\bigg)
     \nonumber\\
     &\!\!+\!\!&\hskip 0cm
     \frac{\Gamma\big(\frac{n+2-D}{2}\big)\Gamma\big(\frac{n+3-D}{2}\big)
        \Gamma\big(\frac{D}{2}\!-\!n\big) \Gamma\big(1\!+\!n\!-\!\frac{D}{2}\big)}
          {2^n\Gamma\big(\frac{D}{2}\big)\Gamma\big(\frac{3-D}{2}\!+\!n\big)}
          \times\hypgeo{2}{1}\bigg(n,2\!+\!n\!-\!D;\frac{3\!-\!D}{2}\!+\!n;\frac14\bigg)
     \Bigg \} 
,\qquad\;\;
\label{appendix D: I1*: 3n} 
\end{eqnarray}
where we also used Legendre's duplication formula, 
 $\Gamma(2z) = \frac{2^{2z-1}}{\sqrt{\pi}}\Gamma(z)\Gamma\big(z+\frac12\big)$.
 
 In the remainder of this subsection we show how to expand $I_1^*$, $I_2^*$,
 $I_3^*$, and  $I_4^*$ around the pole at $D=4$ to the order needed for 
 dimensional regularization of the effective action.
 
 \bigskip
 {$\mathbf{I_1^*}$.}
For $n=1$ Eq.~(\ref{appendix D: I1*: 3n}) simplifies to, 
\begin{eqnarray}
     I_1^{*}&\!=\!&-\frac{(m^2)^{D-3}}{(4 \pi)^D}
 \Gamma\Big(1-\frac{D}{2}\Big) \Gamma\Big(2-\frac{D}{2}\Big)
    \Bigg[\hypgeo{2}{1}\bigg(1,2-\frac{D}{2};\frac32;\frac14\bigg)
     \nonumber\\
     &\!\!&\hskip 6cm
     +\,\frac{1}{D-3}\times\hypgeo{2}{1}\bigg(1,3-D;\frac{5-D}{2};\frac14\bigg)
     \Bigg]   
\,.\qquad
\label{appendix D: I1*: 3} 
\end{eqnarray}

 The prefactor in~(\ref{appendix D: I1*: 3})  is quadratically divergent, so we need 
 to expand the hypergeometric functions in~(\ref{appendix D: I1*: 3})
 to ${\cal O}\big((D\!-\!4)^2\big)$. It is convenient to first rewrite 
 these hypergeometric functions as, 
\begin{eqnarray}
 \hypgeo{2}{1}\bigg(1,2\!-\!\frac{D}{2};\frac32;\frac14\bigg)
 &\!\!=\!\!&   1 - \frac{D-4}{12}\times\hypgeo{2}{1}\bigg(1,3\!-\!\frac{D}{2};\frac52;\frac14\bigg)
 \label{appendix D: rewriting 2F1} \\
\hypgeo{2}{1}\bigg(1,3\!-\!D;\frac{5-D}{2};\frac14\bigg)
 &\!\!=\!\!& 1 \!-\! \frac{D-3}{2(5-D)}
 \!+\!\frac{(D-3)(D-4)}{4(5-D)(7-D)}
 \times\hypgeo{2}{1}\bigg(1,5\!-\!D;\frac{9-D}{2};\frac14\bigg)
,\qquad
   \nonumber
 \end{eqnarray}
 such that the hypergeometric functions on the right-hand side need to be expanded only to 
 linear order in $D\!-\!4$, 
\begin{eqnarray}
\hypgeo{2}{1}\bigg(1,3\!-\!\frac{D}{2};\frac52;\frac14\bigg)
 &\!\!=\!\!&  3\sqrt{\pi}\sum_{n=1}^\infty
         \frac{\Gamma(n)}{\Gamma\big(n+\frac32\big)}
 \left(\frac{1}{4}\right)^n\left[1\!-\!\frac{D\!-\!4}{2}\left(\psi(n)\!-\!\psi(1)\right)\right]
 \!+\!{\cal O}\big((D\!-\!4)^2\big),
     \nonumber\\
\hypgeo{2}{1}\bigg(1,5\!-\!D;\frac{9-D}{2};\frac14\bigg)
 &\!\!=\!\!& 3\sqrt{\pi}\sum_{n=1}^\infty
         \frac{\Gamma(n)}{\Gamma\big(n+\frac32\big)}
 \left(\frac{1}{4}\right)^n
 \nonumber\\
  &\!\!\!\!& \hskip -3.5cm
  \times\left[1\!-\!(D\!-\!4)\left(\psi(n)\!-\!\psi(1)\right)
  \!+\!\frac{D\!-\!4}{2}\left(\psi\Big(n\!+\!\frac32\Big)\!-\!\psi\Big(\frac32\Big)\!-\!\frac23\right)\right]
 \!+\!{\cal O}\big((D\!-\!4)^2\big).
\qquad
\label{appendix D: rewriting 2F1 2}
 \end{eqnarray}
 Inserting these results into~(\ref{appendix D: I1*: 3}), the square brackets 
 in~(\ref{appendix D: I1*: 3}) can be written as, 
\begin{eqnarray}
\Bigg[\cdots
     \Bigg]  &\!=\!&\frac32 - \frac{3}{2}(D\!-\!4)
     +\frac{(D\!-\!4)^2}{2}
  \nonumber\\
  &\!\!\!\!& \hskip -1.2cm
  \times    
    \Bigg\{1 + \frac{\sqrt{\pi}}{4}\sum_{n=1}^\infty
         \frac{\Gamma(n)}{\Gamma\big(n+\frac32\big)}
 \left(\frac{1}{4}\right)^n\bigg[2 -\Big(\psi(n)-\psi(1)\Big)
              + \left(\psi\Big(n\!+\!\frac32\Big)\!-\!\psi\Big(\frac32\Big)\right)\bigg]
     \Bigg\}
\,.\qquad
\label{appendix D: I1*: 4} 
\end{eqnarray}
 It is now convenient to introduce a finite constant, 
 \begin{eqnarray}
 c_1^* &\!\!\equiv\!\!&  \frac{\sqrt{\pi}}{4}\sum_{n=1}^\infty
         \frac{\Gamma(n)}{\Gamma\big(n+\frac32\big)}
 \left(\frac{1}{4}\right)^n\bigg[2 -\Big(\psi(n)-\psi(1)\Big)
              + \left(\psi\Big(n\!+\!\frac32\Big)\!-\!\psi\Big(\frac32\Big)\right)\bigg]
\nonumber\\
             &\!\! \simeq \!\!& 0.24206957098290097
\,.
 \label{appendix D: c1*}
 \end{eqnarray}
 When these results are combined with, 
\begin{equation}
\Gamma\Big(1\!-\!\frac{D}{2}\Big)\Gamma\Big(2\!-\!\frac{D}{2}\Big)
   \simeq -\frac{4}{(D-4)^2}+\frac{2}{D-4}\big(1+2\psi(1)\big)
   -\bigg(1  + 2 \psi(1) +2 \psi^2(1) +\frac{\pi^2}{6}\bigg)
\,.
\end{equation}
and
\begin{equation}
 \frac{(m^2)^{D-3}}{(4 \pi)^D} \simeq \frac{m^2 \mu^{2(D-4)}}{(4\pi)^4}
 \left[1+(D-4)\ln\left(\frac{m^2}{4\pi\mu^2}\right)
   + \frac{(D-4)^2}{2}\ln^2\left(\frac{m^2}{4\pi\mu^2}\right)\right]
 \,.
\end{equation}
 one finally obtains,
\begin{eqnarray}
     I_1^{*}&\!\!=\!\!&\frac{6m^2}{(4 \pi)^4}
     \Bigg\{\frac{\mu^{2(D-4)}}{(D-4)^2}
     +\frac{\mu^{D-4}}{D-4}\left[\ln\left(\frac{m^2}{4\pi\mu^2}\right)\!+\!\gamma_E 
            \!-\!\frac32\right]
  \nonumber\\
  &\!\!\!\!& \hskip 1.2cm
    +\frac12\left[\ln^2\left(\frac{m^2}{4\pi\mu^2}\right)
    -(3\!-\!2\gamma_E)\ln\left(\frac{m^2}{4\pi\mu^2}\right)
     \!+\!\tilde c_1^* 
            \right]
             \Bigg\}
\!+\!{\cal O}\big(D\!-\!4\big)
\,,\qquad
\label{appendix D: I1*: 5} 
\end{eqnarray}
 where the constant $\tilde c_1^*$ is given by,
\begin{eqnarray}
\tilde c_1^* = \gamma_E^2 \!-\!3\gamma_E +\frac{\pi^2}{12}+\frac{13}{6}
 +\frac{2}{3}c_1^*
    \simeq 1.7520443431825004\,,
\label{appendix D: tilde c1*}
\end{eqnarray} 
 where we opted for a numerical representation of $c_0$, even though
the first two sums in~(\ref{appendix D: c1*}) can be rather easily evaluated.

 \bigskip
 {$\mathbf{I_2^*}$.}
For $n=2$ Eq.~(\ref{appendix D: I1*: 3n}) reduces to, 
\begin{eqnarray}
     I_2^{*}&\!=\!&-\frac{(m^2)^{D-4}}{(4 \pi)^D}
 \Gamma^2\Big(2-\frac{D}{2}\Big)
    \Bigg[\hypgeo{3}{2}\bigg(1,\frac{D}{2},2-\frac{D}{2};\frac{D}{2}-1,\frac32;\frac14\bigg)
     \nonumber\\
     &\!\!&\hskip 5cm
     -\,\frac{1}{(D-2)(5-D)}\times\hypgeo{2}{1}\bigg(2,4-D;\frac{7-D}{2};\frac14\bigg)
     \Bigg]   
\,.\qquad
\label{appendix D: I2*: 3} 
\end{eqnarray}
 The prefactor in~(\ref{appendix D: I2*: 3})  is quadratically divergent, so we need 
 to expand the hypergeometric functions in~(\ref{appendix D: I2*: 3})
 to ${\cal O}\big((D\!-\!4)^2\big)$. We begin with
\begin{eqnarray}
 \hypgeo{3}{2}\bigg(1,\frac{D}{2},2-\frac{D}{2};\frac{D}{2}-1,\frac32;\frac14\bigg)
 &\!\!=\!\!&   1 - \frac{D(D-4)}{12(D-2)}\times\hypgeo{3}{2}\bigg(1,\frac{D}{2}+1,3-\frac{D}{2};\frac{D}{2},\frac52;\frac14\bigg)\qquad\;\;
 \label{appendix D: I2*: rewriting 2F1} \\
\hypgeo{2}{1}\bigg(2,4-D;\frac{7-D}{2};\frac14\bigg)
 &\!\!=\!\!& 1 \!-\! \frac{D-4}{7-D}
 \times\hypgeo{3}{2}\bigg(1,3,5-D;2,\frac{9-D}{2};\frac14\bigg)
,\qquad
   \nonumber
 \end{eqnarray}
 such that the hypergeometric functions on the right-hand side need to be expanded only to 
 linear order in $D\!-\!4$, 
\begin{eqnarray}
\hypgeo{3}{2}\bigg(1,\frac{D}{2}+1,3-\frac{D}{2};\frac{D}{2},\frac52;\frac14\bigg)
 &\!\!=\!\!&  \frac{3\sqrt{\pi}}{2}\sum_{n=1}^\infty
         \frac{\Gamma(n)(n+1)}{\Gamma\big(n+\frac32\big)}
 \left(\frac{1}{4}\right)^n \!+\!{\cal O}(D\!-\!4),
     \nonumber\\
\hypgeo{3}{2}\bigg(1,3,5-D;2,\frac{9-D}{2};\frac14\bigg)
 &\!\!=\!\!&  \frac{3\sqrt{\pi}}{2}\sum_{n=1}^\infty
         \frac{\Gamma(n)(n+1)}{\Gamma\big(n+\frac32\big)}
 \left(\frac{1}{4}\right)^n \!+\!{\cal O}(D\!-\!4),
\qquad
\label{appendix D: I2*: rewriting 2F1 2}
 \end{eqnarray}

Inserting these results into~(\ref{appendix D: I2*: 3}) gives,
\begin{eqnarray}
     I_2^{*}&\!=\!&-\frac{2}{(4 \pi)^4}
    \Bigg[\frac{\mu^{2(D-4)}}{(D-4)^2}
    + \frac{\mu^{D-4}}{D-4}\bigg[
      \ln\left(\frac{m^2}{4\pi\mu^2}\right) +\gamma_E -\frac12 \bigg] 
      \Bigg]
       \!+\!{\cal O}\big((D\!-\!4)^0\big)
\,,\qquad
\label{appendix D: I2*: 4} 
\end{eqnarray}
 where we have not included the ${\cal O}\big((D\!-\!4)^0\big)$ contribution as we do not need it.

\vskip 0.3cm

 \bigskip
 {$\mathbf{I_3^*}$.}
For $n=3$ Eq.~(\ref{appendix D: I1*: 3n}) reduces to, 
\begin{eqnarray}
     I_3^{*}&\!=\!&-\frac{(m^2)^{D-5}}{(4 \pi)^D}
 \frac{\Gamma\Big(2-\frac{D}{2}\Big) \Gamma\Big(3-\frac{D}{2}\Big)}{2}
    \Bigg[\hypgeo{3}{2}\bigg(1,\frac{D}{2},2-\frac{D}{2};\frac{D}{2}-2,\frac32;\frac14\bigg)
     \nonumber\\
     &\!\!&\hskip 3.5cm
     +\,\frac{2}{(D-2)(5-D)(7-D)}\times\hypgeo{2}{1}\bigg(3,5-D;\frac{9-D}{2};\frac14\bigg)
     \Bigg]   
\,.\qquad
\label{appendix D: I3*: 3} 
\end{eqnarray}
 The prefactor in~(\ref{appendix D: I3*: 3})  is linearly divergent, so we need 
 to expand the hypergeometric functions in~(\ref{appendix D: I3*: 3})
 to ${\cal O}\big((D\!-\!4)\big)$. We begin with,
\begin{eqnarray}
 \hypgeo{3}{2}\bigg(1,\frac{D}{2},2-\frac{D}{2};\frac{D}{2}-2,\frac32;\frac14\bigg)
 &\!\!=\!\!&   1 - \frac{D}{12}\times\hypgeo{3}{2}\bigg(1,\frac{D}{2}+1,3-\frac{D}{2};\frac{D}{2}-1,\frac52;\frac14\bigg)\;,
 \label{appendix D: I3*: rewriting 2F1}
   \nonumber
 \end{eqnarray}
 such that the hypergeometric functions on the right-hand side of~(\ref{appendix D: I3*: 3}) 
 need to be expanded to linear order in $D\!-\!4$, 
\begin{eqnarray}
\hypgeo{3}{2}\bigg(1,\frac{D}{2}+1,3-\frac{D}{2};\frac{D}{2}-1,\frac52;\frac14\bigg)
 &\!\!=\!\!&  \frac{3\sqrt{\pi}}{2}\sum_{n=1}^\infty
         \frac{\Gamma(n+2)}{\Gamma\big(n+\frac32\big)}
 \left(\frac{1}{4}\right)^n
 \nonumber\\
 &\!\!\!\!&\hskip -3cm
 \times\left[1\!+\!\frac{D\!-\!4}{2}\left(\psi(2\!+\!n)\!-\!\psi(2)
            \!-\!\frac12\!-\!2\psi(n)\!+\!2\psi(1)\right)\right]
 \!+\!{\cal O}\big((D\!-\!4)^2\big),
     \nonumber\\
\hypgeo{2}{1}\bigg(3,5-D;\frac{9-D}{2};\frac14\bigg)
&\!\!=\!\!& \frac{3\sqrt{\pi}}{2}\sum_{n=1}^\infty
         \frac{\Gamma(n+2)}{\Gamma\big(n+\frac32\big)}
 \left(\frac{1}{4}\right)^n
 \nonumber\\
  &\!\!\!\!& \hskip -4.8cm
  \times\left[1
  \!+\!\frac{D\!-\!4}{2}\left(-2\psi(n)\!+\!2\psi(1)\!+\!\psi\Big(n\!+\!\frac32\Big)\!-\!\psi\Big(\frac32\Big)\!-\!\frac23\right)\right]
 \!+\!{\cal O}\big((D\!-\!4)^2\big).
\qquad
\label{appendix D: I3*: rewriting 2F1 2}
 \end{eqnarray}
Upon inserting these into~(\ref{appendix D: I3*: 3}) gives, 
\begin{eqnarray}
     I_3^{*}&\!\!=\!\!&\frac{(m^2)^{-1}\mu^{D-4}}{(4 \pi)^4}
    \Bigg\{\frac{\mu^{D-4}}{D\!-\!4} 
    +\bigg[\ln\left(\frac{m^2}{4\pi\mu^2}\right)+\gamma_E + c_3^*
    \bigg] \Bigg\}
\,,\qquad\;
\label{appendix D: I3*: 4}
\end{eqnarray}
 where we have introduced a constant $c_3^*$ ({\it cf.} Eq.~(\ref{appendix D: c1*})) defined by,
 \begin{eqnarray}
 c_3^* &\!\!\equiv\!\!&  \frac{\sqrt{\pi}}{4}\sum_{n=1}^\infty
         \frac{\Gamma(n+2)}{\Gamma\big(n+\frac32\big)}
 \left(\frac{1}{4}\right)^n 
 \bigg(\psi\Big(\frac32\!+\!n\Big)\!-\!\psi\Big(\frac32\Big)\!-\!\big(\psi(2\!+\!n)\!-\!\psi(2)\big)
    \!+\!1\bigg)
\nonumber\\
             &\!\! \simeq \!\!& 0.2813024128985736
\,.
 \label{appendix D: c3*}
 \end{eqnarray}

\vskip 0.3cm
 \bigskip
 {$\mathbf{I_4^*}$.}
For $n=4$ Eq.~(\ref{appendix D: I1*: 3n}) reduces to, 
\begin{eqnarray}
     I_4^{*}&\!=\!&-\frac{(m^2)^{D-6}}{(4 \pi)^D}
 \frac{\Gamma\Big(2-\frac{D}{2}\Big) \Gamma\Big(4-\frac{D}{2}\Big)}{6}
    \Bigg[\hypgeo{3}{2}\bigg(1,\frac{D}{2},2-\frac{D}{2};\frac{D}{2}-3,\frac32;\frac14\bigg)
     \nonumber\\
     &\!\!&\hskip 2.cm
  -\,\frac{6}{(D-2)(6-D)(7-D)(9-D)}\times\hypgeo{2}{1}\bigg(4,6-D;\frac{11-D}{2};\frac14\bigg)
     \Bigg]   
\,.\qquad\;
\label{appendix D: I4*: 3} 
\end{eqnarray}

 The prefactor in~(\ref{appendix D: I4*: 3})  is linearly divergent, so we need 
 to expand the hypergeometric functions in~(\ref{appendix D: I4*: 3})
 to ${\cal O}\big((D\!-\!4)\big)$. We begin with,
\begin{eqnarray}
 \hypgeo{3}{2}\bigg(\!1,\frac{D}{2},2\!-\!\frac{D}{2};\frac{D}{2}\!-\!3,\frac32;\frac14\bigg)
 &\!\!=\!\!&   1 \!+\!\frac{D(D\!-\!4)}{12(6\!-\!D)} \!+\! \frac{D(D\!+\!2)}{15\times 16}
\! \times\!\hypgeo{3}{2}\bigg(\!1,\frac{D}{2}\!+\!2,4\!-\!\frac{D}{2};\frac{D}{2}\!-\!1,\frac72;\frac14\bigg),\qquad
 \label{appendix D: I4*: rewriting 2F1}
   \nonumber
 \end{eqnarray}

 such that the hypergeometric functions on the right-hand side of~(\ref{appendix D: I4*: 3}) 
can be expanded as, 
\begin{eqnarray}
\hypgeo{3}{2}\bigg(1,\frac{D}{2}+2,4-\frac{D}{2};\frac{D}{2}-1,\frac72;\frac14\bigg)
 &\!\!=\!\!&  \frac{5\sqrt{\pi}}{4}\sum_{n=1}^\infty
         \frac{\Gamma(n+3)n}{\Gamma\big(n+\frac52\big)}
 \left(\frac{1}{4}\right)^n
 \nonumber\\
 &\!\!\!\!&\hskip -5cm
 \times\left[1\!+\!\frac{D\!-\!4}{2}\left(\psi(3\!+\!n)\!-\!\psi(3)
            \!-\!\psi(n\!+\!1)\!-\!\psi(n)\!+\!2\psi(1)\!+\!\frac23\right)\right]
 \!+\!{\cal O}\big((D\!-\!4)^2\big),
     \nonumber\\
\hypgeo{2}{1}\bigg(4,6-D;\frac{11-D}{2};\frac14\bigg)
&\!\!=\!\!& \frac{5\sqrt{\pi}}{4}\sum_{n=1}^\infty
         \frac{\Gamma(n+3)n}{\Gamma\big(n+\frac52\big)}
 \left(\frac{1}{4}\right)^n
 \nonumber\\
  &\!\!\!\!& \hskip -4.8cm
  \times\left[1
  \!+\!\frac{D\!-\!4}{2}\left(\!-2\psi(n\!+\!1)\!+\!2\psi(1)\!+\!\psi\Big(n\!+\!\frac52\Big)\!-\!\psi\Big(\frac52\Big)\!+\!\frac85\right)\right]
 \!+\!{\cal O}\big((D\!-\!4)^2\big).
\qquad
\label{appendix D: I4*: rewriting 2F1 2}
 \end{eqnarray}

Inserting these into~(\ref{appendix D: I4*: 3}) gives, 
\begin{eqnarray}
     I_4^{*}&\!\!=\!\!&\frac{(m^2)^{-2}\mu^{D-4}}{3(4 \pi)^4}
    \Bigg\{\frac{\mu^{D-4}}{D\!-\!4} 
    +\bigg[\ln\left(\frac{m^2}{4\pi\mu^2}\right)\!+\!\gamma_E \!-\! \frac13\!+\! c_4^*
 \bigg] \Bigg\}
\,.\qquad\;
\label{appendix D: I4*: 4} 
\end{eqnarray}
  where we have introduced a new constant,
 \begin{eqnarray}
 c_4^* &\!\!\equiv\!\!& \frac{\sqrt{\pi}}{16}
    \sum_{n=1}^\infty \frac{\Gamma(3+n)n}{\Gamma\big(\frac52+n\big)}\left(\frac14\right)^n
    \bigg(\big(\psi(3\!+\!n)\!-\!\psi(3)\big)
        \!-\!\Big(\psi\Big(\frac52\!+\!n\Big)\!-\!\psi\Big(\frac52\Big)\Big)\!+\!\frac1n
        \!-\!\frac{7}{6}\bigg)
    \bigg]
\nonumber\\
             &\!\! \simeq \!\!& -0.0520309
\,.
 \label{appendix D: c4*}
 \end{eqnarray}

\subsection{Leading order contributions to the 2-loop effective action}
\label{Leading order contributions to the 2-loop effective action}

There are two contributions -- (\ref{appendix D: I1a: two loop integral}) 
and~(\ref{appendix D: I1b: simple integral}) -- to the effective action 
at zeroth order in gradients~(\ref{appendix D: Gamma 20: 0th derivative}), 
which have a simple representation (\ref{appendix D: I1* to Sigma20})--(\ref{appendix D: I1 to Sigma20})  in terms of the integrals  evaluated above
 in Appendix~\ref{Evaluation of the integrals}.

Upon inserting~(\ref{appendix D: I1*: 5}) into~(\ref{appendix D: I1* to Sigma20})
 one immediately gets  the first contribution $\Gamma^{(2,0)}_a$, 
 \begin{eqnarray}
     \Gamma^{(2,0)}_a&\!\!=\!\!&-\frac{\lambda}{2(4 \pi)^4}\int {\rm d}^DX 
     \lambda\phi^2(X)m^2
     \Bigg\{\frac{\mu^{2(D-4)}}{(D-4)^2}
     +\frac{\mu^{D-4}}{D-4}\left[\ln\left(\frac{m^2}{4\pi\mu^2}\right)\!+\!\gamma_E 
            \!-\!\frac32\right]
  \nonumber\\
  &\!\!\!\!& \hskip -.6cm
    \!+\,\frac12\left[\ln^2\!\left(\frac{m^2}{4\pi\mu^2}\right)
   \! -\!(3\!-\!2\gamma_E)\ln\left(\frac{m^2}{4\pi\mu^2}\right)
 \!+\!\gamma_E^2 \!-\!3\gamma_E\!+\!\frac{\pi^2}{12}\!+\!\frac{13}{6}\!+\!\frac{2}{3}c_1^*
            \right]
             \!\Bigg\}
\!\!+\!{\cal O}\big(D\!-\!4\big)
,\qquad\;
\label{appendix D: Gamma 20a final} 
\end{eqnarray}
 where $c_1^*$ 
  is the numerical constant given in~(\ref{appendix D: c1*}).
 
 The second contribution is obtained by inserting~(\ref{appendix D: easy integral: In}) into~(\ref{appendix D: I1 to Sigma20}), 
\begin{eqnarray}
\Gamma^{(2,0)}_b &\!\!=\!\!& -\frac{\lambda}{8}\int {\rm d}^DX \frac{(m^2)^{D-2}}{(4\pi)^D}
    \Gamma^2\Big(1-\frac{D}{2}\Big)
\nonumber\\
&\!\!=\!\!& -\frac{\lambda}{2(4\pi)^4}\int {\rm d}^DX m^4(X)
\Bigg\{\frac{\mu^{2(D-4)}}{(D-4)^2}
+\frac{\mu^{D-4}}{D-4}\bigg[\ln\left(\frac{m^2}{4\pi\mu^2}\right)\!+\!\gamma_E 
            \!-\!1\bigg]
\label{appendix D: Gamma 20b final} \\
&\!\!\!\!&
+ \frac12\bigg[\ln^2\left(\frac{m^2}{4\pi\mu^2}\right)
 \!-\!2(1-\gamma_E)\ln\left(\frac{m^2}{4\pi\mu^2}\right)
 +\gamma_E^2-2\gamma_E+\frac{\pi^2}{12}+\frac{3}{2}\bigg]            
\Bigg\}
\nonumber
\!+\!{\cal O}\big(D\!-\!4\big)
\,.\qquad
\end{eqnarray} 

\subsection{Gradient corrections to the 2-loop effective action}

We are now ready to calculate the two-derivative contributions to the two-loop effective action.
These corrections can be split into three parts~(\ref{splitting Gamma 22}): 
\begin{equation}
\Gamma^{(2,2)}=\Gamma^{(2,2)}_a+\Gamma^{(2,2)}_b+\Gamma^{(2,2)}_c
\,.
\label{appendix D: splitting Gamma 22}
\end{equation}
given in Eqs.~(\ref{appendix D: Gamma 22a: vertex}),
 (\ref{appendix D: Gamma 22b: prop}) and~(\ref{appendix D: Gamma 22c: prop}),
 respectively, where they are written in terms of the integrals we have evaluated above in 
Appendix~\ref{Evaluation of the integrals}. 

Let us first consider the first contribution $\Gamma^{(2,2)}_a$. 
Making use of (\ref{appendix D: I3*: 4}) and~(\ref{appendix D: I3*: 4}) 
in~(\ref{appendix D: Gamma 22a: vertex}) one obtains, 
\begin{eqnarray}
\Gamma^{(2,2)}_a 
&\!\!=\!\!&\frac{\lambda^2 \mu^{D-4}}{48(4 \pi)^4} \int {\rm d}^DX 
\Big[(\partial_X\phi)^2 -\phi (\partial_X^2\phi)\Big]
\left(\frac{\mu^{D-4}}{D-4}+\log\bigg(\frac{m^2}{4\pi\mu^2}\bigg)+\gamma_E +\frac14 + 2c_3^*\right)
    ,\qquad\;\;
\label{appendix D: Gamma 22a: final}
\end{eqnarray}
%

Next we consider $\Gamma^{(2,2)}_b$.
Making use of Eq.~(\ref{appendix D: easy integral: In})
 in  (\ref{appendix D: Gamma 22b: prop}) gives, 
\begin{eqnarray}
   \Gamma^{(2,2)}_b &\!\!=\!\!& \frac{\lambda^2}{24(4\pi)^D}
   \Gamma\left(\!1\!-\!\frac{D}{2}\right)\Gamma\left(\!3\!-\!\frac{D}{2}\right)
\nonumber\\
       &\!\!\times\!\!&       
       \int {\rm d}^DX\big(m^2\big)^{D-4}
       \Bigg\{\Big[(\partial_X\phi)^2 \!+\!\phi(\partial_X^2\phi)\Big]
                -\frac{(6\!-\!D)}{2}\frac{\lambda\phi^2}{2m^2}(\partial_X\phi)^2
               \Bigg\}
\,.
\label{appendix D: Gamma 22b: primitive}
\end{eqnarray}
This is linearly divergent in $(D\!-\!4)$,~\footnote{This is a general feature of perturbation theory.
Namely, at 
 $n$ loops, the leading order divergence in the effective potential is 
of the type 
$\sim 1/(D-4)^n$, 
such that the corresponding leading order logarithm is
$\ln^n\big(m^2/(4\pi\mu^2)\big)$. 
On the other hand, the leading 
divergence to the two-derivative (kinetic) term is  
$\sim 1/(D\!\!-4)^{n-1}$,
implying that the leading order logarithm is 
$\ln^{n-1}\big(m^2/(4\pi\mu^2)\big)$.
We suspect that this pattern continues for higher order derivative contributions to the effective
action.
} 
and so the expansion around the pole is straightforward,
\begin{equation}
    \begin{split}
          \Gamma^{(2,2)}_b = \frac{\lambda^2\mu^{D-4}}{12(4\pi)^4}\int {\rm d}^DX
&\Bigg\{\Bigg[\frac{\mu^{D-4}}{D\!-\!4}
+\!\ln\left(\!\frac{m^2}{4\pi\mu^2}\!\right)\!-\!\psi(1)\!-\!\frac12\Bigg]\\ 
  &\times\! \left[\Big[(\partial_X\phi)^2 \!+\!\phi(\partial_X^2\phi)\Big]
     \!-\!\frac{\lambda\phi^2}{2m^2}(\partial_X\phi)^2\right]
\!+\!\frac{\lambda\phi^2}{4m^2}(\partial_X\phi)^2
    \!\Bigg\}
.\qquad 
\label{appendix D: Gamma 22b: final}
    \end{split}
\end{equation}

Finally, to calculate  $\Gamma^{(2,2)}_c$ one inserts 
(\ref{appendix D: I3*: 4}  and~(\ref{appendix D: I3*: 4})
  into~(\ref{appendix D: Gamma 22c: prop}) to obtain, 
\begin{eqnarray}
   \Gamma^{(2,2)}_c &\!\!=\!\!& 
 \frac{\lambda^2\mu^{D-4}}{12(4 \pi)^4}\!\int\!{\rm d}^D{X}\,
 \frac{\lambda\phi^2}{m^2}\bigg\{\! 
   \big[(\partial_{X}\phi)^2\!+\!\phi (\partial_{X}^2 \phi)\big] 
   \bigg[\frac{\mu^{D-4}}{D\!-\!4} 
    \!+\!\bigg(\!\ln\Big(\frac{m^2}{4\pi\mu^2}\Big)\!+\!\gamma_E\!+\!\frac{1}{24}
    \!+\! \frac{3c_3^* \!+\! c_4^*}{4}
    \bigg)\bigg]
    \nonumber\\
  &\!\!\!\!& \hskip 3.2cm
   \!-\,\frac{\lambda \phi^2}{2m^2} (\partial_{X}\phi)^2 \bigg[\frac{\mu^{D-4}}{D\!-\!4} 
    \!+\!\bigg(\!\ln\Big(\frac{m^2}{4\pi\mu^2}\Big)\!+\!\gamma_E \!-\! \frac13\!+\! c_4^*
 \bigg) \bigg]
 \! \bigg\}
\,,\qquad\;
 \label{appendix D: Gamma 22c: final}
\end{eqnarray}
where the constants $c_3^*$ and $c_4^*$ are defined in~(\ref{appendix D: c3*})
and~(\ref{appendix D: c4*}), respectively.

Summing all three contributions yields, 
\begin{eqnarray}
   \Gamma^{(2,2)} &\!\!=\!\!& 
 \frac{\lambda^2\mu^{D-4}}{12(4 \pi)^4}\!\int\!\!{\rm d}^D{X}\!
\Bigg\{\! \Bigg[
   \bigg[(\partial_{X}\phi)^2\!+\!\phi (\partial_{X}^2 \phi)
    \!-\!\frac{\lambda \phi^2}{2m^2} (\partial_{X}\phi)^2\bigg] 
    \bigg(\!1\!+\!\frac{\lambda\phi^2}{m^2}\bigg)
    \!+\! \frac14\bigg[(\partial_{X}\phi)^2\!-\!\phi (\partial_{X}^2 \phi)\bigg] \Bigg]
\nonumber\\    
 &\!\!\!\!& \hskip 2.95cm    
  \times \bigg[\frac{\mu^{D-4}}{D\!-\!4} 
    \!+\!\ln\Big(\frac{m^2}{4\pi\mu^2}\Big)\!+\!\gamma_E\bigg]
\nonumber\\    
 &\!\!\!\!& \hskip .7cm
 +\,\big[(\partial_{X}\phi)^2\!+\!\phi (\partial_{X}^2 \phi)\big]
    \bigg[\!\!-\!\frac{1}{2}\!+\!\frac{\lambda \phi^2}{m^2}
    \Big(\frac1{24}\!+\!\frac{3c_3^*\!+\!c_4^*}{4}\Big)\!\bigg]
    \!+\!\big[(\partial_{X}\phi)^2\!-\!\phi (\partial_{X}^2 \phi)\big]
              \bigg[\!\frac1{16}\!+\!\frac{c_3^*}{2}\bigg]
\nonumber\\    
 &\!\!\!\!& \hskip .7cm
    \!+\, (\partial_{X}\phi)^2  \bigg(\frac{\lambda \phi^2}{m^2} \bigg)
    \bigg[\frac{1}{4}\!+\!\frac{\lambda \phi^2}{m^2}\Big(\frac16\!-\!\frac{c_4^*}{2}\Big)\bigg]
 \! \Bigg\}
.\qquad\;
 \label{appendix D: Gamma 22: final}
\end{eqnarray}

This is the primitive result for the two-derivative contribution to the 2-loop effective action
used in the main text as a starting point for renormalization.


\end{document}